\documentclass[aps,prl,graphicx,twocolumn,a4paper,10pt,superscriptaddress]{revtex4-1}

\usepackage{graphicx}
\usepackage{hyperref}
\usepackage{xcolor}

\begin{document}

\title{Extreme Fermi surface smearing in a maximally disordered concentrated solid solution}

\author{Hannah C. Robarts}
\author{Thomas E. Millichamp}
\author{Daniel A. Lagos}
\author{Jude Laverock}
\affiliation{H.H. Wills Physics Laboratory, University of Bristol, Tyndall Avenue, Bristol BS8 1TL, UK}

\author{David Billington}
\affiliation{Japan Synchrotron Radiation Research Institute, SPring-8, Sayo, 679-5198, Japan}
\affiliation{School of Physics and Astronomy, Cardiff University, Queen's Building, The Parade, Cardiff, CF24 3AA, United Kingdom}

\author{Jonathan A.~Duffy}
\author{Daniel O'Neill}
\affiliation{Department of Physics, University of Warwick, Coventry, CV4 7AL, United Kingdom}

\author{Sean R.~Giblin}
\affiliation{School of Physics and Astronomy, Cardiff University, Queen's Building, The Parade, Cardiff, CF24 3AA, United Kingdom}

\author{Jonathan W.~Taylor}
\affiliation{DMSC - European Spallation Source,  Universitetsparken 1, Copenhagen 2100, Denmark}

\author{Grazyna Kontrym-Sznajd}
\author{Ma{\l}gorzata Samsel-Czeka{\l}a}
\affiliation{Institute of Low Temperature and Structure Research, Polish Academy of Sciences, PO Box 1410, 50-950 Wroc{\l}aw 2, Poland}

\author{Hongbin Bei}
\author{Sai Mu}
\author{German D.~Samolyuk}
\author{G.~Malcolm Stocks}
\affiliation{Materials Science and Technology Division, Oak Ridge National Laboratory, Oak Ridge, TN 37831, USA}

\author{Stephen B.~Dugdale}
\email{s.b.dugdale@bristol.ac.uk}
\affiliation{H.H. Wills Physics Laboratory, University of Bristol, Tyndall Avenue, Bristol BS8 1TL, UK}

\date{\today}
\begin{abstract}
We show that the Fermi surface can survive the presence of extreme compositional disorder in the equiatomic alloy Ni$_{0.25}$Fe$_{0.25}$Co$_{0.25}$Cr$_{0.25}$.
Our high-resolution Compton scattering experiments reveal a Fermi surface which is smeared across a significant fraction of the Brillouin zone (up to 40\% of $\frac{2\pi}{a}$).
The extent of this smearing and its variation on and between different sheets of the Fermi surface has been determined, and estimates
of the electron mean-free-path and residual resistivity have been made by connecting this smearing with the coherence length of the quasiparticle states.
\end{abstract}

\maketitle{}

The emergence of the Fermi surface (FS) from the theory of the electronic structure of metals, together with the pioneering 
experimental determinations of its shape, stand proudly among the greatest achievements of twentieth century physics \cite{hoch:83}. 
The FS, defined by the discontinuity in the momentum distribution, exists even
 for interacting electrons \cite{luttinger:60,huotari:10}, and here we demonstrate its remarkable ability to survive maximal compositional disorder 
in which the electron mean-free-path (which we can also extract from our measurements) is comparable to the lattice spacing.
In disordered systems, the Mott-Ioffe-Regel (MIR) limit describes the semiclassical upper bound for coherent transport in a metal,
occuring when the electron mean-free-path becomes comparable with the interatomic spacing \cite{hussey:04}.
The modern description of the electronic structure of crystalline solids --- the band theory of electrons --- depends on the notion of 
perfect crystals exhibiting long-range order. The Bloch wavefunctions which emerge are a direct consequence of the discrete translational invariance of
the potential experienced by the electrons traveling through the ionic lattice. This premise is strongly
challenged in substitutionally disordered random alloys (concentrated solid solutions) where there is no such periodicity.
Abandoning the familiar concepts associated with a well-defined reciprocal lattice, such as the Brillouin zone (BZ) and indeed
the FS, seems inevitable. 
However, there is considerable theoretical and experimental evidence (e.g. \cite{gyorffy:83,wilkinson:01})
 that by considering an {\it ordered} system comprising suitably chosen effective scatterers
to restore periodicity, the BZ and FS can be resurrected. The resulting electron states, however,
have finite lifetimes due to the presence of disorder, and the ``bands'' are smeared in both energy, $E$ (resulting in a finite electron lifetime) 
and crystal momentum, ${\mathbf k}$ (finite mean-free-path). This also means that the
discontinuity in the momentum distribution associated with the FS is also smeared out in both $E$ and ${\mathbf k}$, with correspondingly
reduced Fermi energy electron lifetimes and short mean-free-paths.  A sharp FS in an ultra-pure metal at cryogenic temperatures is
 associated with electron mean-free-paths of more than a centimeter \cite{yaqub:65}.         
While bulk resistivities of metals are rather well known, comparatively little {\it direct} information exists about
electron mean-free-paths \cite{gall:16}.

A new class of metallic alloys, referred to as ``high entropy alloys'' (with the terms ``multi-principal element alloys'' and ``complex concentrated alloys''
being used in a broader sense \cite{miracle:17}), has recently been introduced.
These alloys consist of approximately equiatomic concentrations of multiple metallic elements and are thus fundamentally different from
traditional alloys which have one principal component.
The Cantor-Wu alloys are particularly interesting examples \cite{cantor:04,wu:14}. 
Whilst these alloys form on a face-centered-cubic lattice and the atoms have relatively small deviations from their ideal lattice positions (small 
displacement fluctuations \cite{song:17}), the chemical (compositional) disorder is considerable leading to a smeared electronic structure \cite{samolyuk:18,mu:18}.
The lattice vibrations are also expected to be scattered by
the disorder, leading to broadening of the phonon linewidths due to a shortened lifetime \cite{kormann:17}. In spite of such disorder, 
superconductivity was recently observed in a refractory (bcc) high entropy alloy \cite{heasuper}, implying the existence of a FS. 

In this Letter we present an experimental measurement of the FS of a maximally disordered 
(in the sense that the ideal entropy of mixing is maximal at equiatomic composition) medium entropy alloy,
equiatomic Ni$_{0.25}$Fe$_{0.25}$Co$_{0.25}$Cr$_{0.25}$ (henceforth referred to as NiFeCoCr), using high-resolution Compton scattering. 
Compton scattering is a bulk-sensitive probe of the occupied momentum states in a solid and is thus a particularly powerful technique for measuring the 
FSs of disordered alloys, it not being limited by short electron mean-free-paths \cite{dugdale:14}. 
We show that a meaningful FS can be identified, in spite of it being smeared across a substantial fraction of the BZ, and that studying the smeared-out FS discontinuity can be used
 to deduce the electron mean-free-path. Furthermore, we show that state-of-the-art theory for disordered systems, which generalizes the notion
of the FS to systems dominated by disorder, is able to predict correctly the FS topology.

\begin{figure}
        \begin{center}
        \includegraphics[trim={1.5cm 2cm 4cm 2cm},angle=0,width=1.0\linewidth]{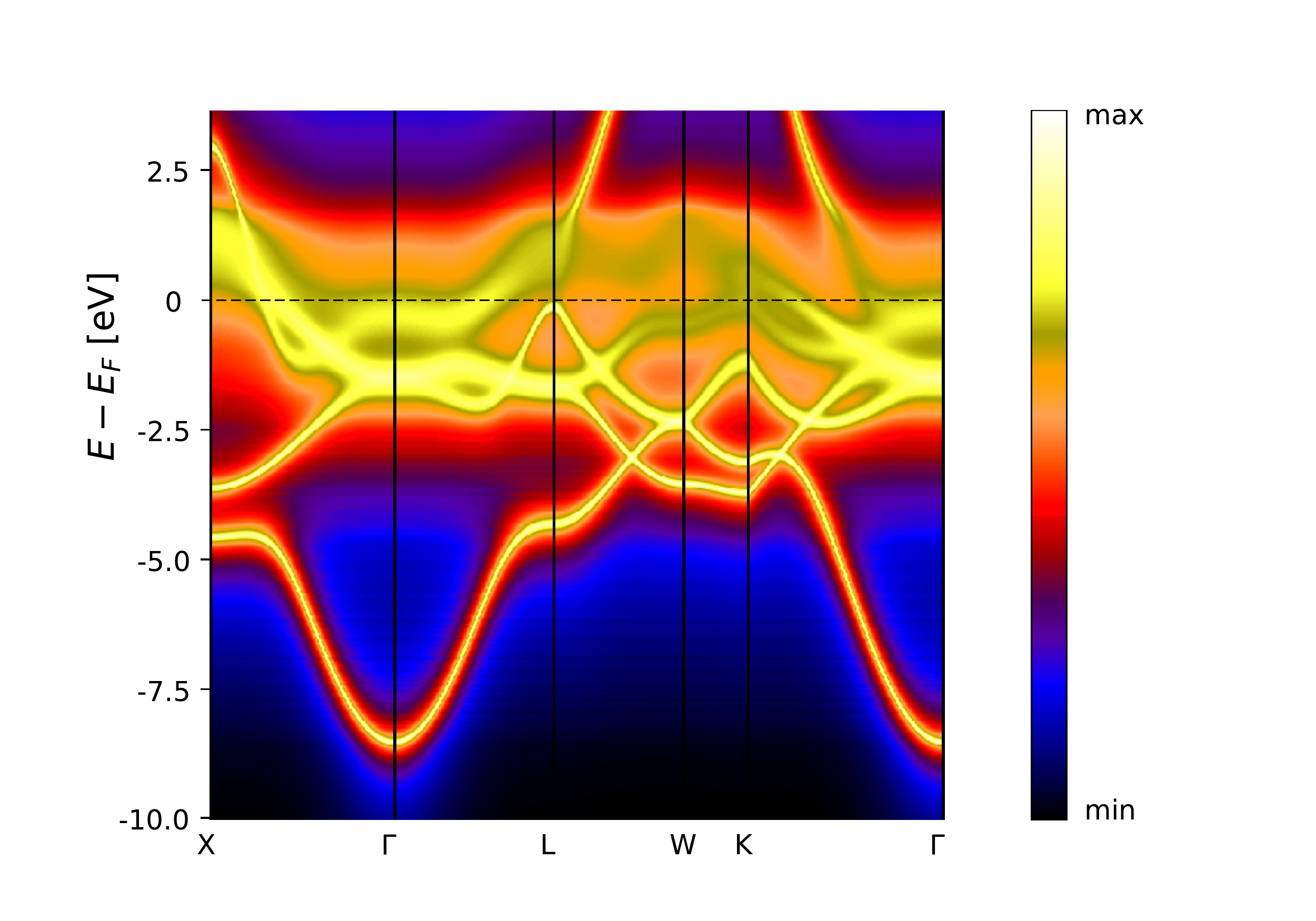}
        \end{center}
\caption{Logarithm of the Bloch spectral function of NiFeCoCr calculated within the KKR-CPA-DLM framework.}
        \label{bands}
\end{figure}

Korringa-Kohn-Rostoker (KKR) calculations which describe both the chemical disorder 
and the magnetic disorder above the Curie temperature ($T_C\sim$120K, the disordered local moment (DLM) state \cite{gyorffy:85,staunton:85}) 
with the mean-field coherent potential approximation (CPA) \cite{sprkkr} complement our experiments.
The electronic structure can most easily be described through 
the ${\mathbf k}$-resolved density-of-states known as the Bloch spectral function (BSF), $A_B(E,{\mathbf k})$ \cite{faulkner:80}.
For a perfectly ordered system, the behavior of $A_B(E,{\mathbf k})$
on passing through a band (in either ${\mathbf k}$ or $E$) would be a $\delta$-function (implying infinite quasiparticle lifetime), but
 in the presence of weak disorder the lineshape is broadened into an approximately Lorentzian form. The KKR-CPA BSF is valid for arbitrarily
strong disorder and deals with both the real part of the self-energy (leading to shifts in the band energies) and the imaginary part (which broadens the
energy bands). The electronic band structure can be visualized through the BSF, which for
NiFeCoCr (Fig.~\ref{bands}) is particularly smeared for states close to the Fermi energy 
where the energy separation between the flat $d$-bands of the component elements is large on the scale of the corresponding $d$-band width.

Compton profiles were measured on a single crystal of NiFeCoCr at room temperature along 15 special 
crystallographic directions \cite{fehlner:76} 
using the spectrometer on beamline BL08W at SPring-8 \cite{hiraoka:01}. 
The full 3D electron momentum density, $\rho ({\mathbf p})$ was reconstructed from the 1D Compton profiles ($J(p_z)$) by lattice harmonic expansion in the method
proposed by Kontrym-Sznajd and Samsel-Czeka{\l}a \cite{sznajd:00}. The resulting distribution in ${\mathbf p}$-space was then translated back into the first Brillouin zone (1BZ)
by summing contributions separated by a reciprocal lattice vector as prescribed by the Lock--Crisp--West (LCW) theorem \cite{lcw:73}, thus giving an occupation number $n({\mathbf k})$ in the 1BZ.
The contributions of electrons in fully occupied bands sum to give a constant contribution across the BZ (with an intensity
proportional to the number of occupied bands), whereas the signatures of the FS due to electrons in partially occupied bands
reinforce constructively, producing features in $n({\mathbf k})$ which are steplike discontinuities for a perfectly ordered system, but
which are smeared out in a disordered alloy.

\begin{figure}
        \begin{center}
        \includegraphics[trim={0 0cm 0 0cm},clip,angle=0,width=1.00\linewidth]{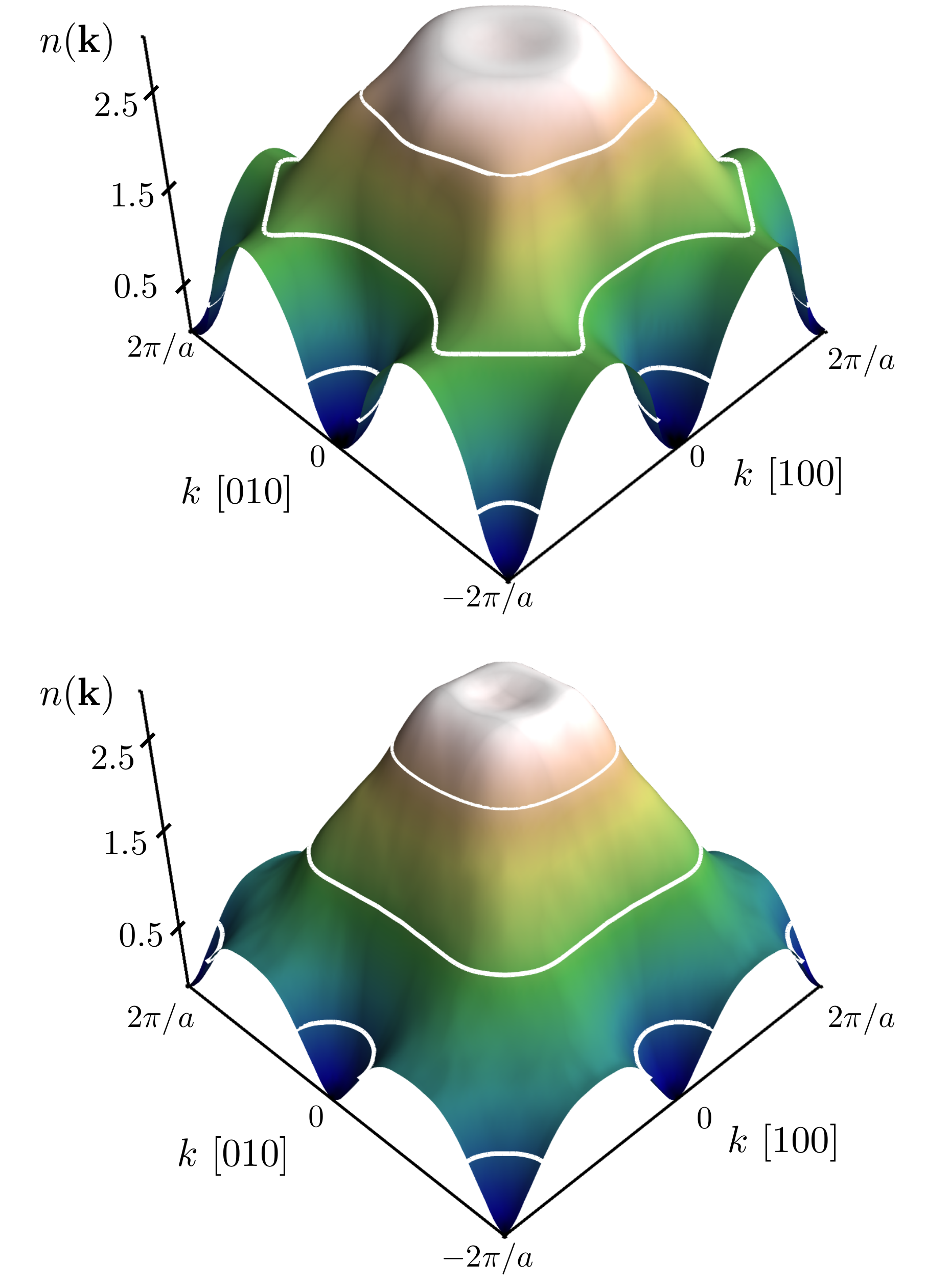}
        \end{center}
\caption{Occupation $n({\mathbf k})$ in the (001) plane through $\Gamma$, showing the KKR-CPA-DLM theory (top) and experiment 
(bottom). $\frac{2\pi}{a} \approx 0.93$ a.u., and the white lines indicate the isodensities associated with the three 
FS sheets.}
        \label{nk}
\end{figure}

From the KKR-CPA-DLM calculations, $n({\mathbf k})$ can be obtained from an energy integral of the BSF:
\begin{eqnarray}
n({\mathbf k}) = \int_{-\infty}^{E_F} A_B (E,{\mathbf k}) dE.
\end{eqnarray}
In Fig.~\ref{nk}, we show $n({\mathbf k})$ on the (001) plane (through $\Gamma$), from both the calculation (top) and experiment (bottom). 
The topology of the FS can immediately be deduced, with occupied states around the $\Gamma$ point at the center and hole pockets at the $\rm X$ points (located in the middle of the edges and at the corners). 
The KKR-CPA-DLM theory predicts three FS sheets, and given that there are regions of the BZ where none of these
three bands are occupied and regions where all three are occupied, the dynamic range (minimum to maximum) of $n({\mathbf k})$ can be normalized to
a value of three.
%
For an arbitrary ${\mathbf k}$-point in the BZ, if $n({\mathbf k})=0$, then none of the three bands are occupied, while $n({\mathbf k})=3$ would imply that all three bands are occupied. In this scheme, a set of isodensities at 0.5, 1.5 and 2.5 could be used to visualize the three FS 
sheets, and these are shown as white lines in Fig.~\ref{nk}. 
The smearing (there are no sharp steps) is intrinsic to the disorder, the impact of the finite experimental resolution (0.10~a.u.) being very small (see
\cite{supp}). However, for ease of comparison, all the theoretical distributions have been convoluted with a function representing the experimental resolution.  
The qualitative level of agreement is remarkable, with even a small dip around the $\Gamma$ point (due to a grazing band \cite{supp})
appearing in both.

\begin{figure}
        \begin{center}
        \includegraphics[trim={5cm 1.5cm 0.0cm 2.0cm},clip,angle=0,width=1.00\linewidth]{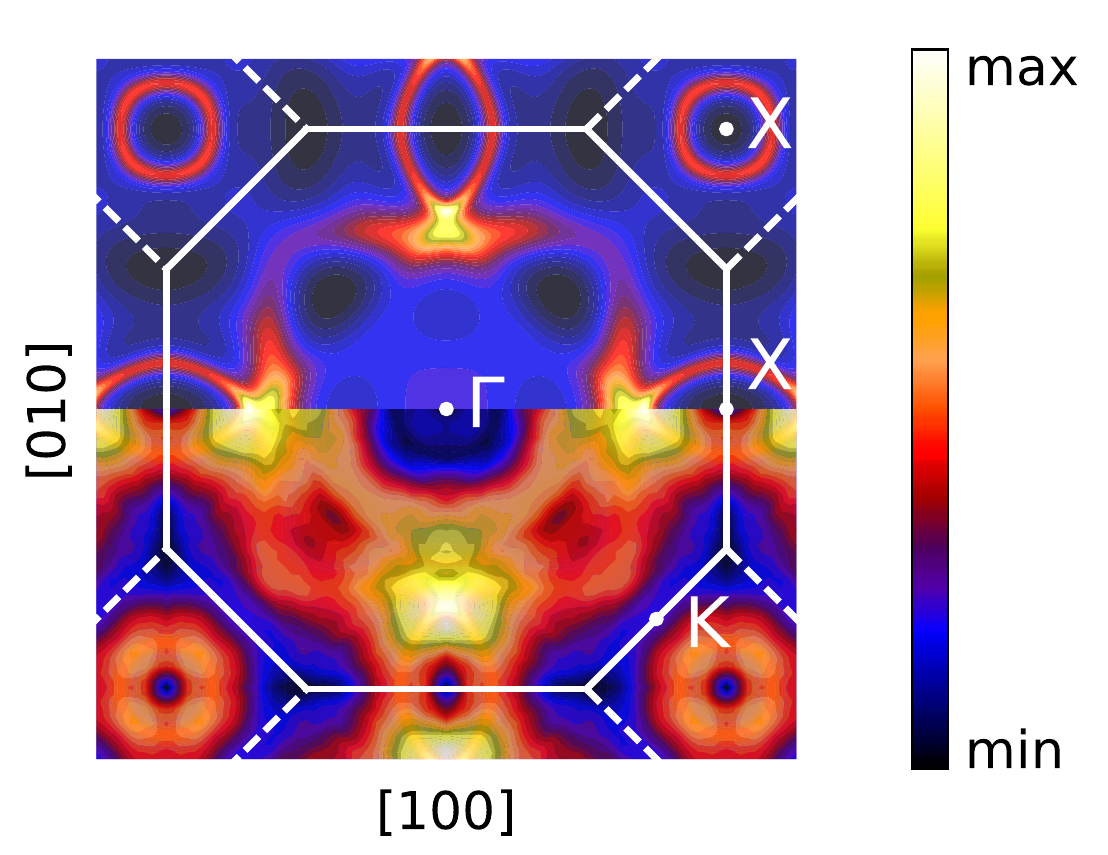}
        \end{center}
\caption{Bloch spectral function $A_B(E=E_F,{\mathbf k})$ in the (001) plane through $\Gamma$ from the KKR-CPA-DLM theory (top), with the
absolute square of the gradient of the experimental data (bottom). The BZ and some symmetry points are also plotted.}
        \label{bsfgrad}
\end{figure}

The BSF from the KKR-CPA-DLM calculation is shown in the top half of Fig.~\ref{bsfgrad}, with the hole pockets at the $\rm X$ points 
clearly visible, and the larger $\Gamma$-centred electron sheet being rather smeared out, particularly 
along the $\langle 1 1 0 \rangle$ directions.
To identify both the location and sharpness of the 
FS, it can be useful to look at a function related to the gradient of $n({\mathbf k})$  \cite{weber:17}.
This is applied to the experimental data and
displayed in the bottom half of Fig.~\ref{bsfgrad} where it can be seen that the least smeared regions (largest derivative) in the experimental data closely match the sharper BSF
peaks in the KKR-CPA-DLM calculation shown in the top half of Fig.~\ref{bsfgrad}. The hole pockets at the $\rm X$ points are also revealed in the experimental data, the locus of the maximum of
the gradient function showing that the sizes of the hole pockets are comparable with that predicted by the
KKR-CPA-DLM theory, but even more smeared.

\begin{figure}[ht]
        \begin{center}
        \includegraphics[trim={1cm 0cm 0cm 0cm},clip,angle=0,width=0.82\linewidth]{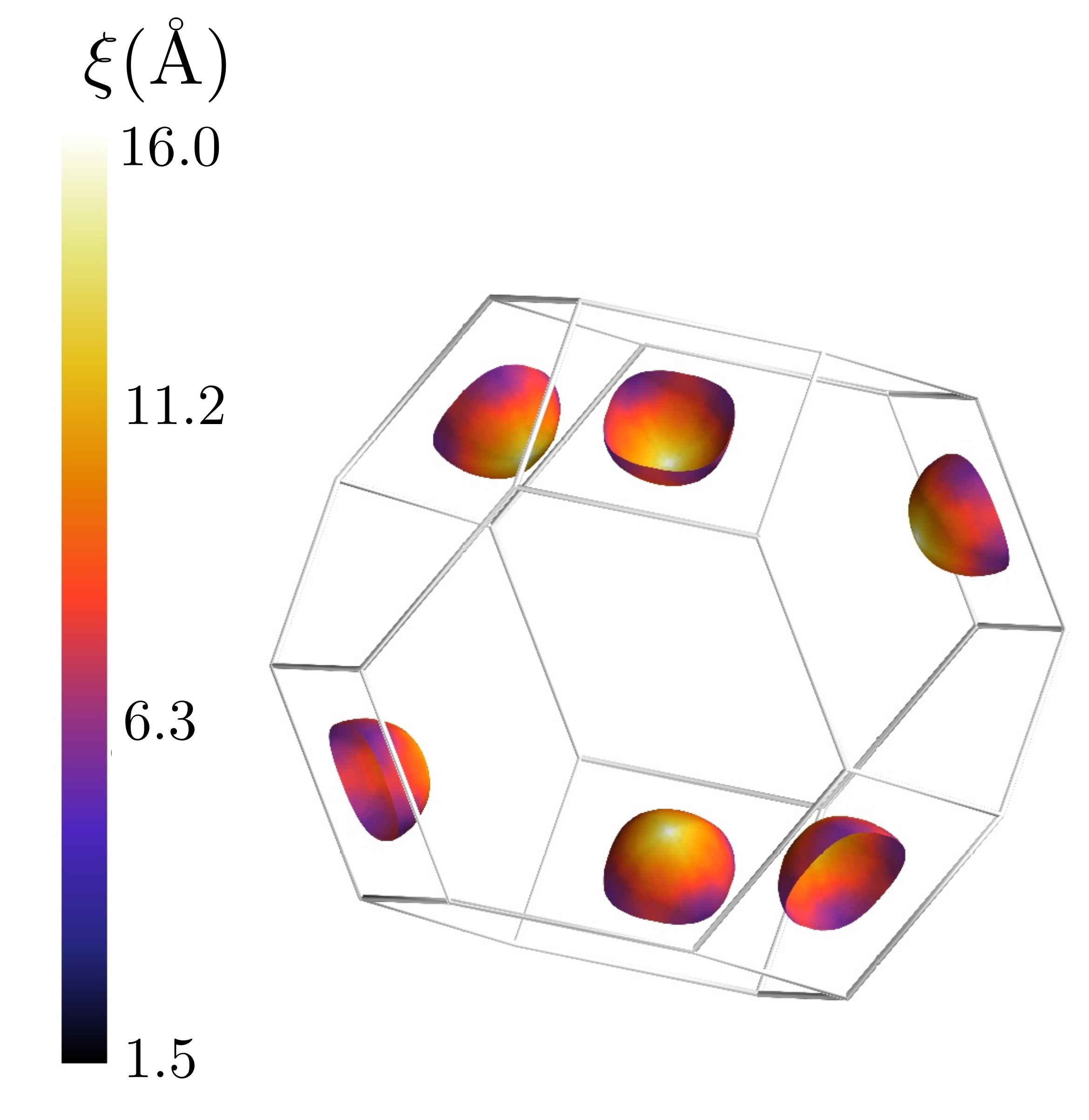}
        \includegraphics[trim={0 6cm 0 6cm},clip,angle=0,width=1.00\linewidth]{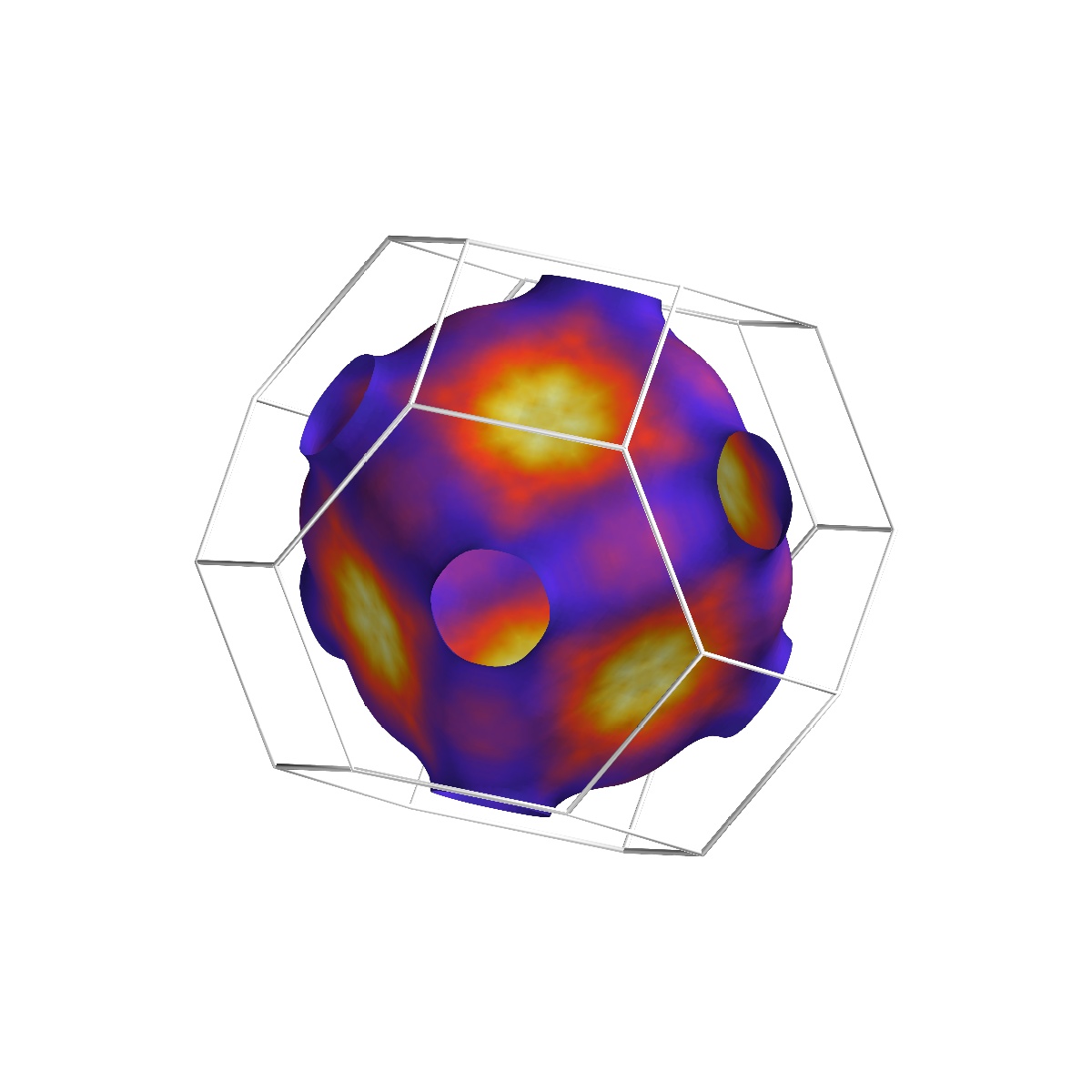}
        \includegraphics[trim={0 6cm 0 6cm},clip,angle=0,width=1.00\linewidth]{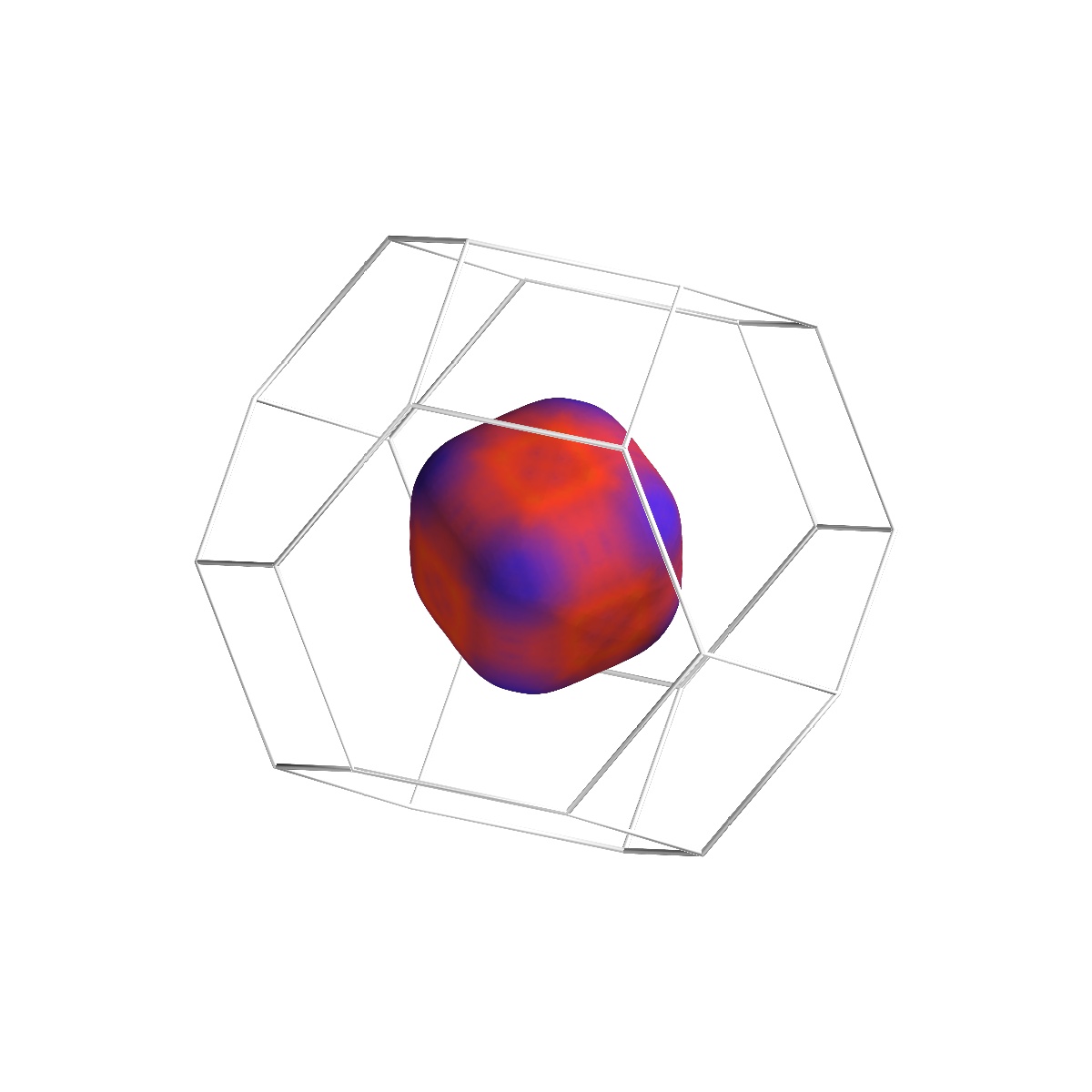}
        \end{center}
\caption{The hole (top) and two electron (middle and bottom) FS sheets of NiFeCoCr obtained from the Compton experiment.
These surfaces are the iso-densities (0.5, 1.5 and 2.5) plotted from the reconstructed occupation number.
The color shows the variation of the quasiparticle coherence length $\xi$ (in \AA) across the FS extracted from the smearing. The
wireframe box is the first Brillouin zone.}
        \label{FS}
\end{figure}

The isodensities can be used to visualize the full 3D FS topology from the Compton scattering experiment, as shown in 
Fig.~\ref{FS}. 
The larger $\Gamma$-centered electron sheet with necks along the $\langle 111 \rangle$ directions is reminiscent of the well-known FS of Cu
\cite{pippard:57} which originates from a free-electron-like band that emerges from the top of the $d$-bands and intersects with the
BZ boundary.  The color of the isosurfaces corresponds to the coherence length of the quasiparticles which has been evaluated through the method described below.       

\begin{figure}
        \begin{center}
        \includegraphics[angle=0,width=1.00\linewidth]{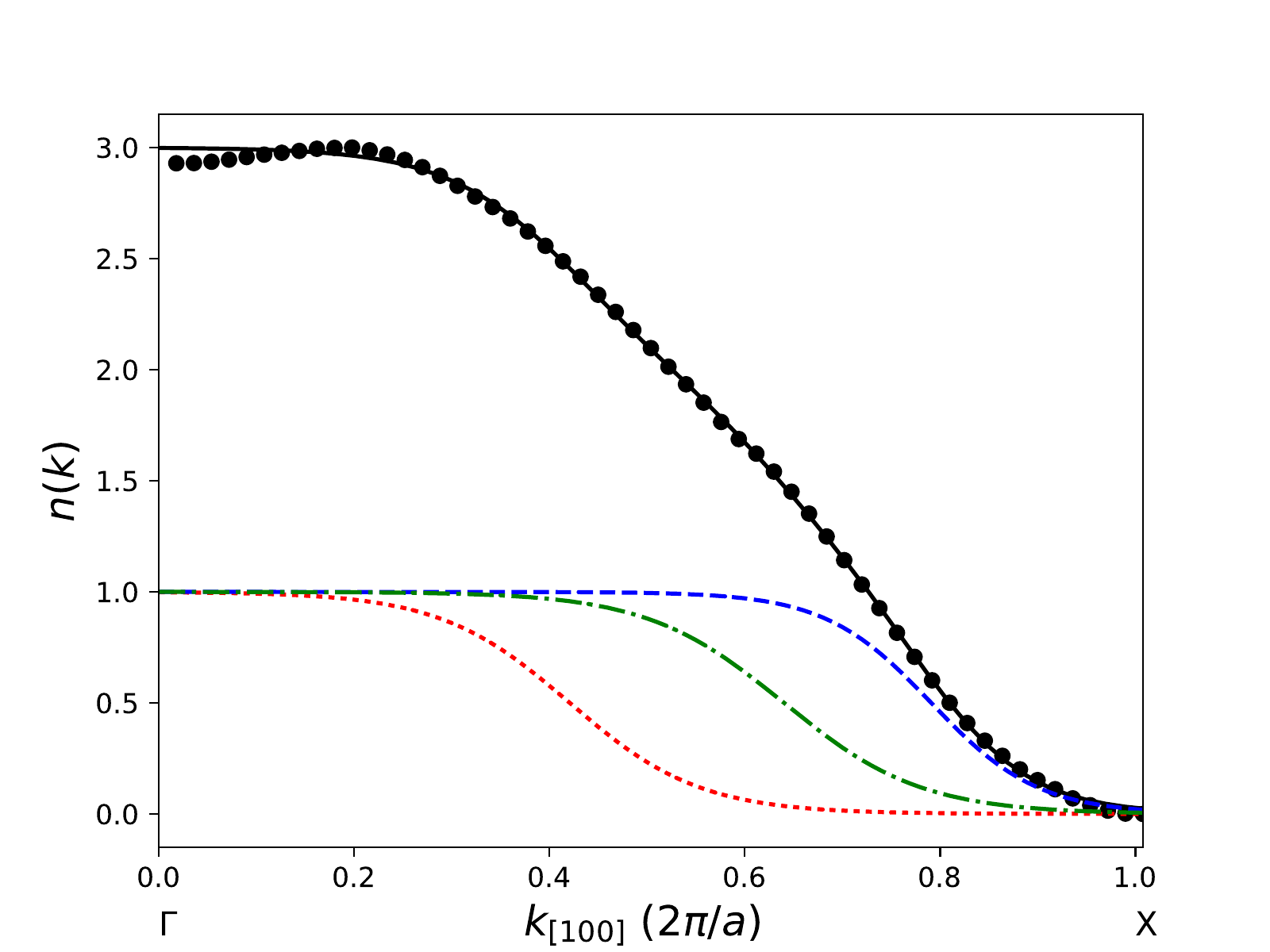}
        \includegraphics[angle=0,width=1.00\linewidth]{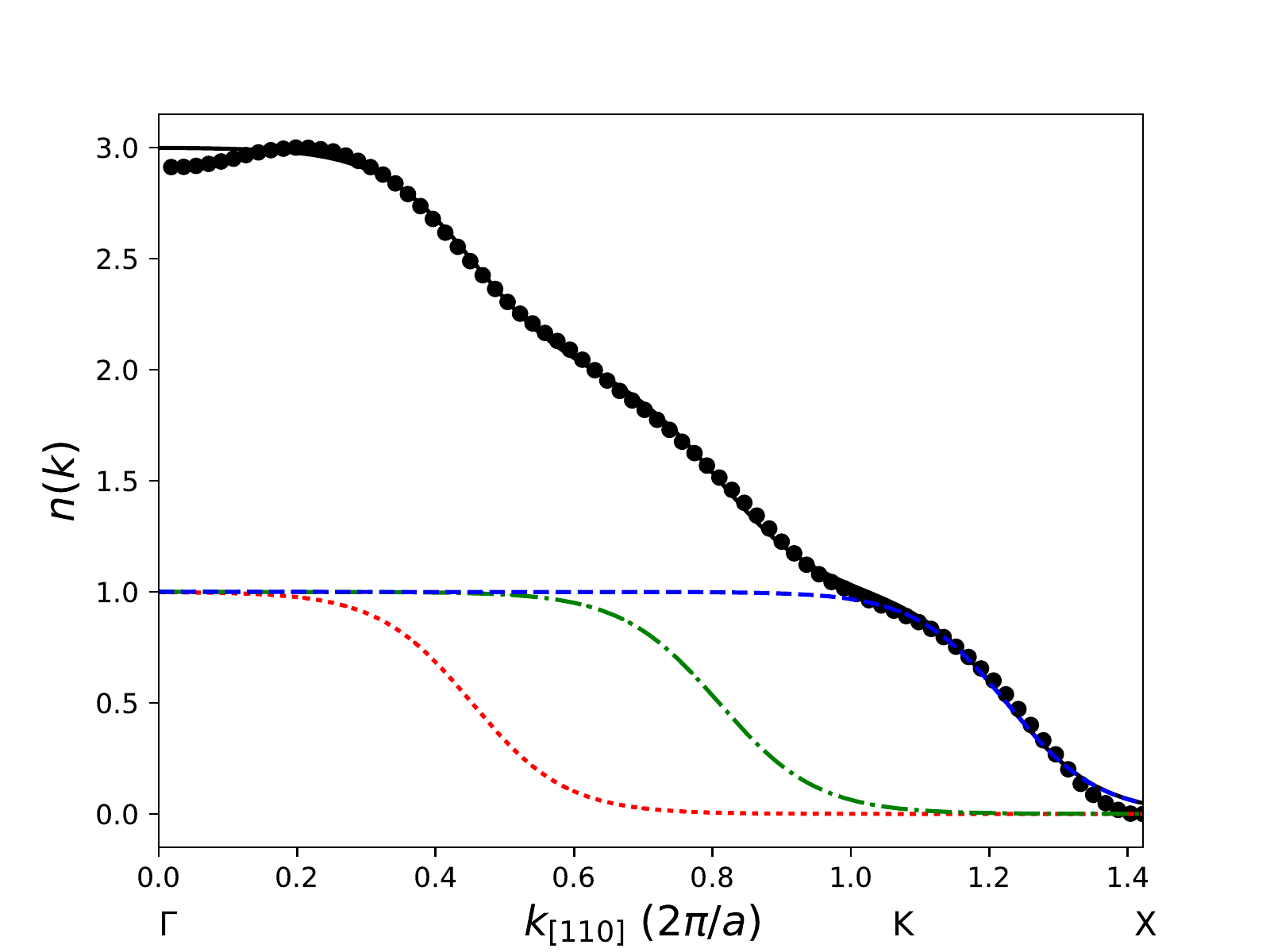}
        \end{center}
\caption{Occupation $n({\mathbf k})$ from the experiment (filled circles) along the [100] (top) and [110] (bottom) directions in the BZ (where
$k$ is in units of $2\pi/a$). 
$n({\mathbf k})$ has been fitted with the sum (black line) of three $\tanh$ functions (red dots, green dash-dots and 
blue dashes).
The edge of the BZ is at 1.00 and 1.06 for the [100] and [110] directions, respectively, and symmetry points such as $\Gamma$, $\rm K$ and $\rm X$ are shown.}
        \label{steps}
\end{figure}

By fitting a series of smeared unit step functions to $n({\mathbf k})$
along a set of special directions \cite{kontrymsznajd:15}, it is possible to extract the degree to which each sheet of FS is smeared along
each direction. Such fits are not expected to capture all the details (e.g. the dip close to
the $\Gamma$ point \cite{supp}), but are useful for quantifying the smearing (used to estimate the quasiparticle coherence
length \cite{supp}). When visualizing the individual FS sheets, the isodensities 
extracted from the step function fits (0.45, 1.48 and 2.47) are very close to the canonical values used above.
Corrections were made to the coherence lengths extracted from the fits by simulating the impact of finite experimental resolution on different degrees of smearing.
 The FS-averaged mean-free-paths were calculated from a set of special directions using the approach detailed by Kontrym-Sznajd and
Dugdale \cite{kontrymsznajd:15}. Examples of fits to the $[100]$ and $[110]$ directions 
(high symmetry directions that are not part of the set of special directions) are shown in Fig.~\ref{steps}, 
where the experimental $n({\mathbf k})$ has been fitted with three smeared unit step functions (represented by $\tanh$ functions) 
with the position and width of each being free parameters \cite{supp}.
The smeared steps in Fig.~\ref{steps} span up to 40\% of $\frac{2\pi}{a}$.

A summary of the FS-averaged mean-free-paths, FS areas and resistivities can be found in Table~\ref{mfptable}. 
 The conductivity is
dominated by the second (Cu-like) FS sheet, and the variation of the quasiparticle coherence length
across this FS sheet strongly resembles the variation of the Fermi velocity across the FS of Cu \cite{gradhand:11}. The resistivity scales inversely with the 
area of the FS and thus the differences between the sheets is dominated by the differences in their surface areas. 
Combining the contribution from all three sheets
in parallel gives a resistivity of 61~$\mu\Omega{\rm cm}$. As a crosscheck, this value compares favorably with the residual resistivity
of 77~$\mu\Omega{\rm cm}$ measured by Jin {\it et al.} \cite{jin:16b}.
The residual resistivity predicted directly by the KKR-CPA calculations using the Kubo-Greenwood formalism \cite{supp} is 51~$\mu\Omega{\rm cm}$, 
increasing to 67~$\mu\Omega{\rm cm}$ for the DLM state.
%

\begin{table}[ht]
  \begin{ruledtabular}
     \begin{tabular}{c c c c}
     FS sheet & mean-free-path $\langle \lambda \rangle $ & FS area $\mathcal{A}$ & Resistivity $\rho$ \\
              &  (nm) & ($\times 10^{17}$ cm$^{-2}$) & $(\mu\Omega{\rm cm})$ \\
      \hline
       1 & 0.91 & 0.75 & 225  \\
       2 & 0.73 & 2.16 &  97 \\
       3 & 0.70 & 0.39 & 565  \\
     \end{tabular}
  \end{ruledtabular}
  \caption{The surface area, mean-free-path (inferred from the smearing \cite{supp}) and resistivity ($\frac{12\pi^3 \hbar}{e^2 \lambda \mathcal{A}}$, see \cite{supp}) of each FS sheet.}
  \label{mfptable}
\end{table}

In conclusion, we have measured the FS of the equiatomic disordered 
alloy NiFeCoCr which exhibits significant smearing of up to
$\sim 40\%$ of $\frac{2\pi}{a}$. Across some regions of the FS the quasiparticle 
coherence length is very close to the nearest-neighbor distance, implying
close proximity to the MIR limit where the usual picture of ballistically propagating quasiparticles 
is invalid \cite{hussey:04}.
Both strong electron-electron scattering (due to strong
electron correlations) and strong electron scattering due to atomic disorder 
can push a material towards the MIR limit. Treating the former is
challenging for theory, but the latter can be incorporated within the CPA \cite{samolyuk:19}. 
High-entropy alloys might thus provide an environment in which some important aspects of transport physics, 
normally associated strongly correlated electrons, could be simulated by disorder physics. 
NiFeCoCr appears close to a regime where
the conventional picture of a quasiparticle with a well-defined momentum being subjected to occasional scattering events breaks
down, signalling the demise of the Fermi liquid.    
It is interesting to note that NiCoCr, NiFeCoCrMn and NiFeCoCrPd all have residual resistivities
which are higher than NiFeCoCr and all exhibit similar non-Fermi-liquid behavior \cite{jin:16b}. 
An electron mean free path, obtained by averaging the coherence length of the quasiparticles over each FS sheet
as inferred from the FS smearing, gives an estimated residual resistivity of $\sim$61~$\mu \Omega {\rm cm}$. 
We have confirmed the presence of significant FS smearing which has been predicted for some Cantor-Wu alloys \cite{mu:18}.
We have shown that the FS in a metal whose properties are dominated by disorder can be accurately described by
state-of-the-art theoretical approaches (density functional theory combined with the coherent potential approximation)  
which generalize the concept of a metallic FS to such materials.
Finally, magnetic Compton scattering \cite{duffy:13} could be a useful probe of the nature of magnetic order in these
and other alloys (for example, probing ferrimagnetism \cite{haynes:12} in hcp NiFeCoCrW \cite{lizarraga:18}).

\begin{acknowledgments}
The Compton scattering experiment was performed with the approval of the Japan Synchrotron Radiation Research Institute (JASRI, proposal no. 2016A1323).
H.A.R. and D.A.L. gratefully acknowledge the financial support of the UK EPSRC (EP/L015544/1), and the National Secretariat of Higher Education, 
Science, Technology and Innovation of Ecuador (SENESCYT), respectively.
S.M., G.D.S., G.M.S. acknowledge funding support by the Energy Dissipation and Defect Evolution (EDDE), an Energy Frontier Research Center funded by the U. S. Department of Energy (DOE), Office of Science, Basic Energy Sciences under contract number DE-AC05-00OR22725. S.B.D. would like to thank Dr Christopher Bell (Bristol) for 
valuable discussions.  The original research data are available at the DOI specified in Ref.~\cite{researchdata}.
\end{acknowledgments}

\bibliography{paper}

\begin{thebibliography}{36}%
\makeatletter
\providecommand \@ifxundefined [1]{%
 \@ifx{#1\undefined}
}%
\providecommand \@ifnum [1]{%
 \ifnum #1\expandafter \@firstoftwo
 \else \expandafter \@secondoftwo
 \fi
}%
\providecommand \@ifx [1]{%
 \ifx #1\expandafter \@firstoftwo
 \else \expandafter \@secondoftwo
 \fi
}%
\providecommand \natexlab [1]{#1}%
\providecommand \enquote  [1]{``#1''}%
\providecommand \bibnamefont  [1]{#1}%
\providecommand \bibfnamefont [1]{#1}%
\providecommand \citenamefont [1]{#1}%
\providecommand \href@noop [0]{\@secondoftwo}%
\providecommand \href [0]{\begingroup \@sanitize@url \@href}%
\providecommand \@href[1]{\@@startlink{#1}\@@href}%
\providecommand \@@href[1]{\endgroup#1\@@endlink}%
\providecommand \@sanitize@url [0]{\catcode `\\12\catcode `\$12\catcode
  `\&12\catcode `\#12\catcode `\^12\catcode `\_12\catcode `\%12\relax}%
\providecommand \@@startlink[1]{}%
\providecommand \@@endlink[0]{}%
\providecommand \url  [0]{\begingroup\@sanitize@url \@url }%
\providecommand \@url [1]{\endgroup\@href {#1}{\urlprefix }}%
\providecommand \urlprefix  [0]{URL }%
\providecommand \Eprint [0]{\href }%
\providecommand \doibase [0]{http://dx.doi.org/}%
\providecommand \selectlanguage [0]{\@gobble}%
\providecommand \bibinfo  [0]{\@secondoftwo}%
\providecommand \bibfield  [0]{\@secondoftwo}%
\providecommand \translation [1]{[#1]}%
\providecommand \BibitemOpen [0]{}%
\providecommand \bibitemStop [0]{}%
\providecommand \bibitemNoStop [0]{.\EOS\space}%
\providecommand \EOS [0]{\spacefactor3000\relax}%
\providecommand \BibitemShut  [1]{\csname bibitem#1\endcsname}%
\let\auto@bib@innerbib\@empty
\bibitem [{\citenamefont {Hoch}(1983)}]{hoch:83}%
  \BibitemOpen
  \bibfield  {author} {\bibinfo {author} {\bibfnamefont {P.}~\bibnamefont
  {Hoch}},\ }\href@noop {} {\bibfield  {journal} {\bibinfo  {journal}
  {Contemporary Physics}\ }\textbf {\bibinfo {volume} {24}},\ \bibinfo {pages}
  {3} (\bibinfo {year} {1983})}\BibitemShut {NoStop}%
\bibitem [{\citenamefont {Luttinger}(1960)}]{luttinger:60}%
  \BibitemOpen
  \bibfield  {author} {\bibinfo {author} {\bibfnamefont {J.~M.}\ \bibnamefont
  {Luttinger}},\ }\href@noop {} {\bibfield  {journal} {\bibinfo  {journal}
  {Phys. Rev.}\ }\textbf {\bibinfo {volume} {119}},\ \bibinfo {pages} {1153}
  (\bibinfo {year} {1960})}\BibitemShut {NoStop}%
\bibitem [{\citenamefont {Huotari}\ \emph {et~al.}(2010)\citenamefont
  {Huotari}, \citenamefont {Soininen}, \citenamefont {Pylkk\"anen},
  \citenamefont {H\"am\"al\"ainen}, \citenamefont {Issolah}, \citenamefont
  {Titov}, \citenamefont {McMinis}, \citenamefont {Kim}, \citenamefont {Esler},
  \citenamefont {Ceperley}, \citenamefont {Holzmann},\ and\ \citenamefont
  {Olevano}}]{huotari:10}%
  \BibitemOpen
  \bibfield  {author} {\bibinfo {author} {\bibfnamefont {S.}~\bibnamefont
  {Huotari}}, \bibinfo {author} {\bibfnamefont {J.~A.}\ \bibnamefont
  {Soininen}}, \bibinfo {author} {\bibfnamefont {T.}~\bibnamefont
  {Pylkk\"anen}}, \bibinfo {author} {\bibfnamefont {K.}~\bibnamefont
  {H\"am\"al\"ainen}}, \bibinfo {author} {\bibfnamefont {A.}~\bibnamefont
  {Issolah}}, \bibinfo {author} {\bibfnamefont {A.}~\bibnamefont {Titov}},
  \bibinfo {author} {\bibfnamefont {J.}~\bibnamefont {McMinis}}, \bibinfo
  {author} {\bibfnamefont {J.}~\bibnamefont {Kim}}, \bibinfo {author}
  {\bibfnamefont {K.}~\bibnamefont {Esler}}, \bibinfo {author} {\bibfnamefont
  {D.~M.}\ \bibnamefont {Ceperley}}, \bibinfo {author} {\bibfnamefont
  {M.}~\bibnamefont {Holzmann}}, \ and\ \bibinfo {author} {\bibfnamefont
  {V.}~\bibnamefont {Olevano}},\ }\href@noop {} {\bibfield  {journal} {\bibinfo
   {journal} {Phys. Rev. Lett.}\ }\textbf {\bibinfo {volume} {105}},\ \bibinfo
  {pages} {086403} (\bibinfo {year} {2010})}\BibitemShut {NoStop}%
\bibitem [{\citenamefont {Hussey}\ \emph {et~al.}(2004)\citenamefont {Hussey},
  \citenamefont {Takenaka},\ and\ \citenamefont {Takagi}}]{hussey:04}%
  \BibitemOpen
  \bibfield  {author} {\bibinfo {author} {\bibfnamefont {N.~E.}\ \bibnamefont
  {Hussey}}, \bibinfo {author} {\bibfnamefont {K.}~\bibnamefont {Takenaka}}, \
  and\ \bibinfo {author} {\bibfnamefont {H.}~\bibnamefont {Takagi}},\ }\href
  {\doibase 10.1080/14786430410001716944} {\bibfield  {journal} {\bibinfo
  {journal} {Philosophical Magazine}\ }\textbf {\bibinfo {volume} {84}},\
  \bibinfo {pages} {2847} (\bibinfo {year} {2004})}\BibitemShut {NoStop}%
\bibitem [{\citenamefont {Gyorffy}\ and\ \citenamefont
  {Stocks}(1983)}]{gyorffy:83}%
  \BibitemOpen
  \bibfield  {author} {\bibinfo {author} {\bibfnamefont {B.~L.}\ \bibnamefont
  {Gyorffy}}\ and\ \bibinfo {author} {\bibfnamefont {G.~M.}\ \bibnamefont
  {Stocks}},\ }\href@noop {} {\bibfield  {journal} {\bibinfo  {journal} {Phys.
  Rev. Lett.}\ }\textbf {\bibinfo {volume} {50}},\ \bibinfo {pages} {374}
  (\bibinfo {year} {1983})}\BibitemShut {NoStop}%
\bibitem [{\citenamefont {Wilkinson}\ \emph {et~al.}(2001)\citenamefont
  {Wilkinson}, \citenamefont {Hughes}, \citenamefont {Major}, \citenamefont
  {Dugdale}, \citenamefont {Alam}, \citenamefont {Bruno}, \citenamefont
  {Ginatempo},\ and\ \citenamefont {Giuliano}}]{wilkinson:01}%
  \BibitemOpen
  \bibfield  {author} {\bibinfo {author} {\bibfnamefont {I.}~\bibnamefont
  {Wilkinson}}, \bibinfo {author} {\bibfnamefont {R.~J.}\ \bibnamefont
  {Hughes}}, \bibinfo {author} {\bibfnamefont {Z.}~\bibnamefont {Major}},
  \bibinfo {author} {\bibfnamefont {S.~B.}\ \bibnamefont {Dugdale}}, \bibinfo
  {author} {\bibfnamefont {M.~A.}\ \bibnamefont {Alam}}, \bibinfo {author}
  {\bibfnamefont {E.}~\bibnamefont {Bruno}}, \bibinfo {author} {\bibfnamefont
  {B.}~\bibnamefont {Ginatempo}}, \ and\ \bibinfo {author} {\bibfnamefont
  {E.~S.}\ \bibnamefont {Giuliano}},\ }\href@noop {} {\bibfield  {journal}
  {\bibinfo  {journal} {Phys. Rev. Lett.}\ }\textbf {\bibinfo {volume} {87}},\
  \bibinfo {pages} {216401} (\bibinfo {year} {2001})}\BibitemShut {NoStop}%
\bibitem [{\citenamefont {Yaqub}\ and\ \citenamefont
  {Cochran}(1965)}]{yaqub:65}%
  \BibitemOpen
  \bibfield  {author} {\bibinfo {author} {\bibfnamefont {M.}~\bibnamefont
  {Yaqub}}\ and\ \bibinfo {author} {\bibfnamefont {J.~F.}\ \bibnamefont
  {Cochran}},\ }\href@noop {} {\bibfield  {journal} {\bibinfo  {journal} {Phys.
  Rev.}\ }\textbf {\bibinfo {volume} {137}},\ \bibinfo {pages} {A1182}
  (\bibinfo {year} {1965})}\BibitemShut {NoStop}%
\bibitem [{\citenamefont {Gall}(2016)}]{gall:16}%
  \BibitemOpen
  \bibfield  {author} {\bibinfo {author} {\bibfnamefont {D.}~\bibnamefont
  {Gall}},\ }\href@noop {} {\bibfield  {journal} {\bibinfo  {journal} {Journal
  of Applied Physics}\ }\textbf {\bibinfo {volume} {119}},\ \bibinfo {pages}
  {085101} (\bibinfo {year} {2016})}\BibitemShut {NoStop}%
\bibitem [{\citenamefont {Miracle}\ and\ \citenamefont
  {Senkov}(2017)}]{miracle:17}%
  \BibitemOpen
  \bibfield  {author} {\bibinfo {author} {\bibfnamefont {D.}~\bibnamefont
  {Miracle}}\ and\ \bibinfo {author} {\bibfnamefont {O.}~\bibnamefont
  {Senkov}},\ }\href@noop {} {\bibfield  {journal} {\bibinfo  {journal} {Acta
  Materialia}\ }\textbf {\bibinfo {volume} {122}},\ \bibinfo {pages} {448 }
  (\bibinfo {year} {2017})}\BibitemShut {NoStop}%
\bibitem [{\citenamefont {Cantor}\ \emph {et~al.}(2004)\citenamefont {Cantor},
  \citenamefont {Chang}, \citenamefont {Knight},\ and\ \citenamefont
  {Vincent}}]{cantor:04}%
  \BibitemOpen
  \bibfield  {author} {\bibinfo {author} {\bibfnamefont {B.}~\bibnamefont
  {Cantor}}, \bibinfo {author} {\bibfnamefont {I.}~\bibnamefont {Chang}},
  \bibinfo {author} {\bibfnamefont {P.}~\bibnamefont {Knight}}, \ and\ \bibinfo
  {author} {\bibfnamefont {A.}~\bibnamefont {Vincent}},\ }\href@noop {}
  {\bibfield  {journal} {\bibinfo  {journal} {Mater Sci Eng A}\ }\textbf
  {\bibinfo {volume} {213}},\ \bibinfo {pages} {375} (\bibinfo {year}
  {2004})}\BibitemShut {NoStop}%
\bibitem [{\citenamefont {Wu}\ \emph {et~al.}(2014)\citenamefont {Wu},
  \citenamefont {Bei}, \citenamefont {Otto}, \citenamefont {Pharr},\ and\
  \citenamefont {George}}]{wu:14}%
  \BibitemOpen
  \bibfield  {author} {\bibinfo {author} {\bibfnamefont {Z.}~\bibnamefont
  {Wu}}, \bibinfo {author} {\bibfnamefont {H.}~\bibnamefont {Bei}}, \bibinfo
  {author} {\bibfnamefont {F.}~\bibnamefont {Otto}}, \bibinfo {author}
  {\bibfnamefont {G.}~\bibnamefont {Pharr}}, \ and\ \bibinfo {author}
  {\bibfnamefont {E.}~\bibnamefont {George}},\ }\href@noop {} {\bibfield
  {journal} {\bibinfo  {journal} {Intermetallics}\ }\textbf {\bibinfo {volume}
  {46}},\ \bibinfo {pages} {131 } (\bibinfo {year} {2014})}\BibitemShut
  {NoStop}%
\bibitem [{\citenamefont {Song}\ \emph {et~al.}(2017)\citenamefont {Song},
  \citenamefont {Tian}, \citenamefont {Hu}, \citenamefont {Vitos},
  \citenamefont {Wang}, \citenamefont {Shen},\ and\ \citenamefont
  {Chen}}]{song:17}%
  \BibitemOpen
  \bibfield  {author} {\bibinfo {author} {\bibfnamefont {H.}~\bibnamefont
  {Song}}, \bibinfo {author} {\bibfnamefont {F.}~\bibnamefont {Tian}}, \bibinfo
  {author} {\bibfnamefont {Q.-M.}\ \bibnamefont {Hu}}, \bibinfo {author}
  {\bibfnamefont {L.}~\bibnamefont {Vitos}}, \bibinfo {author} {\bibfnamefont
  {Y.}~\bibnamefont {Wang}}, \bibinfo {author} {\bibfnamefont {J.}~\bibnamefont
  {Shen}}, \ and\ \bibinfo {author} {\bibfnamefont {N.}~\bibnamefont {Chen}},\
  }\href@noop {} {\bibfield  {journal} {\bibinfo  {journal} {Phys. Rev.
  Materials}\ }\textbf {\bibinfo {volume} {1}},\ \bibinfo {pages} {023404}
  (\bibinfo {year} {2017})}\BibitemShut {NoStop}%
\bibitem [{\citenamefont {Samolyuk}\ \emph {et~al.}(2018)\citenamefont
  {Samolyuk}, \citenamefont {Mu}, \citenamefont {May}, \citenamefont {Sales},
  \citenamefont {Wimmer}, \citenamefont {Mankovsky}, \citenamefont {Ebert},\
  and\ \citenamefont {Stocks}}]{samolyuk:18}%
  \BibitemOpen
  \bibfield  {author} {\bibinfo {author} {\bibfnamefont {G.~D.}\ \bibnamefont
  {Samolyuk}}, \bibinfo {author} {\bibfnamefont {S.}~\bibnamefont {Mu}},
  \bibinfo {author} {\bibfnamefont {A.~F.}\ \bibnamefont {May}}, \bibinfo
  {author} {\bibfnamefont {B.~C.}\ \bibnamefont {Sales}}, \bibinfo {author}
  {\bibfnamefont {S.}~\bibnamefont {Wimmer}}, \bibinfo {author} {\bibfnamefont
  {S.}~\bibnamefont {Mankovsky}}, \bibinfo {author} {\bibfnamefont
  {H.}~\bibnamefont {Ebert}}, \ and\ \bibinfo {author} {\bibfnamefont {G.~M.}\
  \bibnamefont {Stocks}},\ }\href@noop {} {\bibfield  {journal} {\bibinfo
  {journal} {Phys. Rev. B}\ }\textbf {\bibinfo {volume} {98}},\ \bibinfo
  {pages} {165141} (\bibinfo {year} {2018})}\BibitemShut {NoStop}%
\bibitem [{\citenamefont {Mu}\ \emph {et~al.}(2019)\citenamefont {Mu},
  \citenamefont {Samolyuk}, \citenamefont {Wimmer}, \citenamefont
  {Troparevsky}, \citenamefont {Khan}, \citenamefont {Mankovsky}, \citenamefont
  {Ebert},\ and\ \citenamefont {Stocks}}]{mu:18}%
  \BibitemOpen
  \bibfield  {author} {\bibinfo {author} {\bibfnamefont {S.}~\bibnamefont
  {Mu}}, \bibinfo {author} {\bibfnamefont {G.~D.}\ \bibnamefont {Samolyuk}},
  \bibinfo {author} {\bibfnamefont {S.}~\bibnamefont {Wimmer}}, \bibinfo
  {author} {\bibfnamefont {M.~C.}\ \bibnamefont {Troparevsky}}, \bibinfo
  {author} {\bibfnamefont {S.~N.}\ \bibnamefont {Khan}}, \bibinfo {author}
  {\bibfnamefont {S.}~\bibnamefont {Mankovsky}}, \bibinfo {author}
  {\bibfnamefont {H.}~\bibnamefont {Ebert}}, \ and\ \bibinfo {author}
  {\bibfnamefont {G.~M.}\ \bibnamefont {Stocks}},\ }\href@noop {} {\bibfield
  {journal} {\bibinfo  {journal} {npj Computational Materials}\ }\textbf
  {\bibinfo {volume} {5}},\ \bibinfo {pages} {1} (\bibinfo {year}
  {2019})}\BibitemShut {NoStop}%
\bibitem [{\citenamefont {K{\"o}rmann}\ \emph {et~al.}(2017)\citenamefont
  {K{\"o}rmann}, \citenamefont {Ikeda}, \citenamefont {Grabowski},\ and\
  \citenamefont {Sluiter}}]{kormann:17}%
  \BibitemOpen
  \bibfield  {author} {\bibinfo {author} {\bibfnamefont {F.}~\bibnamefont
  {K{\"o}rmann}}, \bibinfo {author} {\bibfnamefont {Y.}~\bibnamefont {Ikeda}},
  \bibinfo {author} {\bibfnamefont {B.}~\bibnamefont {Grabowski}}, \ and\
  \bibinfo {author} {\bibfnamefont {M.~H.~F.}\ \bibnamefont {Sluiter}},\
  }\href@noop {} {\bibfield  {journal} {\bibinfo  {journal} {npj Computational
  Materials}\ }\textbf {\bibinfo {volume} {3}},\ \bibinfo {pages} {36}
  (\bibinfo {year} {2017})}\BibitemShut {NoStop}%
\bibitem [{\citenamefont {Ko\ifmmode~\check{z}\else \v{z}\fi{}elj}\ \emph
  {et~al.}(2014)\citenamefont {Ko\ifmmode~\check{z}\else \v{z}\fi{}elj},
  \citenamefont {Vrtnik}, \citenamefont {Jelen}, \citenamefont {Jazbec},
  \citenamefont {Jagli\ifmmode \check{c}\else
  \v{c}\fi{}i\ifmmode~\acute{c}\else \'{c}\fi{}}, \citenamefont {Maiti},
  \citenamefont {Feuerbacher}, \citenamefont {Steurer},\ and\ \citenamefont
  {Dolin\ifmmode~\check{s}\else \v{s}\fi{}ek}}]{heasuper}%
  \BibitemOpen
  \bibfield  {author} {\bibinfo {author} {\bibfnamefont {P.}~\bibnamefont
  {Ko\ifmmode~\check{z}\else \v{z}\fi{}elj}}, \bibinfo {author} {\bibfnamefont
  {S.}~\bibnamefont {Vrtnik}}, \bibinfo {author} {\bibfnamefont
  {A.}~\bibnamefont {Jelen}}, \bibinfo {author} {\bibfnamefont
  {S.}~\bibnamefont {Jazbec}}, \bibinfo {author} {\bibfnamefont
  {Z.}~\bibnamefont {Jagli\ifmmode \check{c}\else
  \v{c}\fi{}i\ifmmode~\acute{c}\else \'{c}\fi{}}}, \bibinfo {author}
  {\bibfnamefont {S.}~\bibnamefont {Maiti}}, \bibinfo {author} {\bibfnamefont
  {M.}~\bibnamefont {Feuerbacher}}, \bibinfo {author} {\bibfnamefont
  {W.}~\bibnamefont {Steurer}}, \ and\ \bibinfo {author} {\bibfnamefont
  {J.}~\bibnamefont {Dolin\ifmmode~\check{s}\else \v{s}\fi{}ek}},\ }\href@noop
  {} {\bibfield  {journal} {\bibinfo  {journal} {Phys. Rev. Lett.}\ }\textbf
  {\bibinfo {volume} {113}},\ \bibinfo {pages} {107001} (\bibinfo {year}
  {2014})}\BibitemShut {NoStop}%
\bibitem [{\citenamefont {Dugdale}(2014)}]{dugdale:14}%
  \BibitemOpen
  \bibfield  {author} {\bibinfo {author} {\bibfnamefont {S.~B.}\ \bibnamefont
  {Dugdale}},\ }\href {\doibase http://dx.doi.org/10.1063/1.4869588} {\bibfield
   {journal} {\bibinfo  {journal} {Low Temperature Physics}\ }\textbf {\bibinfo
  {volume} {40}},\ \bibinfo {pages} {328} (\bibinfo {year} {2014})}\BibitemShut
  {NoStop}%
\bibitem [{\citenamefont {Gyorffy}\ \emph {et~al.}(1985)\citenamefont
  {Gyorffy}, \citenamefont {Pindor}, \citenamefont {Staunton}, \citenamefont
  {Stocks},\ and\ \citenamefont {Winter}}]{gyorffy:85}%
  \BibitemOpen
  \bibfield  {author} {\bibinfo {author} {\bibfnamefont {B.~L.}\ \bibnamefont
  {Gyorffy}}, \bibinfo {author} {\bibfnamefont {A.~J.}\ \bibnamefont {Pindor}},
  \bibinfo {author} {\bibfnamefont {J.}~\bibnamefont {Staunton}}, \bibinfo
  {author} {\bibfnamefont {G.~M.}\ \bibnamefont {Stocks}}, \ and\ \bibinfo
  {author} {\bibfnamefont {H.}~\bibnamefont {Winter}},\ }\href@noop {}
  {\bibfield  {journal} {\bibinfo  {journal} {Journal of Physics F: Metal
  Physics}\ }\textbf {\bibinfo {volume} {15}},\ \bibinfo {pages} {1337}
  (\bibinfo {year} {1985})}\BibitemShut {NoStop}%
\bibitem [{\citenamefont {Staunton}\ \emph {et~al.}(1985)\citenamefont
  {Staunton}, \citenamefont {Gyorffy}, \citenamefont {Pindor}, \citenamefont
  {Stocks},\ and\ \citenamefont {Winter}}]{staunton:85}%
  \BibitemOpen
  \bibfield  {author} {\bibinfo {author} {\bibfnamefont {J.}~\bibnamefont
  {Staunton}}, \bibinfo {author} {\bibfnamefont {B.~L.}\ \bibnamefont
  {Gyorffy}}, \bibinfo {author} {\bibfnamefont {A.~J.}\ \bibnamefont {Pindor}},
  \bibinfo {author} {\bibfnamefont {G.~M.}\ \bibnamefont {Stocks}}, \ and\
  \bibinfo {author} {\bibfnamefont {H.}~\bibnamefont {Winter}},\ }\href@noop {}
  {\bibfield  {journal} {\bibinfo  {journal} {Journal of Physics F: Metal
  Physics}\ }\textbf {\bibinfo {volume} {15}},\ \bibinfo {pages} {1387}
  (\bibinfo {year} {1985})}\BibitemShut {NoStop}%
\bibitem [{\citenamefont {Ebert}(2017)}]{sprkkr}%
  \BibitemOpen
  \bibfield  {author} {\bibinfo {author} {\bibfnamefont {H.}~\bibnamefont
  {Ebert}},\ }\href {http://olymp.cup.uni-muenchen.de/ak/ebert/SPRKKR} {\emph
  {\bibinfo {title} {The Munich SPR-KKR package, version 7}}} (\bibinfo {year}
  {2017})\BibitemShut {NoStop}%
\bibitem [{\citenamefont {Faulkner}\ and\ \citenamefont
  {Stocks}(1980)}]{faulkner:80}%
  \BibitemOpen
  \bibfield  {author} {\bibinfo {author} {\bibfnamefont {J.~S.}\ \bibnamefont
  {Faulkner}}\ and\ \bibinfo {author} {\bibfnamefont {G.~M.}\ \bibnamefont
  {Stocks}},\ }\href@noop {} {\bibfield  {journal} {\bibinfo  {journal} {Phys.
  Rev. B}\ }\textbf {\bibinfo {volume} {21}},\ \bibinfo {pages} {3222}
  (\bibinfo {year} {1980})}\BibitemShut {NoStop}%
\bibitem [{\citenamefont {Fehlner}\ and\ \citenamefont
  {Vosko}(1976)}]{fehlner:76}%
  \BibitemOpen
  \bibfield  {author} {\bibinfo {author} {\bibfnamefont {W.~R.}\ \bibnamefont
  {Fehlner}}\ and\ \bibinfo {author} {\bibfnamefont {S.~H.}\ \bibnamefont
  {Vosko}},\ }\href@noop {} {\bibfield  {journal} {\bibinfo  {journal}
  {Canadian Journal of Physics}\ }\textbf {\bibinfo {volume} {54}},\ \bibinfo
  {pages} {2159} (\bibinfo {year} {1976})}\BibitemShut {NoStop}%
\bibitem [{\citenamefont {Hiraoka}\ \emph {et~al.}(2001)\citenamefont
  {Hiraoka}, \citenamefont {Itou}, \citenamefont {Ohata}, \citenamefont
  {Mizumaki}, \citenamefont {Sakurai},\ and\ \citenamefont
  {Sakai}}]{hiraoka:01}%
  \BibitemOpen
  \bibfield  {author} {\bibinfo {author} {\bibfnamefont {N.}~\bibnamefont
  {Hiraoka}}, \bibinfo {author} {\bibfnamefont {M.}~\bibnamefont {Itou}},
  \bibinfo {author} {\bibfnamefont {T.}~\bibnamefont {Ohata}}, \bibinfo
  {author} {\bibfnamefont {M.}~\bibnamefont {Mizumaki}}, \bibinfo {author}
  {\bibfnamefont {Y.}~\bibnamefont {Sakurai}}, \ and\ \bibinfo {author}
  {\bibfnamefont {N.}~\bibnamefont {Sakai}},\ }\href@noop {} {\bibfield
  {journal} {\bibinfo  {journal} {Journal of Synchrotron Radiation}\ }\textbf
  {\bibinfo {volume} {8}},\ \bibinfo {pages} {26} (\bibinfo {year}
  {2001})}\BibitemShut {NoStop}%
\bibitem [{\citenamefont {Kontrym-Sznajd}\ and\ \citenamefont
  {Samsel-Czeka\l{}a}(2000)}]{sznajd:00}%
  \BibitemOpen
  \bibfield  {author} {\bibinfo {author} {\bibfnamefont {G.}~\bibnamefont
  {Kontrym-Sznajd}}\ and\ \bibinfo {author} {\bibfnamefont {M.}~\bibnamefont
  {Samsel-Czeka\l{}a}},\ }\href@noop {} {\bibfield  {journal} {\bibinfo
  {journal} {Applied Physics A}\ }\textbf {\bibinfo {volume} {70}},\ \bibinfo
  {pages} {89} (\bibinfo {year} {2000})}\BibitemShut {NoStop}%
\bibitem [{\citenamefont {Lock}\ \emph {et~al.}(1973)\citenamefont {Lock},
  \citenamefont {Crisp},\ and\ \citenamefont {West}}]{lcw:73}%
  \BibitemOpen
  \bibfield  {author} {\bibinfo {author} {\bibfnamefont {D.~G.}\ \bibnamefont
  {Lock}}, \bibinfo {author} {\bibfnamefont {V.~H.~C.}\ \bibnamefont {Crisp}},
  \ and\ \bibinfo {author} {\bibfnamefont {R.~N.}\ \bibnamefont {West}},\
  }\href@noop {} {\bibfield  {journal} {\bibinfo  {journal} {Journal of Physics
  F: Metal Physics}\ }\textbf {\bibinfo {volume} {3}},\ \bibinfo {pages} {561}
  (\bibinfo {year} {1973})}\BibitemShut {NoStop}%
\bibitem [{sup()}]{supp}%
  \BibitemOpen
  \href@noop {} {}\bibinfo {note} {See Supplemental Material.}\BibitemShut
  {Stop}%
\bibitem [{\citenamefont {Weber}\ \emph {et~al.}(2017)\citenamefont {Weber},
  \citenamefont {Benea}, \citenamefont {Appelt}, \citenamefont {Ceeh},
  \citenamefont {Kreuzpaintner}, \citenamefont {Leitner}, \citenamefont
  {Vollhardt}, \citenamefont {Hugenschmidt},\ and\ \citenamefont
  {Chioncel}}]{weber:17}%
  \BibitemOpen
  \bibfield  {author} {\bibinfo {author} {\bibfnamefont {J.~A.}\ \bibnamefont
  {Weber}}, \bibinfo {author} {\bibfnamefont {D.}~\bibnamefont {Benea}},
  \bibinfo {author} {\bibfnamefont {W.~H.}\ \bibnamefont {Appelt}}, \bibinfo
  {author} {\bibfnamefont {H.}~\bibnamefont {Ceeh}}, \bibinfo {author}
  {\bibfnamefont {W.}~\bibnamefont {Kreuzpaintner}}, \bibinfo {author}
  {\bibfnamefont {M.}~\bibnamefont {Leitner}}, \bibinfo {author} {\bibfnamefont
  {D.}~\bibnamefont {Vollhardt}}, \bibinfo {author} {\bibfnamefont
  {C.}~\bibnamefont {Hugenschmidt}}, \ and\ \bibinfo {author} {\bibfnamefont
  {L.}~\bibnamefont {Chioncel}},\ }\href@noop {} {\bibfield  {journal}
  {\bibinfo  {journal} {Phys. Rev. B}\ }\textbf {\bibinfo {volume} {95}},\
  \bibinfo {pages} {075119} (\bibinfo {year} {2017})}\BibitemShut {NoStop}%
\bibitem [{\citenamefont {Pippard}(1957)}]{pippard:57}%
  \BibitemOpen
  \bibfield  {author} {\bibinfo {author} {\bibfnamefont {A.~B.}\ \bibnamefont
  {Pippard}},\ }\href {\doibase 10.1098/rsta.1957.0023} {\bibfield  {journal}
  {\bibinfo  {journal} {Philosophical Transactions of the Royal Society of
  London A: Mathematical, Physical and Engineering Sciences}\ }\textbf
  {\bibinfo {volume} {250}},\ \bibinfo {pages} {325} (\bibinfo {year}
  {1957})}\BibitemShut {NoStop}%
\bibitem [{\citenamefont {Kontrym-Sznajd}\ and\ \citenamefont
  {Dugdale}(2015)}]{kontrymsznajd:15}%
  \BibitemOpen
  \bibfield  {author} {\bibinfo {author} {\bibfnamefont {G.}~\bibnamefont
  {Kontrym-Sznajd}}\ and\ \bibinfo {author} {\bibfnamefont {S.~B.}\
  \bibnamefont {Dugdale}},\ }\href@noop {} {\bibfield  {journal} {\bibinfo
  {journal} {Journal of Physics: Condensed Matter}\ }\textbf {\bibinfo {volume}
  {27}},\ \bibinfo {pages} {435501} (\bibinfo {year} {2015})}\BibitemShut
  {NoStop}%
\bibitem [{\citenamefont {Gradhand}\ \emph {et~al.}(2011)\citenamefont
  {Gradhand}, \citenamefont {Fedorov}, \citenamefont {Pientka}, \citenamefont
  {Zahn}, \citenamefont {Mertig},\ and\ \citenamefont
  {Gy\"orffy}}]{gradhand:11}%
  \BibitemOpen
  \bibfield  {author} {\bibinfo {author} {\bibfnamefont {M.}~\bibnamefont
  {Gradhand}}, \bibinfo {author} {\bibfnamefont {D.~V.}\ \bibnamefont
  {Fedorov}}, \bibinfo {author} {\bibfnamefont {F.}~\bibnamefont {Pientka}},
  \bibinfo {author} {\bibfnamefont {P.}~\bibnamefont {Zahn}}, \bibinfo {author}
  {\bibfnamefont {I.}~\bibnamefont {Mertig}}, \ and\ \bibinfo {author}
  {\bibfnamefont {B.~L.}\ \bibnamefont {Gy\"orffy}},\ }\href@noop {} {\bibfield
   {journal} {\bibinfo  {journal} {Phys. Rev. B}\ }\textbf {\bibinfo {volume}
  {84}},\ \bibinfo {pages} {075113} (\bibinfo {year} {2011})}\BibitemShut
  {NoStop}%
\bibitem [{\citenamefont {Jin}\ \emph {et~al.}(2016)\citenamefont {Jin},
  \citenamefont {Sales}, \citenamefont {Stocks}, \citenamefont {Samolyuk},
  \citenamefont {Daene}, \citenamefont {Weber}, \citenamefont {Zhang},\ and\
  \citenamefont {Bei}}]{jin:16b}%
  \BibitemOpen
  \bibfield  {author} {\bibinfo {author} {\bibfnamefont {K.}~\bibnamefont
  {Jin}}, \bibinfo {author} {\bibfnamefont {B.~C.}\ \bibnamefont {Sales}},
  \bibinfo {author} {\bibfnamefont {G.~M.}\ \bibnamefont {Stocks}}, \bibinfo
  {author} {\bibfnamefont {G.~D.}\ \bibnamefont {Samolyuk}}, \bibinfo {author}
  {\bibfnamefont {M.}~\bibnamefont {Daene}}, \bibinfo {author} {\bibfnamefont
  {W.~J.}\ \bibnamefont {Weber}}, \bibinfo {author} {\bibfnamefont
  {Y.}~\bibnamefont {Zhang}}, \ and\ \bibinfo {author} {\bibfnamefont
  {H.}~\bibnamefont {Bei}},\ }\href@noop {} {\bibfield  {journal} {\bibinfo
  {journal} {Scientific Reports}\ }\textbf {\bibinfo {volume} {6}},\ \bibinfo
  {pages} {20159} (\bibinfo {year} {2016})}\BibitemShut {NoStop}%
\bibitem [{\citenamefont {Samolyuk}\ \emph {et~al.}(2019)\citenamefont
  {Samolyuk}, \citenamefont {Homes}, \citenamefont {May}, \citenamefont {Mu},
  \citenamefont {Jin}, \citenamefont {Bei}, \citenamefont {Stocks},\ and\
  \citenamefont {Sales}}]{samolyuk:19}%
  \BibitemOpen
  \bibfield  {author} {\bibinfo {author} {\bibfnamefont {G.~D.}\ \bibnamefont
  {Samolyuk}}, \bibinfo {author} {\bibfnamefont {C.~C.}\ \bibnamefont {Homes}},
  \bibinfo {author} {\bibfnamefont {A.~F.}\ \bibnamefont {May}}, \bibinfo
  {author} {\bibfnamefont {S.}~\bibnamefont {Mu}}, \bibinfo {author}
  {\bibfnamefont {K.}~\bibnamefont {Jin}}, \bibinfo {author} {\bibfnamefont
  {H.}~\bibnamefont {Bei}}, \bibinfo {author} {\bibfnamefont {G.~M.}\
  \bibnamefont {Stocks}}, \ and\ \bibinfo {author} {\bibfnamefont {B.~C.}\
  \bibnamefont {Sales}},\ }\href {\doibase 10.1103/PhysRevB.100.075128}
  {\bibfield  {journal} {\bibinfo  {journal} {Phys. Rev. B}\ }\textbf {\bibinfo
  {volume} {100}},\ \bibinfo {pages} {075128} (\bibinfo {year}
  {2019})}\BibitemShut {NoStop}%
\bibitem [{\citenamefont {Duffy}(2013)}]{duffy:13}%
  \BibitemOpen
  \bibfield  {author} {\bibinfo {author} {\bibfnamefont {J.~A.}\ \bibnamefont
  {Duffy}},\ }\href@noop {} {\bibfield  {journal} {\bibinfo  {journal} {Journal
  of Physics: Conference Series}\ }\textbf {\bibinfo {volume} {443}},\ \bibinfo
  {pages} {012011} (\bibinfo {year} {2013})}\BibitemShut {NoStop}%
\bibitem [{\citenamefont {Haynes}\ \emph {et~al.}(2012)\citenamefont {Haynes},
  \citenamefont {Maskery}, \citenamefont {Butchers}, \citenamefont {Duffy},
  \citenamefont {Taylor}, \citenamefont {Giblin}, \citenamefont {Utfeld},
  \citenamefont {Laverock}, \citenamefont {Dugdale}, \citenamefont {Sakurai},
  \citenamefont {Itou}, \citenamefont {Pfleiderer}, \citenamefont
  {Hirschberger}, \citenamefont {Neubauer}, \citenamefont {Duncan},\ and\
  \citenamefont {Grosche}}]{haynes:12}%
  \BibitemOpen
  \bibfield  {author} {\bibinfo {author} {\bibfnamefont {T.~D.}\ \bibnamefont
  {Haynes}}, \bibinfo {author} {\bibfnamefont {I.}~\bibnamefont {Maskery}},
  \bibinfo {author} {\bibfnamefont {M.~W.}\ \bibnamefont {Butchers}}, \bibinfo
  {author} {\bibfnamefont {J.~A.}\ \bibnamefont {Duffy}}, \bibinfo {author}
  {\bibfnamefont {J.~W.}\ \bibnamefont {Taylor}}, \bibinfo {author}
  {\bibfnamefont {S.~R.}\ \bibnamefont {Giblin}}, \bibinfo {author}
  {\bibfnamefont {C.}~\bibnamefont {Utfeld}}, \bibinfo {author} {\bibfnamefont
  {J.}~\bibnamefont {Laverock}}, \bibinfo {author} {\bibfnamefont {S.~B.}\
  \bibnamefont {Dugdale}}, \bibinfo {author} {\bibfnamefont {Y.}~\bibnamefont
  {Sakurai}}, \bibinfo {author} {\bibfnamefont {M.}~\bibnamefont {Itou}},
  \bibinfo {author} {\bibfnamefont {C.}~\bibnamefont {Pfleiderer}}, \bibinfo
  {author} {\bibfnamefont {M.}~\bibnamefont {Hirschberger}}, \bibinfo {author}
  {\bibfnamefont {A.}~\bibnamefont {Neubauer}}, \bibinfo {author}
  {\bibfnamefont {W.}~\bibnamefont {Duncan}}, \ and\ \bibinfo {author}
  {\bibfnamefont {F.~M.}\ \bibnamefont {Grosche}},\ }\href@noop {} {\bibfield
  {journal} {\bibinfo  {journal} {Phys. Rev. B}\ }\textbf {\bibinfo {volume}
  {85}},\ \bibinfo {pages} {115137} (\bibinfo {year} {2012})}\BibitemShut
  {NoStop}%
\bibitem [{\citenamefont {Liz\'arraga}\ \emph {et~al.}(2018)\citenamefont
  {Liz\'arraga}, \citenamefont {Holmstr\"om},\ and\ \citenamefont
  {Vitos}}]{lizarraga:18}%
  \BibitemOpen
  \bibfield  {author} {\bibinfo {author} {\bibfnamefont {R.}~\bibnamefont
  {Liz\'arraga}}, \bibinfo {author} {\bibfnamefont {E.}~\bibnamefont
  {Holmstr\"om}}, \ and\ \bibinfo {author} {\bibfnamefont {L.}~\bibnamefont
  {Vitos}},\ }\href@noop {} {\bibfield  {journal} {\bibinfo  {journal} {Phys.
  Rev. Materials}\ }\textbf {\bibinfo {volume} {2}},\ \bibinfo {pages} {094407}
  (\bibinfo {year} {2018})}\BibitemShut {NoStop}%
\bibitem [{res()}]{researchdata}%
  \BibitemOpen
  \href {\doibase 10.5523/bris.1a9u533pcxzk23lbwu1p56ica} {\
  10.5523/bris.1a9u533pcxzk23lbwu1p56ica}\BibitemShut {NoStop}%
\end{thebibliography}%


\begin{thebibliography}{31}%
\makeatletter
\providecommand \@ifxundefined [1]{%
 \@ifx{#1\undefined}
}%
\providecommand \@ifnum [1]{%
 \ifnum #1\expandafter \@firstoftwo
 \else \expandafter \@secondoftwo
 \fi
}%
\providecommand \@ifx [1]{%
 \ifx #1\expandafter \@firstoftwo
 \else \expandafter \@secondoftwo
 \fi
}%
\providecommand \natexlab [1]{#1}%
\providecommand \enquote  [1]{``#1''}%
\providecommand \bibnamefont  [1]{#1}%
\providecommand \bibfnamefont [1]{#1}%
\providecommand \citenamefont [1]{#1}%
\providecommand \href@noop [0]{\@secondoftwo}%
\providecommand \href [0]{\begingroup \@sanitize@url \@href}%
\providecommand \@href[1]{\@@startlink{#1}\@@href}%
\providecommand \@@href[1]{\endgroup#1\@@endlink}%
\providecommand \@sanitize@url [0]{\catcode `\\12\catcode `\$12\catcode
  `\&12\catcode `\#12\catcode `\^12\catcode `\_12\catcode `\%12\relax}%
\providecommand \@@startlink[1]{}%
\providecommand \@@endlink[0]{}%
\providecommand \url  [0]{\begingroup\@sanitize@url \@url }%
\providecommand \@url [1]{\endgroup\@href {#1}{\urlprefix }}%
\providecommand \urlprefix  [0]{URL }%
\providecommand \Eprint [0]{\href }%
\providecommand \doibase [0]{http://dx.doi.org/}%
\providecommand \selectlanguage [0]{\@gobble}%
\providecommand \bibinfo  [0]{\@secondoftwo}%
\providecommand \bibfield  [0]{\@secondoftwo}%
\providecommand \translation [1]{[#1]}%
\providecommand \BibitemOpen [0]{}%
\providecommand \bibitemStop [0]{}%
\providecommand \bibitemNoStop [0]{.\EOS\space}%
\providecommand \EOS [0]{\spacefactor3000\relax}%
\providecommand \BibitemShut  [1]{\csname bibitem#1\endcsname}%
\let\auto@bib@innerbib\@empty
\bibitem [{\citenamefont {Wu}\ \emph {et~al.}(2014)\citenamefont {Wu},
  \citenamefont {Bei}, \citenamefont {Otto}, \citenamefont {Pharr},\ and\
  \citenamefont {George}}]{wu:14}%
  \BibitemOpen
  \bibfield  {author} {\bibinfo {author} {\bibfnamefont {Z.}~\bibnamefont
  {Wu}}, \bibinfo {author} {\bibfnamefont {H.}~\bibnamefont {Bei}}, \bibinfo
  {author} {\bibfnamefont {F.}~\bibnamefont {Otto}}, \bibinfo {author}
  {\bibfnamefont {G.}~\bibnamefont {Pharr}}, \ and\ \bibinfo {author}
  {\bibfnamefont {E.}~\bibnamefont {George}},\ }\href@noop {} {\bibfield
  {journal} {\bibinfo  {journal} {Intermetallics}\ }\textbf {\bibinfo {volume}
  {46}},\ \bibinfo {pages} {131 } (\bibinfo {year} {2014})}\BibitemShut
  {NoStop}%
\bibitem [{\citenamefont {Bei}\ and\ \citenamefont {George}(2005)}]{bei:05}%
  \BibitemOpen
  \bibfield  {author} {\bibinfo {author} {\bibfnamefont {H.}~\bibnamefont
  {Bei}}\ and\ \bibinfo {author} {\bibfnamefont {E.}~\bibnamefont {George}},\
  }\href@noop {} {\bibfield  {journal} {\bibinfo  {journal} {Acta Materialia}\
  }\textbf {\bibinfo {volume} {53}},\ \bibinfo {pages} {69 } (\bibinfo {year}
  {2005})}\BibitemShut {NoStop}%
\bibitem [{\citenamefont {Platzman}\ and\ \citenamefont
  {Tzoar}(1965)}]{platzman:65}%
  \BibitemOpen
  \bibfield  {author} {\bibinfo {author} {\bibfnamefont {P.~M.}\ \bibnamefont
  {Platzman}}\ and\ \bibinfo {author} {\bibfnamefont {N.}~\bibnamefont
  {Tzoar}},\ }\href@noop {} {\bibfield  {journal} {\bibinfo  {journal} {Phys.
  Rev.}\ }\textbf {\bibinfo {volume} {139}},\ \bibinfo {pages} {A410} (\bibinfo
  {year} {1965})}\BibitemShut {NoStop}%
\bibitem [{\citenamefont {Eisenberger}\ and\ \citenamefont
  {Platzman}(1970)}]{eisenberger:70}%
  \BibitemOpen
  \bibfield  {author} {\bibinfo {author} {\bibfnamefont {P.}~\bibnamefont
  {Eisenberger}}\ and\ \bibinfo {author} {\bibfnamefont {P.~M.}\ \bibnamefont
  {Platzman}},\ }\href@noop {} {\bibfield  {journal} {\bibinfo  {journal}
  {Phys. Rev. A}\ }\textbf {\bibinfo {volume} {2}},\ \bibinfo {pages} {415}
  (\bibinfo {year} {1970})}\BibitemShut {NoStop}%
\bibitem [{\citenamefont {Lock}\ \emph {et~al.}(1973)\citenamefont {Lock},
  \citenamefont {Crisp},\ and\ \citenamefont {West}}]{lcw:73}%
  \BibitemOpen
  \bibfield  {author} {\bibinfo {author} {\bibfnamefont {D.~G.}\ \bibnamefont
  {Lock}}, \bibinfo {author} {\bibfnamefont {V.~H.~C.}\ \bibnamefont {Crisp}},
  \ and\ \bibinfo {author} {\bibfnamefont {R.~N.}\ \bibnamefont {West}},\
  }\href@noop {} {\bibfield  {journal} {\bibinfo  {journal} {Journal of Physics
  F: Metal Physics}\ }\textbf {\bibinfo {volume} {3}},\ \bibinfo {pages} {561}
  (\bibinfo {year} {1973})}\BibitemShut {NoStop}%
\bibitem [{\citenamefont {Fehlner}\ and\ \citenamefont
  {Vosko}(1976)}]{fehlner:76}%
  \BibitemOpen
  \bibfield  {author} {\bibinfo {author} {\bibfnamefont {W.~R.}\ \bibnamefont
  {Fehlner}}\ and\ \bibinfo {author} {\bibfnamefont {S.~H.}\ \bibnamefont
  {Vosko}},\ }\href@noop {} {\bibfield  {journal} {\bibinfo  {journal}
  {Canadian Journal of Physics}\ }\textbf {\bibinfo {volume} {54}},\ \bibinfo
  {pages} {2159} (\bibinfo {year} {1976})}\BibitemShut {NoStop}%
\bibitem [{\citenamefont {Zukowski}(2004)}]{zukowski:04}%
  \BibitemOpen
  \bibfield  {author} {\bibinfo {author} {\bibfnamefont {E.}~\bibnamefont
  {Zukowski}},\ }in\ \href@noop {} {\emph {\bibinfo {booktitle} {X-ray Compton
  Scattering}}},\ \bibinfo {editor} {edited by\ \bibinfo {editor}
  {\bibfnamefont {M.~J.}\ \bibnamefont {Cooper}}, \bibinfo {editor}
  {\bibfnamefont {P.~E.}\ \bibnamefont {Mijnarends}}, \bibinfo {editor}
  {\bibfnamefont {N.}~\bibnamefont {Shiotani}}, \bibinfo {editor}
  {\bibfnamefont {N.}~\bibnamefont {Sakai}}, \ and\ \bibinfo {editor}
  {\bibfnamefont {A.}~\bibnamefont {Bansil}}}\ (\bibinfo  {publisher} {Oxford
  University Press},\ \bibinfo {address} {Oxford},\ \bibinfo {year} {2004})\
  Chap.\ \bibinfo {chapter} {The processing of experimental data}\BibitemShut
  {NoStop}%
\bibitem [{\citenamefont {Sakai}(1987)}]{sakai:87}%
  \BibitemOpen
  \bibfield  {author} {\bibinfo {author} {\bibfnamefont {N.}~\bibnamefont
  {Sakai}},\ }\href@noop {} {\bibfield  {journal} {\bibinfo  {journal} {Journal
  of the Physical Society of Japan}\ }\textbf {\bibinfo {volume} {56}},\
  \bibinfo {pages} {2477} (\bibinfo {year} {1987})}\BibitemShut {NoStop}%
\bibitem [{\citenamefont {Biggs}\ \emph {et~al.}(1975)\citenamefont {Biggs},
  \citenamefont {Mendelsohn},\ and\ \citenamefont {Mann}}]{biggs:75}%
  \BibitemOpen
  \bibfield  {author} {\bibinfo {author} {\bibfnamefont {F.}~\bibnamefont
  {Biggs}}, \bibinfo {author} {\bibfnamefont {L.}~\bibnamefont {Mendelsohn}}, \
  and\ \bibinfo {author} {\bibfnamefont {J.}~\bibnamefont {Mann}},\ }\href@noop
  {} {\bibfield  {journal} {\bibinfo  {journal} {Atomic Data and Nuclear Data
  Tables}\ }\textbf {\bibinfo {volume} {16}},\ \bibinfo {pages} {201 }
  (\bibinfo {year} {1975})}\BibitemShut {NoStop}%
\bibitem [{\citenamefont {Korringa}(1947)}]{korringa:47}%
  \BibitemOpen
  \bibfield  {author} {\bibinfo {author} {\bibfnamefont {J.}~\bibnamefont
  {Korringa}},\ }\href@noop {} {\bibfield  {journal} {\bibinfo  {journal}
  {Physica}\ }\textbf {\bibinfo {volume} {13}},\ \bibinfo {pages} {392 }
  (\bibinfo {year} {1947})}\BibitemShut {NoStop}%
\bibitem [{\citenamefont {Kohn}\ and\ \citenamefont
  {Rostoker}(1954)}]{kohn:54}%
  \BibitemOpen
  \bibfield  {author} {\bibinfo {author} {\bibfnamefont {W.}~\bibnamefont
  {Kohn}}\ and\ \bibinfo {author} {\bibfnamefont {N.}~\bibnamefont
  {Rostoker}},\ }\href@noop {} {\bibfield  {journal} {\bibinfo  {journal}
  {Phys. Rev.}\ }\textbf {\bibinfo {volume} {94}},\ \bibinfo {pages} {1111}
  (\bibinfo {year} {1954})}\BibitemShut {NoStop}%
\bibitem [{\citenamefont {Soven}(1967)}]{soven:67}%
  \BibitemOpen
  \bibfield  {author} {\bibinfo {author} {\bibfnamefont {P.}~\bibnamefont
  {Soven}},\ }\href@noop {} {\bibfield  {journal} {\bibinfo  {journal} {Phys.
  Rev.}\ }\textbf {\bibinfo {volume} {156}},\ \bibinfo {pages} {809} (\bibinfo
  {year} {1967})}\BibitemShut {NoStop}%
\bibitem [{\citenamefont {Gyorffy}(1972)}]{gyorffy:72}%
  \BibitemOpen
  \bibfield  {author} {\bibinfo {author} {\bibfnamefont {B.~L.}\ \bibnamefont
  {Gyorffy}},\ }\href@noop {} {\bibfield  {journal} {\bibinfo  {journal} {Phys.
  Rev. B}\ }\textbf {\bibinfo {volume} {5}},\ \bibinfo {pages} {2382} (\bibinfo
  {year} {1972})}\BibitemShut {NoStop}%
\bibitem [{\citenamefont {Zunger}\ \emph {et~al.}(1990)\citenamefont {Zunger},
  \citenamefont {Wei}, \citenamefont {Ferreira},\ and\ \citenamefont
  {Bernard}}]{zunger:90}%
  \BibitemOpen
  \bibfield  {author} {\bibinfo {author} {\bibfnamefont {A.}~\bibnamefont
  {Zunger}}, \bibinfo {author} {\bibfnamefont {S.-H.}\ \bibnamefont {Wei}},
  \bibinfo {author} {\bibfnamefont {L.~G.}\ \bibnamefont {Ferreira}}, \ and\
  \bibinfo {author} {\bibfnamefont {J.~E.}\ \bibnamefont {Bernard}},\
  }\href@noop {} {\bibfield  {journal} {\bibinfo  {journal} {Phys. Rev. Lett.}\
  }\textbf {\bibinfo {volume} {65}},\ \bibinfo {pages} {353} (\bibinfo {year}
  {1990})}\BibitemShut {NoStop}%
\bibitem [{\citenamefont {Wei}\ \emph {et~al.}(1990)\citenamefont {Wei},
  \citenamefont {Ferreira}, \citenamefont {Bernard},\ and\ \citenamefont
  {Zunger}}]{wei:90}%
  \BibitemOpen
  \bibfield  {author} {\bibinfo {author} {\bibfnamefont {S.-H.}\ \bibnamefont
  {Wei}}, \bibinfo {author} {\bibfnamefont {L.~G.}\ \bibnamefont {Ferreira}},
  \bibinfo {author} {\bibfnamefont {J.~E.}\ \bibnamefont {Bernard}}, \ and\
  \bibinfo {author} {\bibfnamefont {A.}~\bibnamefont {Zunger}},\ }\href@noop {}
  {\bibfield  {journal} {\bibinfo  {journal} {Phys. Rev. B}\ }\textbf {\bibinfo
  {volume} {42}},\ \bibinfo {pages} {9622} (\bibinfo {year}
  {1990})}\BibitemShut {NoStop}%
\bibitem [{\citenamefont {van~de Walle}(2009)}]{avdw:atat2}%
  \BibitemOpen
  \bibfield  {author} {\bibinfo {author} {\bibfnamefont {A.}~\bibnamefont
  {van~de Walle}},\ }\href {\doibase 10.1016/j.calphad.2008.12.005} {\bibfield
  {journal} {\bibinfo  {journal} {Calphad}\ }\textbf {\bibinfo {volume} {33}},\
  \bibinfo {pages} {266} (\bibinfo {year} {2009})}\BibitemShut {NoStop}%
\bibitem [{\citenamefont {van~de Walle}\ \emph {et~al.}(2002)\citenamefont
  {van~de Walle}, \citenamefont {Asta},\ and\ \citenamefont
  {Ceder}}]{avdw:atat}%
  \BibitemOpen
  \bibfield  {author} {\bibinfo {author} {\bibfnamefont {A.}~\bibnamefont
  {van~de Walle}}, \bibinfo {author} {\bibfnamefont {M.~D.}\ \bibnamefont
  {Asta}}, \ and\ \bibinfo {author} {\bibfnamefont {G.}~\bibnamefont {Ceder}},\
  }\href {\doibase 10.1016/S0364-5916(02)80006-2} {\bibfield  {journal}
  {\bibinfo  {journal} {Calphad}\ }\textbf {\bibinfo {volume} {26}},\ \bibinfo
  {pages} {539} (\bibinfo {year} {2002})}\BibitemShut {NoStop}%
\bibitem [{\citenamefont {van~de Walle}\ \emph {et~al.}(2013)\citenamefont
  {van~de Walle}, \citenamefont {Tiwary}, \citenamefont {de~Jong},
  \citenamefont {Olmsted}, \citenamefont {Asta}, \citenamefont {Dick},
  \citenamefont {Shin}, \citenamefont {Wang}, \citenamefont {Chen},\ and\
  \citenamefont {Liu}}]{avdw:mcsqs}%
  \BibitemOpen
  \bibfield  {author} {\bibinfo {author} {\bibfnamefont {A.}~\bibnamefont
  {van~de Walle}}, \bibinfo {author} {\bibfnamefont {P.}~\bibnamefont
  {Tiwary}}, \bibinfo {author} {\bibfnamefont {M.~M.}\ \bibnamefont {de~Jong}},
  \bibinfo {author} {\bibfnamefont {D.~L.}\ \bibnamefont {Olmsted}}, \bibinfo
  {author} {\bibfnamefont {M.~D.}\ \bibnamefont {Asta}}, \bibinfo {author}
  {\bibfnamefont {A.}~\bibnamefont {Dick}}, \bibinfo {author} {\bibfnamefont
  {D.}~\bibnamefont {Shin}}, \bibinfo {author} {\bibfnamefont {Y.}~\bibnamefont
  {Wang}}, \bibinfo {author} {\bibfnamefont {L.-Q.}\ \bibnamefont {Chen}}, \
  and\ \bibinfo {author} {\bibfnamefont {Z.-K.}\ \bibnamefont {Liu}},\ }\href
  {\doibase 10.1016/j.calphad.2013.06.006} {\bibfield  {journal} {\bibinfo
  {journal} {Calphad}\ }\textbf {\bibinfo {volume} {42}},\ \bibinfo {pages}
  {13} (\bibinfo {year} {2013})}\BibitemShut {NoStop}%
\bibitem [{\citenamefont {Dewhurst}(2013)}]{elk}%
  \BibitemOpen
  \bibfield  {author} {\bibinfo {author} {\bibfnamefont {J.~K.}\ \bibnamefont
  {Dewhurst}},\ }\href {http://elk.sourceforge.net} {\emph {\bibinfo {title}
  {The Elk FP-LAPW code, version 2.3.22}}} (\bibinfo {year} {2013})\BibitemShut
  {NoStop}%
\bibitem [{\citenamefont {Ernsting}\ \emph {et~al.}(2014)\citenamefont
  {Ernsting}, \citenamefont {Billington}, \citenamefont {Haynes}, \citenamefont
  {Millichamp}, \citenamefont {Taylor}, \citenamefont {Duffy}, \citenamefont
  {Giblin}, \citenamefont {Dewhurst},\ and\ \citenamefont
  {Dugdale}}]{ernsting:14}%
  \BibitemOpen
  \bibfield  {author} {\bibinfo {author} {\bibfnamefont {D.}~\bibnamefont
  {Ernsting}}, \bibinfo {author} {\bibfnamefont {D.}~\bibnamefont
  {Billington}}, \bibinfo {author} {\bibfnamefont {T.~D.}\ \bibnamefont
  {Haynes}}, \bibinfo {author} {\bibfnamefont {T.~E.}\ \bibnamefont
  {Millichamp}}, \bibinfo {author} {\bibfnamefont {J.~W.}\ \bibnamefont
  {Taylor}}, \bibinfo {author} {\bibfnamefont {J.~A.}\ \bibnamefont {Duffy}},
  \bibinfo {author} {\bibfnamefont {S.~R.}\ \bibnamefont {Giblin}}, \bibinfo
  {author} {\bibfnamefont {J.~K.}\ \bibnamefont {Dewhurst}}, \ and\ \bibinfo
  {author} {\bibfnamefont {S.~B.}\ \bibnamefont {Dugdale}},\ }\href@noop {}
  {\bibfield  {journal} {\bibinfo  {journal} {Journal of Physics: Condensed
  Matter}\ }\textbf {\bibinfo {volume} {26}},\ \bibinfo {pages} {495501}
  (\bibinfo {year} {2014})}\BibitemShut {NoStop}%
\bibitem [{\citenamefont {Perdew}\ and\ \citenamefont
  {Wang}(1992)}]{perdew:92}%
  \BibitemOpen
  \bibfield  {author} {\bibinfo {author} {\bibfnamefont {J.~P.}\ \bibnamefont
  {Perdew}}\ and\ \bibinfo {author} {\bibfnamefont {Y.}~\bibnamefont {Wang}},\
  }\href@noop {} {\bibfield  {journal} {\bibinfo  {journal} {Phys. Rev. B}\
  }\textbf {\bibinfo {volume} {45}},\ \bibinfo {pages} {13244} (\bibinfo {year}
  {1992})}\BibitemShut {NoStop}%
\bibitem [{\citenamefont {Vosko}\ \emph {et~al.}(1980)\citenamefont {Vosko},
  \citenamefont {Wilk},\ and\ \citenamefont {Nusair}}]{vosko:80}%
  \BibitemOpen
  \bibfield  {author} {\bibinfo {author} {\bibfnamefont {S.~H.}\ \bibnamefont
  {Vosko}}, \bibinfo {author} {\bibfnamefont {L.}~\bibnamefont {Wilk}}, \ and\
  \bibinfo {author} {\bibfnamefont {M.}~\bibnamefont {Nusair}},\ }\href
  {\doibase 10.1139/p80-159} {\bibfield  {journal} {\bibinfo  {journal}
  {Canadian Journal of Physics}\ }\textbf {\bibinfo {volume} {58}},\ \bibinfo
  {pages} {1200} (\bibinfo {year} {1980})}\BibitemShut {NoStop}%
\bibitem [{\citenamefont {Kontrym-Sznajd}\ and\ \citenamefont
  {Samsel-Czeka{\l}a}(2011)}]{sznajd:11}%
  \BibitemOpen
  \bibfield  {author} {\bibinfo {author} {\bibfnamefont {G.}~\bibnamefont
  {Kontrym-Sznajd}}\ and\ \bibinfo {author} {\bibfnamefont {M.}~\bibnamefont
  {Samsel-Czeka{\l}a}},\ }\href@noop {} {\bibfield  {journal} {\bibinfo
  {journal} {Journal of Applied Crystallography}\ }\textbf {\bibinfo {volume}
  {44}},\ \bibinfo {pages} {1246} (\bibinfo {year} {2011})}\BibitemShut
  {NoStop}%
\bibitem [{\citenamefont {Kontrym-Sznajd}\ and\ \citenamefont
  {Samsel-Czeka\l{}a}(2000)}]{sznajd:00}%
  \BibitemOpen
  \bibfield  {author} {\bibinfo {author} {\bibfnamefont {G.}~\bibnamefont
  {Kontrym-Sznajd}}\ and\ \bibinfo {author} {\bibfnamefont {M.}~\bibnamefont
  {Samsel-Czeka\l{}a}},\ }\href@noop {} {\bibfield  {journal} {\bibinfo
  {journal} {Applied Physics A}\ }\textbf {\bibinfo {volume} {70}},\ \bibinfo
  {pages} {89} (\bibinfo {year} {2000})}\BibitemShut {NoStop}%
\bibitem [{\citenamefont {Ebert}(2017)}]{sprkkr}%
  \BibitemOpen
  \bibfield  {author} {\bibinfo {author} {\bibfnamefont {H.}~\bibnamefont
  {Ebert}},\ }\href {http://olymp.cup.uni-muenchen.de/ak/ebert/SPRKKR} {\emph
  {\bibinfo {title} {The Munich SPR-KKR package, version 7}}} (\bibinfo {year}
  {2017})\BibitemShut {NoStop}%
\bibitem [{\citenamefont {Tulip}\ \emph {et~al.}(2008)\citenamefont {Tulip},
  \citenamefont {Staunton}, \citenamefont {Lowitzer}, \citenamefont
  {K\"odderitzsch},\ and\ \citenamefont {Ebert}}]{tulip:08}%
  \BibitemOpen
  \bibfield  {author} {\bibinfo {author} {\bibfnamefont {P.~R.}\ \bibnamefont
  {Tulip}}, \bibinfo {author} {\bibfnamefont {J.~B.}\ \bibnamefont {Staunton}},
  \bibinfo {author} {\bibfnamefont {S.}~\bibnamefont {Lowitzer}}, \bibinfo
  {author} {\bibfnamefont {D.}~\bibnamefont {K\"odderitzsch}}, \ and\ \bibinfo
  {author} {\bibfnamefont {H.}~\bibnamefont {Ebert}},\ }\href@noop {}
  {\bibfield  {journal} {\bibinfo  {journal} {Phys. Rev. B}\ }\textbf {\bibinfo
  {volume} {77}},\ \bibinfo {pages} {165116} (\bibinfo {year}
  {2008})}\BibitemShut {NoStop}%
\bibitem [{\citenamefont {Kubo}(1957)}]{kubo:57}%
  \BibitemOpen
  \bibfield  {author} {\bibinfo {author} {\bibfnamefont {R.}~\bibnamefont
  {Kubo}},\ }\href@noop {} {\bibfield  {journal} {\bibinfo  {journal} {Journal
  of the Physical Society of Japan}\ }\textbf {\bibinfo {volume} {12}},\
  \bibinfo {pages} {570} (\bibinfo {year} {1957})}\BibitemShut {NoStop}%
\bibitem [{\citenamefont {Greenwood}(1958)}]{greenwood:58}%
  \BibitemOpen
  \bibfield  {author} {\bibinfo {author} {\bibfnamefont {D.~A.}\ \bibnamefont
  {Greenwood}},\ }\href@noop {} {\bibfield  {journal} {\bibinfo  {journal}
  {Proceedings of the Physical Society}\ }\textbf {\bibinfo {volume} {71}},\
  \bibinfo {pages} {585} (\bibinfo {year} {1958})}\BibitemShut {NoStop}%
\bibitem [{\citenamefont {Szotek}\ \emph {et~al.}(1984)\citenamefont {Szotek},
  \citenamefont {Gyorffy}, \citenamefont {Stocks},\ and\ \citenamefont
  {Temmerman}}]{szotek:84}%
  \BibitemOpen
  \bibfield  {author} {\bibinfo {author} {\bibfnamefont {Z.}~\bibnamefont
  {Szotek}}, \bibinfo {author} {\bibfnamefont {B.~L.}\ \bibnamefont {Gyorffy}},
  \bibinfo {author} {\bibfnamefont {G.~M.}\ \bibnamefont {Stocks}}, \ and\
  \bibinfo {author} {\bibfnamefont {W.~M.}\ \bibnamefont {Temmerman}},\
  }\href@noop {} {\bibfield  {journal} {\bibinfo  {journal} {Journal of Physics
  F: Metal Physics}\ }\textbf {\bibinfo {volume} {14}},\ \bibinfo {pages}
  {2571} (\bibinfo {year} {1984})}\BibitemShut {NoStop}%
\bibitem [{\citenamefont {Kao}\ \emph {et~al.}(2011)\citenamefont {Kao},
  \citenamefont {Chen}, \citenamefont {Chen}, \citenamefont {Chu},
  \citenamefont {Yeh},\ and\ \citenamefont {Lin}}]{kao:11}%
  \BibitemOpen
  \bibfield  {author} {\bibinfo {author} {\bibfnamefont {Y.-F.}\ \bibnamefont
  {Kao}}, \bibinfo {author} {\bibfnamefont {S.-K.}\ \bibnamefont {Chen}},
  \bibinfo {author} {\bibfnamefont {T.-J.}\ \bibnamefont {Chen}}, \bibinfo
  {author} {\bibfnamefont {P.-C.}\ \bibnamefont {Chu}}, \bibinfo {author}
  {\bibfnamefont {J.-W.}\ \bibnamefont {Yeh}}, \ and\ \bibinfo {author}
  {\bibfnamefont {S.-J.}\ \bibnamefont {Lin}},\ }\href@noop {} {\bibfield
  {journal} {\bibinfo  {journal} {Journal of Alloys and Compounds}\ }\textbf
  {\bibinfo {volume} {509}},\ \bibinfo {pages} {1607 } (\bibinfo {year}
  {2011})}\BibitemShut {NoStop}%
\bibitem [{\citenamefont {Jin}\ \emph {et~al.}(2017)\citenamefont {Jin},
  \citenamefont {Mu}, \citenamefont {An}, \citenamefont {Porter}, \citenamefont
  {Samolyuk}, \citenamefont {Stocks},\ and\ \citenamefont {Bei}}]{jin:16}%
  \BibitemOpen
  \bibfield  {author} {\bibinfo {author} {\bibfnamefont {K.}~\bibnamefont
  {Jin}}, \bibinfo {author} {\bibfnamefont {S.}~\bibnamefont {Mu}}, \bibinfo
  {author} {\bibfnamefont {K.}~\bibnamefont {An}}, \bibinfo {author}
  {\bibfnamefont {W.}~\bibnamefont {Porter}}, \bibinfo {author} {\bibfnamefont
  {G.}~\bibnamefont {Samolyuk}}, \bibinfo {author} {\bibfnamefont
  {G.}~\bibnamefont {Stocks}}, \ and\ \bibinfo {author} {\bibfnamefont
  {H.}~\bibnamefont {Bei}},\ }\href@noop {} {\bibfield  {journal} {\bibinfo
  {journal} {Materials \& Design}\ }\textbf {\bibinfo {volume} {117}},\
  \bibinfo {pages} {185 } (\bibinfo {year} {2017})}\BibitemShut {NoStop}%
\end{thebibliography}%

\end{document}


\title{Supplemental material: Extreme Fermi surface smearing in a maximally disordered concentrated solid solution}

\author{Hannah C. Robarts}
\author{Thomas E. Millichamp}
\author{Daniel A. Lagos}
\author{Jude Laverock}
\affiliation{H.H. Wills Physics Laboratory, University of Bristol, Tyndall Avenue, Bristol BS8 1TL, UK}

\author{David Billington}
\affiliation{Japan Synchrotron Radiation Research Institute, SPring-8, Sayo, 679-5198, Japan}

\author{Jonathan A.~Duffy}
\author{Daniel O'Neill}
\affiliation{Department of Physics, University of Warwick, Coventry, CV4 7AL, United Kingdom}

\author{Sean R.~Giblin}
\affiliation{School of Physics and Astronomy, Cardiff University, Queen's Building, The Parade, Cardiff, CF24 3AA, United Kingdom}

\author{Jonathan W.~Taylor}
\affiliation{DMSC - European Spallation Source,  Universitetsparken 1, Copenhagen 2100, Denmark}

\author{Grazyna Kontrym-Sznajd}
\author{Ma{\l}gorzata Samsel-Czeka{\l}a}
\affiliation{Institute of Low Temperature and Structure Research, Polish Academy of Sciences, PO Box 1410, 50-950 Wroc{\l}aw 2, Poland}

\author{Hongbin Bei}
\author{Sai Mu}
\author{German D.~Samolyuk}
\author{G.~Malcolm Stocks}
\affiliation{Materials Science and Technology Division, Oak Ridge National Laboratory, Oak Ridge, TN 37831, USA}

\author{Stephen B.~Dugdale}
\email{s.b.dugdale@bristol.ac.uk}
\affiliation{H.H. Wills Physics Laboratory, University of Bristol, Tyndall Avenue, Bristol BS8 1TL, UK}

\date{\today}
\begin{abstract}
\end{abstract}

\maketitle{}

\beginsupplement

\section*{Materials preparation}

Details of the single crystal growth can be found in Refs.\cite{wu:14,bei:05}. An ingot was produced by arc melting Ni, Fe, Co and Cr
in a water-cooled copper hearth, under an Ar atmosphere. The arc-melted buttons were flipped and remelted five times in order to improve
the compositional homogeneity, before being drop cast into square cross-section copper moulds. These polycrystalline
ingots were then loaded into an optical floating zone furnace to produce a single crystal which was subsequently cut into a disc suitable for the
Compton scattering experiment using electro-discharge machining before being electrolytically polished to remove any damage caused by the cutting.

\section*{Compton scattering}

Within the impulse approximation \cite{platzman:65,eisenberger:70}, a Compton profile, $J(p_z)$, is an integral over two momentum components (orthogonal to the scattering vector parallel to $p_z$) of the three dimensional electron momentum density, $\rho ({\mathbf p})$ :

\begin{eqnarray}
J(p_z) = \int_{-\infty}^\infty \int_{-\infty}^\infty \rho ({\mathbf p}) dp_x dp_y .
\label{compton}
\end{eqnarray}
Since the Compton profile contains information about the occupied momentum states, in metals it can be used to infer the shape of
the Fermi surface. Applying the so-called Lock--Crisp--West (LCW) prescription to transform from real momentum ${\mathbf p}$ to
crystal momentum ${\mathbf k}$ reduces (or folds)
the momentum density back into the first Brillouin zone (1BZ) using reciprocal lattice vectors \cite{lcw:73} and
produces a periodic occupation number, $n({\mathbf k})$. For a perfectly ordered system, $n({\mathbf k})$ is simply the number of occupied bands at a particular ${\mathbf k}$. Following such LCW folding, the contributions of electrons in fully occupied bands sum to give a constant contribution across the Brillouin zone (with an intensity
proportional to the number of occupied bands), whereas the signatures of the Fermi surface (which are initally distributed throughout ${\mathbf p}$-space) due to electrons in partially occupied bands
reinforce constructively, producing discontinuities (which are step-like for a pure material, but
smeared out in an alloy) in $n({\mathbf k})$.
Compton profiles were measured at room temperature along 15 special crystallographic directions \cite{fehlner:76} using the
spectrometer on beamline BL08W at the SPring-8 synchrotron in
Japan. The incident x-ray energy was 115~keV, the momentum resolution was estimated to be 0.10~a.u., and each profile contained of the order of 800,000 counts in the Compton peak. The profiles were corrected for background, absorption, cross-section and multiple scattering \cite{zukowski:04,sakai:87}.
A core contribution (comprising
relativistic Hartree-Fock $1s^2 2s^2 2p^6 3s^2 3p^6$ Compton profiles for each element \cite{biggs:75}) was subtracted from each profile. The contributions of these
core electrons to the momentum distribution would anyway sum to give a constant contribution in the LCW-folded distribution \cite{lcw:73}.

\section*{Treating the substitutional disorder}

Substitutional disorder was treated theoretically in two different approaches. Most of the calculations presented here use the
KKR method \cite{korringa:47,kohn:54} for calculating the electronic structure, and the coherent potential approximation (CPA) \cite{soven:67,gyorffy:72}
for the chemical and magnetic disorder. The alternative method involves creating and populating a supercell, and some of the calculations presented
here used the special quasirandom structure (SQS) method to generate a special supercell in which the physically most relevant atomic correlation functions
match most closely the perfectly random solid solution \cite{zunger:90,wei:90}. A 32-atom SQS supercell was generated using the ATAT code
\cite{avdw:atat2,avdw:atat,avdw:mcsqs}.

\section*{Electronic structure calculations}

\begin{figure}
        \begin{center}
        \includegraphics[angle=0,width=1.00\linewidth]{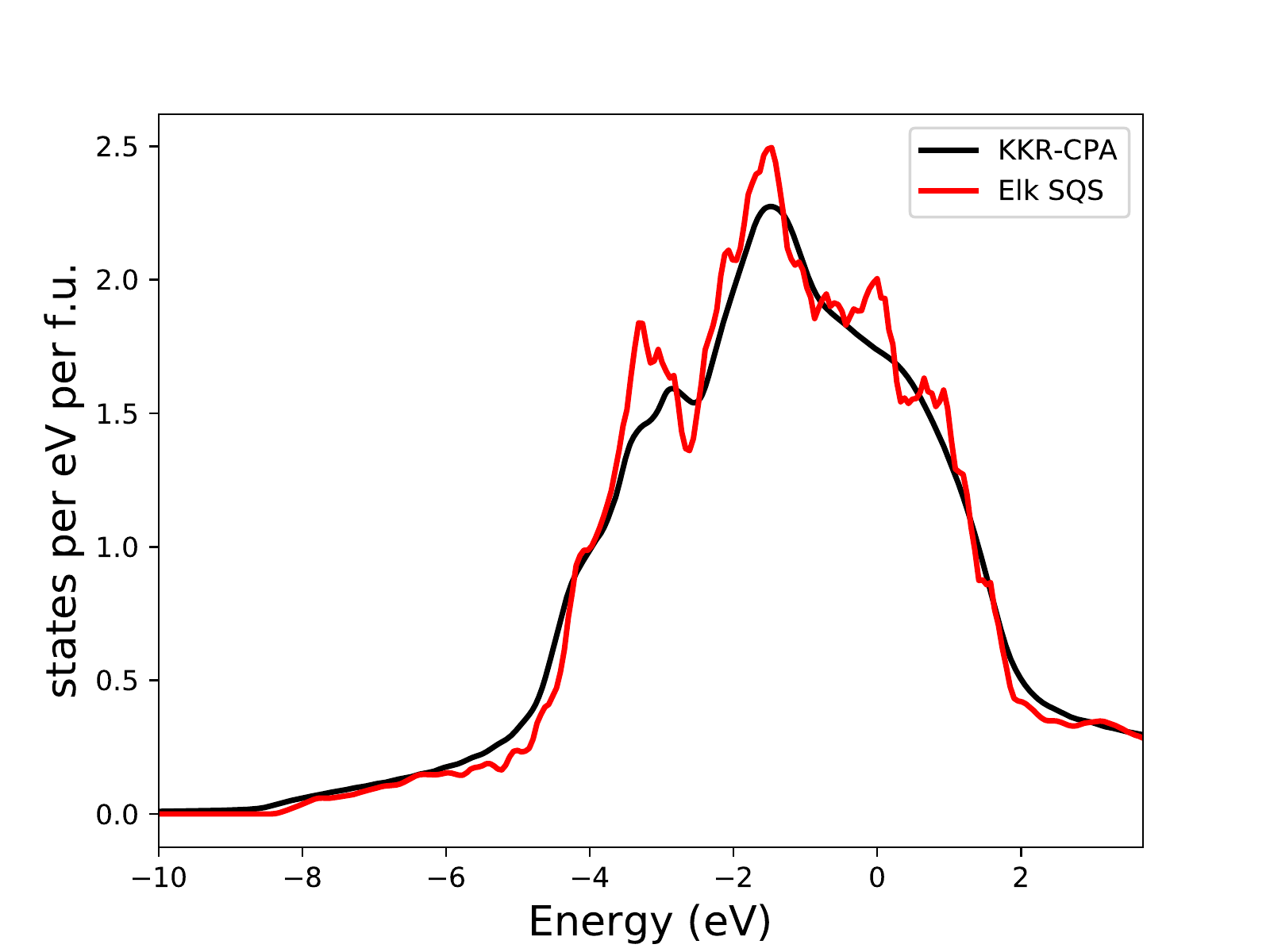}
        \end{center}
\caption{The electron density of states (DOS) per atom for NiFeCoCr calculated within the KKR-CPA and the ELK FP-APW+lo SQS frameworks.}
        \label{dos}
\end{figure}

The supercell calculations were performed with the full-potential augmented planewave + local orbital (FP-APW+lo) Elk code \cite{elk}, 
The SQS 32-atom supercell calculations were performed with 256 $k$-points within the irreducible Brillouin zone. The core electron configurations
were $1s^2 2s^2 2p^6 3s^2$ for Ni and Co and $1s^2 2s^2 2p^6$ for Fe and Cr. The muffin-tin radii were 2.4 a.u. and the plane-wave cutoff in the interstitial region was determined by $\vert \mathbf{G}
+ \mathbf{k} \vert_{\rm max} = 7.0 / R_{\rm mt}$, where $R_{\rm mt}$ is the average muffin-tin radius. The momentum density was calculated
up to a maximum momentum of 12 a.u. \cite{ernsting:14}.
The local density approximation to the exchange-correlation energy functional was Perdew-Wang (PW92) \cite{perdew:92}.
For the KKR-CPA calculation, performed within the atomic sphere approximation (ASA), the core configuration for all elements was $1s^2 2s^2 2p^6 3s^2 3p^6$. 
Note that the $3p$ states were located between 35 and 70 eV below the Fermi energy. 
The muffin-tin radii were 2.38 a.u. and 1200 $k$-points were used to sample the irreducible Brillouin zone. The local density approximation to the exchange-correlation energy functional was Vosko-Wilk-Nusair (VWN) \cite{vosko:80}.
For all calculations the lattice constant used was 3.57\AA.
The densities of states (DOS) from the two different calculations are shown in Fig.~\ref{dos}. 

The motivation for the Elk SQS calculations was as a check on the KKR-CPA results as well as facilitating comparisons
with the Compton profile measurements in $\bf p$-space (see Fig.~\ref{directionalaniso} and Fig.~\ref{reconstructed_plane}).

\section*{Special directions}

\begin{figure}
        \begin{center}
        \includegraphics[angle=0,width=1.00\linewidth]{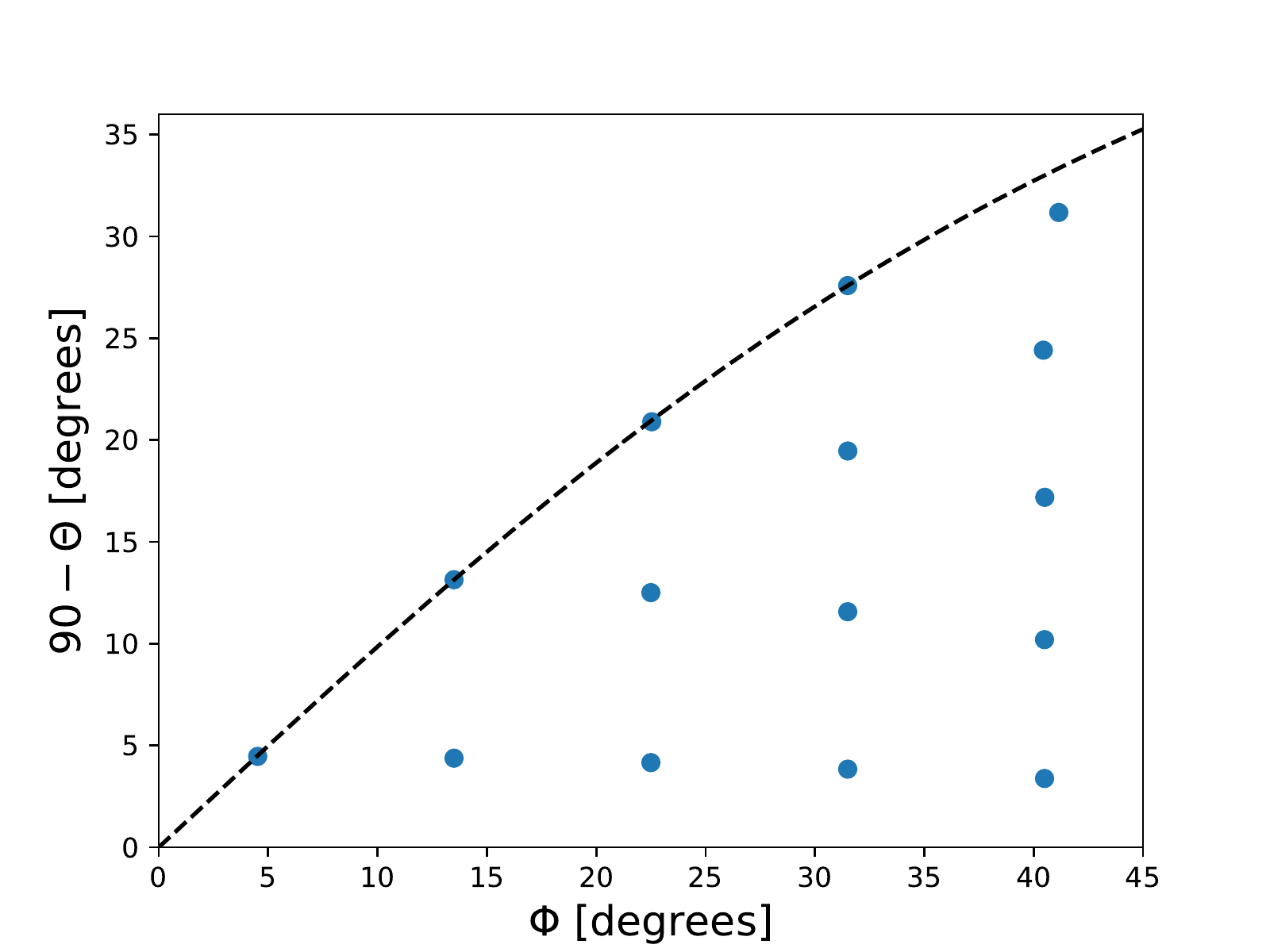}
        \end{center}
\caption{The set of 15 special directions measured in the experiment, as described in Table~\ref{sdtable}. The [100] direction is specified by $(\Theta,\phi)$ being $(90^{\circ},0)$ and [110] by $(90^{\circ},45)$. The dashed line indicates the irreducible wedge of ${\mathbf p}$-space for cubic symmetry.}
        \label{sds}
\end{figure}

When reconstructing 3D electron momentum densities from Compton profiles, it is more profitable to make the measurements along particular sets of crystallographic
directions \cite{sznajd:11}, known as ``special directions'' within the Brillouin zone. 
The particular set of 15 special directions \cite{fehlner:76} are described in Table~\ref{sdtable} and 
shown in Fig.~\ref{sds}.

\begin{table}[ht]
  \begin{ruledtabular}
     \begin{tabular}{c c c}
     Number &  $\phi$ & $\Theta$  \\
     & (degrees) & (degrees) \\
      \hline
     1 & 4.52 & 85.54 \\
     2& 13.50 & 85.62 \\
     3& 13.50 & 76.86 \\ 
     4& 22.50 & 85.84 \\
     5& 22.50 & 77.49 \\
     6& 22.54 & 69.10 \\
     7& 31.50 & 86.16 \\
     8& 31.50 & 78.43 \\
     9& 31.50 & 70.54 \\
    10& 31.50 & 62.41 \\
    11& 40.50 & 86.62 \\
    12& 40.50 & 79.80 \\
    13& 40.51 & 72.81 \\
    14& 40.44 & 65.59 \\
    15& 41.15 & 58.82 
     \end{tabular}
  \end{ruledtabular}
  \caption{The set of 15 special directions measured in the experiment, taken from \cite{fehlner:76}. The [100] 
direction is specified by $(\Theta,\phi)$ being $(90^{\circ},0)$ and [110] by $(90^{\circ},45^{\circ})$.}
  \label{sdtable}
\end{table}

\begin{figure}
        \begin{center}
        \includegraphics[angle=0,width=1.00\linewidth]{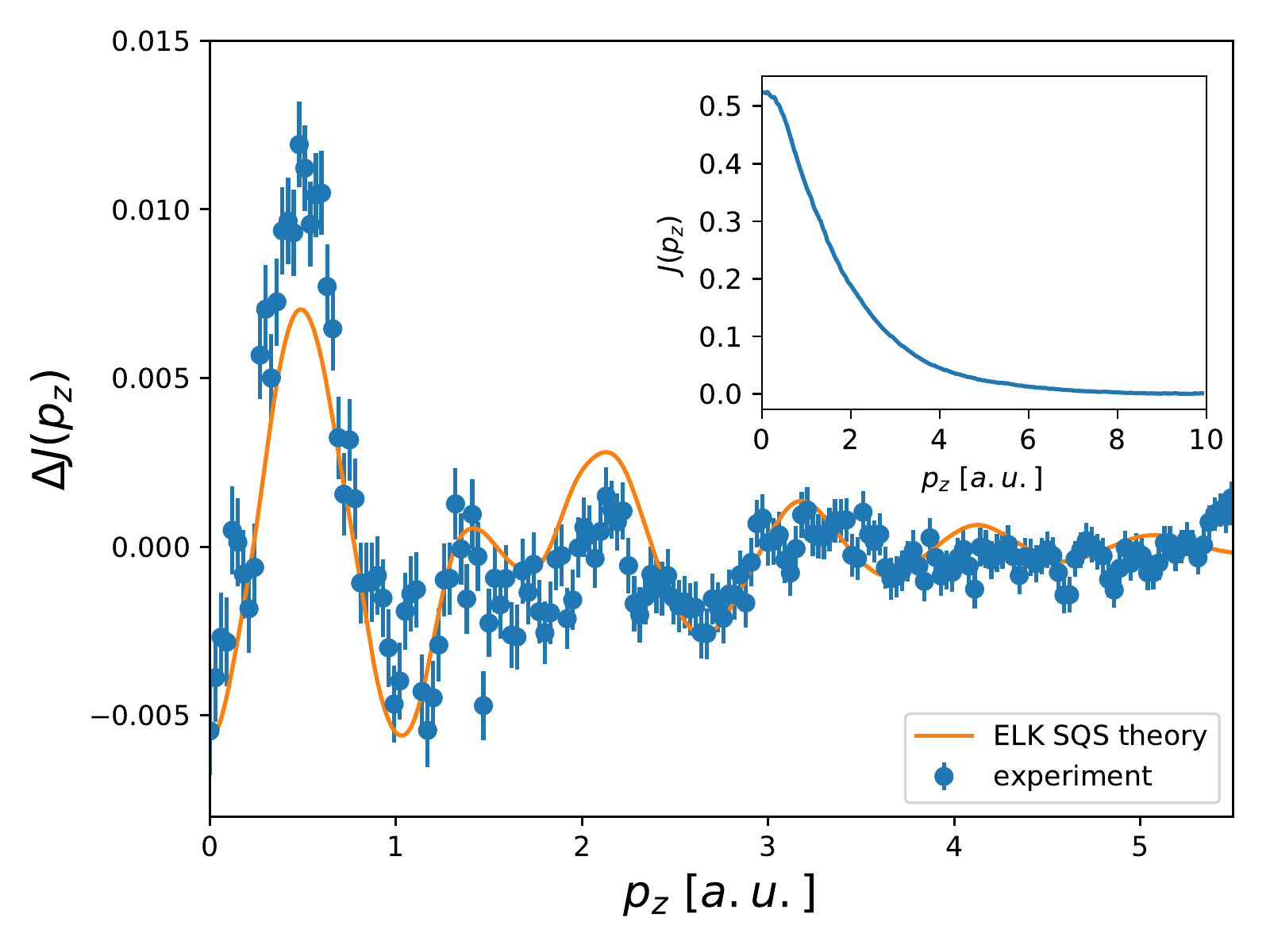}
        \end{center}
\caption{Anisotropy between the two most well-separated special directions (1 and 15) from Table~\ref{sdtable} for both the
experimental data and the Elk SQS calculation (convoluted with the experimental resolution). The inset shows
the Compton profile $J(p_z)$ along special direction 1.}
        \label{directionalaniso}
\end{figure}

The directional anisotropy (difference between Compton profiles measured (and calculated) with the momentum resolved along two different crystallographic
directions) is shown in Fig.~\ref{directionalaniso} for directions 1 and 15. The inset shows the actual Compton profile for direction 1.
 
\section*{Reconstruction}                       

The task is to reconstruct the 3D electron momentum density, $\rho ({\mathbf p})$, from the measured 1D Compton profiles, $J(p_z)$,
which are projections as described in Eq.~\ref{compton}.

Both the $J(p_z)$ and $ \rho ({\mathbf p})$ can be expanded into lattice harmonics.
Details of the three-dimensional reconstruction procedure can be found in \cite{sznajd:00}.  

\begin{figure}
        \begin{center}
        \includegraphics[angle=0,width=1.00\linewidth]{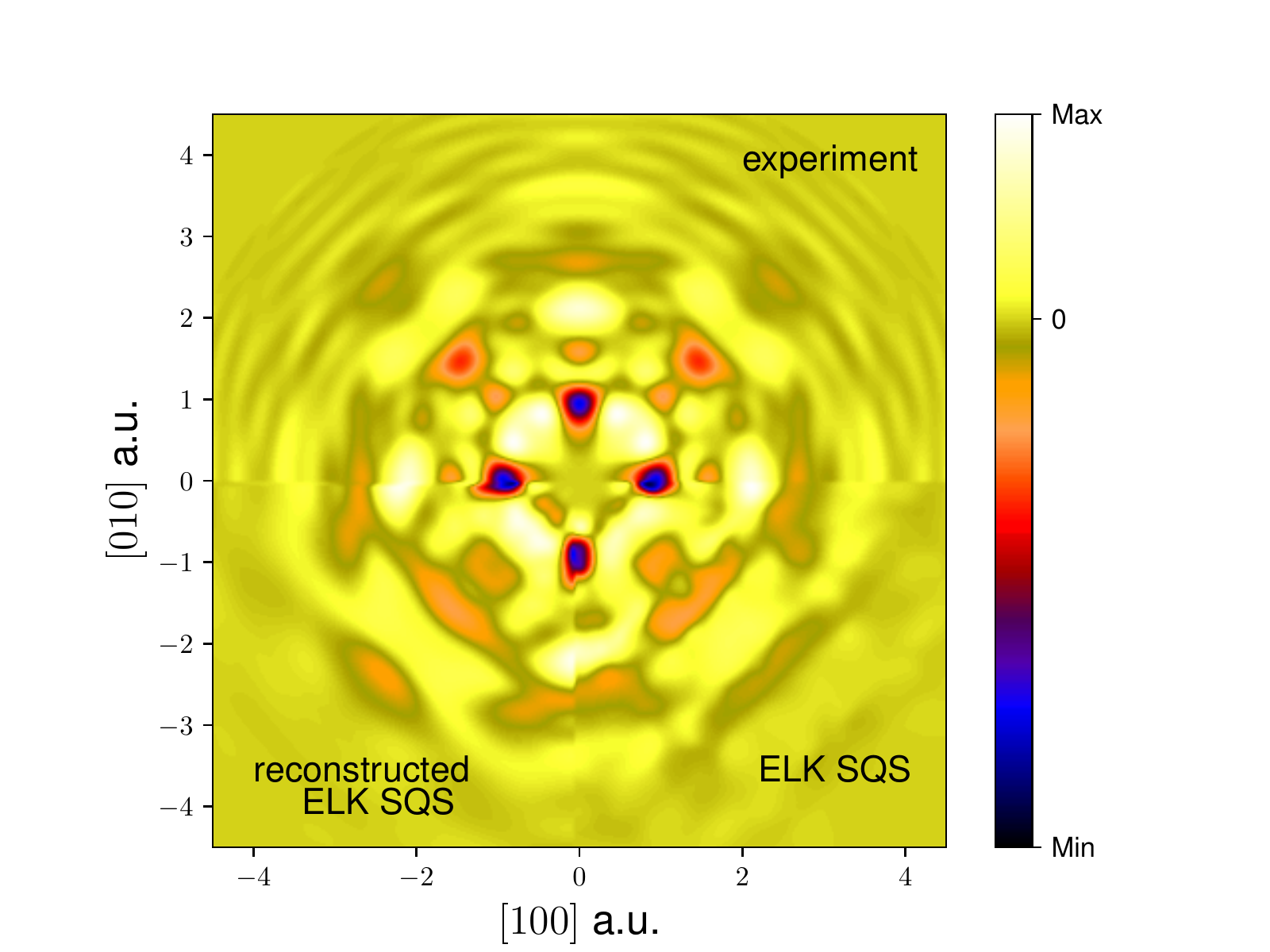}
        \end{center}
\caption{Anisotropy of the reconstructed momentum distribution in the (001) plane through $\Gamma$. The reconstruction
from the experimentally measured Compton profiles is shown at the top, and from the same set of Elk 
SQS Compton profiles (obtained by integrating the theoretical electron momentum distribution over two
momentum components) is shown
at the bottom left. The directly calculated distribution in the (001) plane (which has not gone through any reconstruction) is shown at the bottom
right. The theoretical distributions have been convoluted with the experimental resolution.}
        \label{reconstructed_plane}
\end{figure}

By subtracting an isotropic distribution (either a fitted Gaussian or the simple
angular average of the spectrum itself) from the calculated spectrum, $N(p_x,p_y)$, the anisotropy of the momentum 
distribution is highlighted. This is often referred to as the radial anisotropy,
$R(p_x,p_y)$, and is defined as
\begin{eqnarray}
R(p_x,p_y) = N(p_x,p_y) - \overline{N(p_x,p_y)}|_{p={\rm const.}}
\end{eqnarray}
In Fig.~\ref{reconstructed_plane}, this quantity is plotted (bottom right), together with the anisotropic components obtained via
3D reconstruction of Compton profiles obtained from the calculation and the experiment. In this way, the ability of reconstrution to
capture the anisotropy of the momentum distribution is demonstrated.

\section*{Obtaining the occupation in k-space}

\begin{figure}
        \begin{center}
        \includegraphics[angle=0,width=1.08\linewidth]{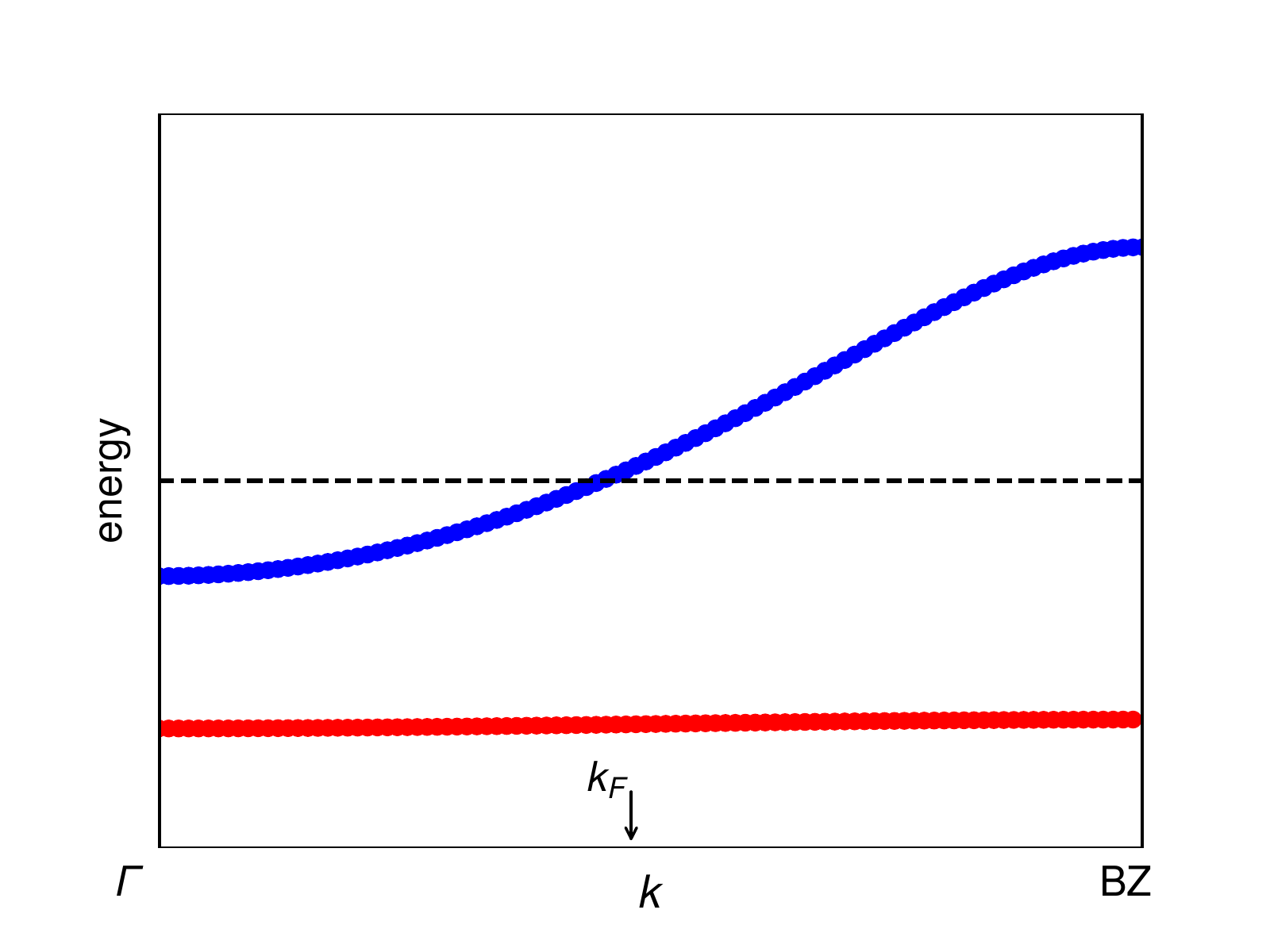}
        \includegraphics[angle=0,width=1.00\linewidth]{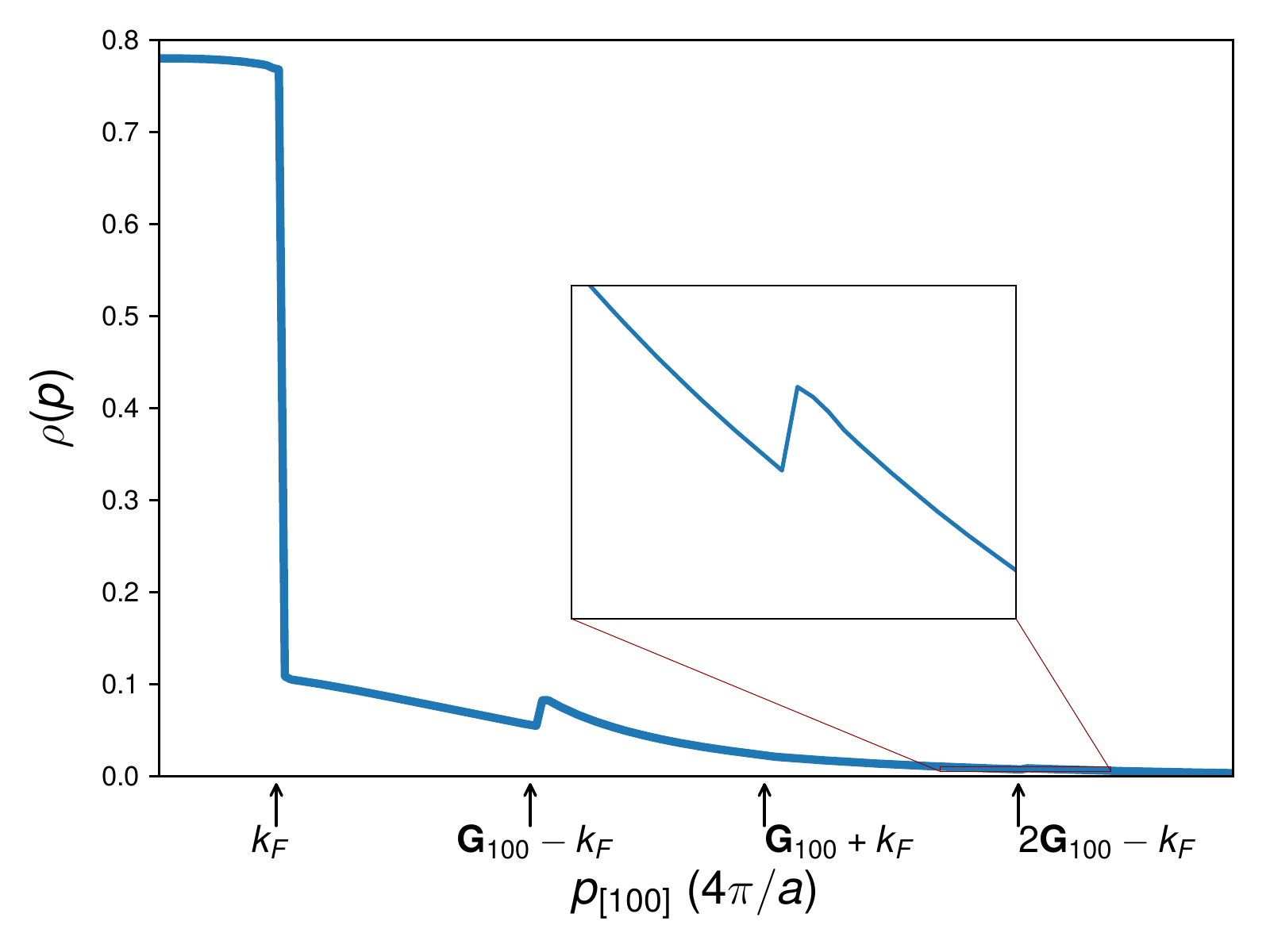}
        \includegraphics[angle=0,width=1.00\linewidth]{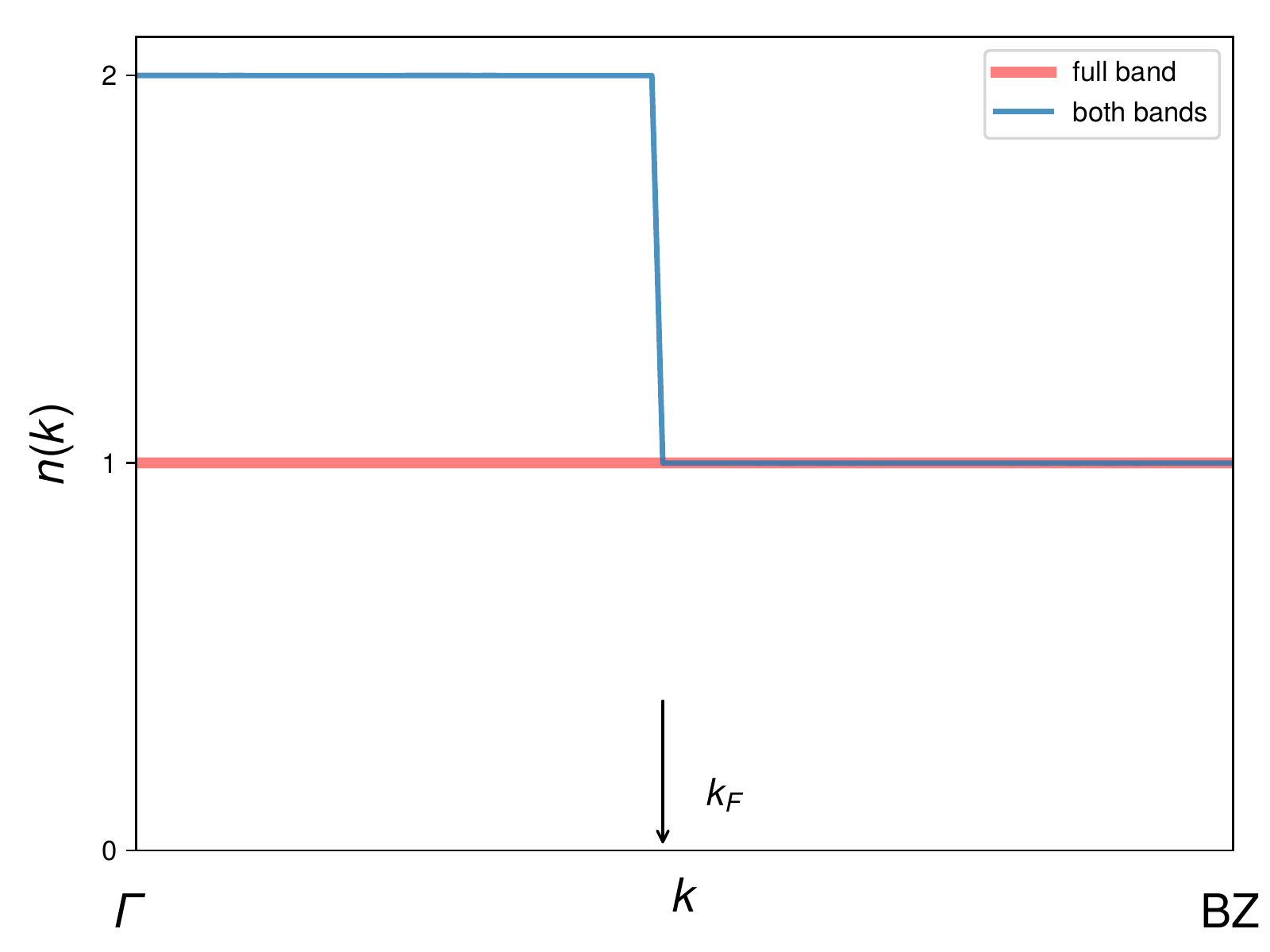}
        \end{center}
\caption{(Top) Hypothetical band structure showing two bands, one fully occupied (red) and one crossing the Fermi energy (blue) and thus partially filled. (Middle) The momentum distribution $\rho ({\mathbf p})$ associated with this band structure. As the partially filled band crosses the Fermi energy (at ${\mathbf k_F}$), the contribution to the momentum distribution from this band disappears, only to reappear (and then disappear again) in the second and subsequent Brillouin zones (at ${\mathbf p}={\mathbf k_F}\pm{\mathbf G}$) where ${\mathbf G}$ is a reciprocal lattice vector.
(Bottom) By summing contributions from all the Brillouin zones, the contribution from
the fully occupied band is constant across the zone, while that from the partially occupied band 
generates the familiar step function; this is the so-called ``LCW'' prescription \cite{lcw:73}.}
        \label{cartoonnk}
\end{figure}

Having reconstructed a full three dimensional momentum distribution, the occupation number within the Brillouin zone can be
obtained by ``folding'' that distribution back into the first Brillouin zone. It is important to recall that crystal momentum
${\mathbf k}$ is not real momentum ${\mathbf p}$, and a band which contributes at some  ${\mathbf k}$ will also contribute
at ${\mathbf p}={\mathbf k}+{\mathbf G}$ where ${\mathbf G}$ is a reciprocal lattice vector. Lock, Crisp and West \cite{lcw:73}
noted that by summing the contributions from each Brillouin zone, the discontinuities reinforce and the smooth contribution from
fully occupied bands sum to give a (possibly) large, but constant background. This is schematically illustrated in Fig.~\ref{cartoonnk} for a pair of hypothetical bands (one fully occupied and one crossing the Fermi energy).

\section*{Dip in occupation close to $\Gamma$ point}

\begin{figure}
        \begin{center}
        \includegraphics[trim={0 0cm 0 0cm},clip,angle=0,width=1.00\linewidth]{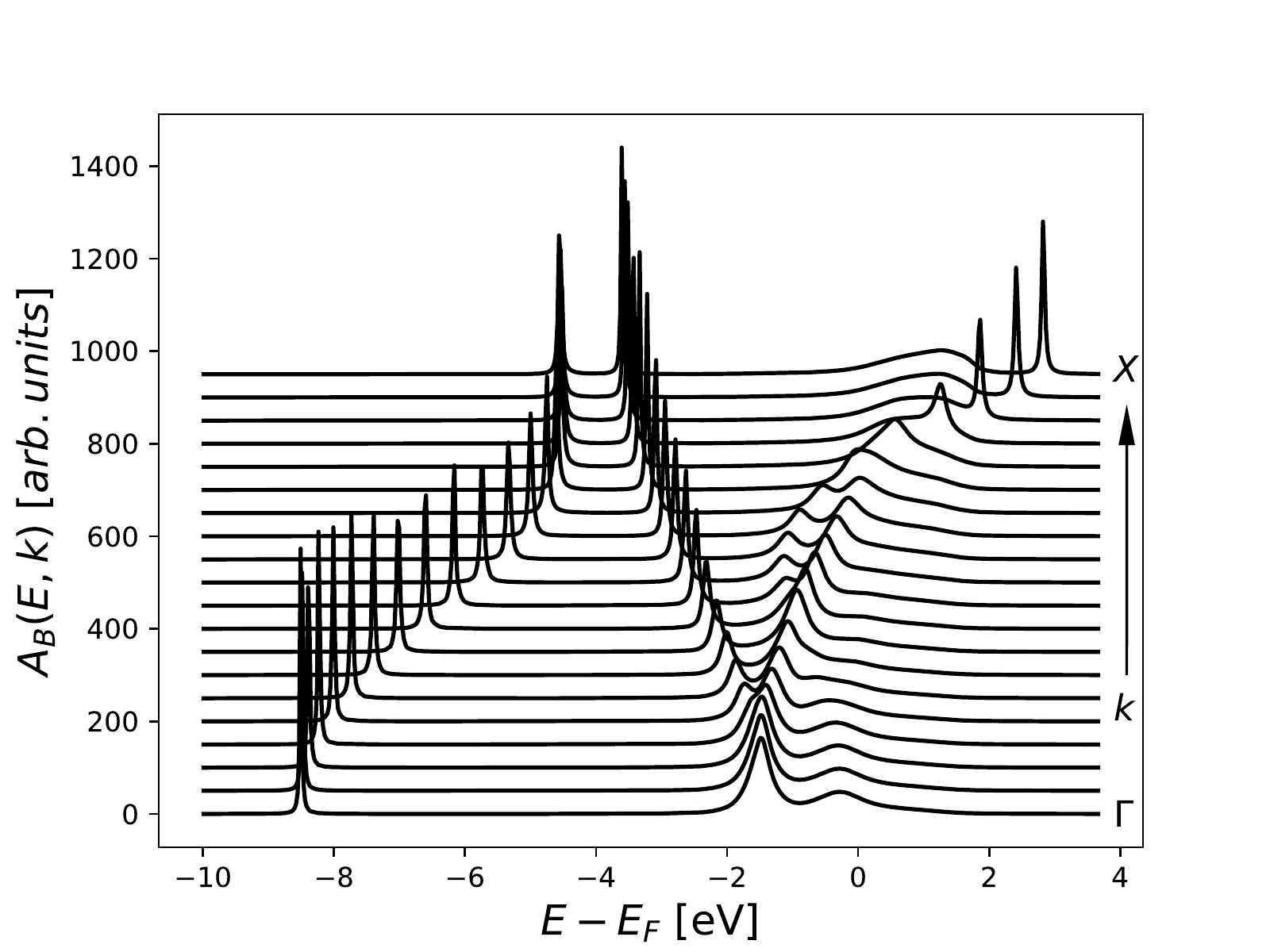}
        \end{center}
\caption{A set of Bloch spectral functions $A_B(E,{\mathbf k})$ for a selection of ${\mathbf k}$ between $\Gamma$ and $X$. Note the
presence of a peak, centred just below $E_F$ but which has a tail which extends above $E_F$ at the $\Gamma$-point. This peak initially disperses to lower energies as ${\mathbf k}$ moves towards $X$ such that spectral weight is below $E_F$.}
        \label{dip}
\end{figure}

It is interesting
to note the small dip in $n({\mathbf k})$ around the $\Gamma$ point in both the KKR-CPA-DLM theory and the experimental data (see
Fig.~2 and Fig.~4 in the main paper). This appears to be due to 
a smeared band with a tail which extends in energy across the $E_F$ close to $\Gamma$, but which initially disperses to
lower energies such that the tail eventually drops below $E_F$ leading to an increase in the occupancy as more of the ``band'' is occupied, as
shown in Fig.~\ref{dip}.  

\section*{Calculation of residual resistivity from KKR}

These calculations were performed with the Munich SPR-KKR code which applies the KKR method to the Kubo-Greenwood
formalism \cite{sprkkr,tulip:08,kubo:57,greenwood:58}.

\section*{Evaluating the mean-free-path and extracting the resistivity}

The electrical conductivity of a metal, $\sigma$, can be expressed as an integral over the Fermi surface:

\begin{eqnarray}
\sigma = \frac{e^2}{8\pi^3 \hbar} \int_{E=E_F} \frac{v^2_x({\mathbf k})}{\vert v({\mathbf k}) \vert} \tau({\mathbf k}) dS,
\end{eqnarray}
where $v({\mathbf k})$ and $\tau({\mathbf k})$ are the Fermi velocity and lifetime, respectively, of a state on the Fermi surface.
The product $v({\mathbf k})\tau({\mathbf k})$ is interpreted as the coherence length of the state.
These quantities ($v({\mathbf k})$ and $\tau({\mathbf k})$) will vary over the Fermi surface, but the quantity $\langle v^2_x ({\mathbf k})) \tau({\mathbf k})/ \vert v({\mathbf k}) \vert \rangle_{E_F}$ (i.e. a Fermi surface average) can be taken outside the integral and be approximated by $\lambda/3$ (for a cubic system,
$v^2_x ({\mathbf k})/ \vert v({\mathbf k}) \vert = v({\mathbf k})/3$), where $\lambda$ is the electronic 
mean-free-path. We also have,
\begin{eqnarray}
 \int_{E=E_F} dS = 2\mathcal{A},
\end{eqnarray}
where $\mathcal{A}$ is the Fermi surface area, and the factor of 2 is related to spin.

Thus, the contribution to the conductivity from each sheet of Fermi surface becomes:

\begin{eqnarray}
\sigma = \frac{e^2}{12\pi^3 \hbar} \lambda \mathcal{A}.
\label{fsconductivity}
\end{eqnarray}

Following Szotek {\it et al.} \cite{szotek:84},
consider a Lorentzian momentum spectral function,
\begin{eqnarray}
\rho({\mathbf k},\epsilon) = \frac{1}{\pi} \left( \frac{\hbar}{\tau_{\mathbf k}} \right)
 \left[ \frac{1}{(\epsilon-\epsilon_{\mathbf k})^2 + \left(\hbar/\tau_{\mathbf k}\right)^2}\right],
\end{eqnarray}
where $\epsilon_{\mathbf k}$ is the location in energy and $\tau_{\mathbf k}$ the lifetime of a state near the Fermi energy.  
Integrating over energy, 
\begin{eqnarray}
n({\mathbf k}) &=& \int_{-\infty}^{E_F} \rho({\mathbf k},\epsilon) d\epsilon\\
               &=&  \frac{1}{2} + \frac{1}{\pi} \arctan{\left[ \frac{\epsilon_F-\epsilon_{\mathbf k}}
{(\hbar/\tau_{\mathbf k})} \right]}. 
\end{eqnarray}
Note that there is a typographical error in Eq.~5.1 of Ref.~\cite{szotek:84} where the denominator of the argument of the $\arctan$ has been written as $\hbar\tau_{\mathbf k}$.

Near the Fermi surface, $\tau_{\mathbf k}$ can be considered to vary slowly with ${\mathbf k}$,
and can be replaced by $\tau_{\mathbf k_{F}}$. Also, Taylor expanding $\epsilon_{\mathbf k}$ around
$\epsilon_{F}$,
\begin{eqnarray}
\epsilon_{\mathbf k} &=& \epsilon_{F} + \frac{\partial \epsilon_{\mathbf k}}{\partial k} ({\mathbf k}-{\mathbf k}_{F}) \\
&=& \epsilon_{F} + \hbar v_{\mathbf k} ({\mathbf k}-{\mathbf k}_{F})
\end{eqnarray}
and therefore,
\begin{eqnarray}
n({\mathbf k}) =  \frac{1}{2} + \frac{1}{\pi} \arctan{\left[ \xi_{{\mathbf k}_{F}} ({\mathbf k} - {\mathbf k}_F ) \right]},
\end{eqnarray}
where $\xi_{{\mathbf k}_{F}}$ represents the coherence length of the state at a point ${\mathbf k}_{F}$ on
the Fermi surface.
The derivative of this function at ${\mathbf k}={\mathbf k}_F$ is,
\begin{eqnarray}
\frac{\partial n({\mathbf k})}{\partial k} =  \frac{\xi_{{\mathbf k}_{F}}}{\pi}.
\label{xi}
\end{eqnarray}

Although the spectral function lineshapes are not perfectly Lorentzian \cite{szotek:84}, it is still possible
to use the gradient to achieve a good estimate of the quasiparticle coherence length, and by extension,
the electronic mean-free-path $\lambda$ once it has been averaged over the Fermi surface
(i.e. $\lambda = \langle \xi_{{\mathbf k}_{F}} \rangle_{FS}$). 

In practice, we  found that the $n({\mathbf k})$ curves were fitted much
better by $\tanh$ functions rather than $\arctan$.
Specifically, we fitted $n({\mathbf k})$ (via two parameters, ${\mathbf k}_F$ and $b$, for each sheet of FS) with a function:
\begin{eqnarray}
n({\mathbf k}) = \frac{1}{2} \left[ 1 - \tanh{\left( \frac{{\mathbf k}-{\mathbf k}_F}{b}\right)} \right].
\label{tanh}
\end{eqnarray}
The magnitude of the gradient of $n({\mathbf k})$ at ${\mathbf k}={\mathbf k}_F$ is just $\frac{1}{2b}$.
Thus we can identify the gradient with $\frac{\xi}{\pi}$ (from Eq.~\ref{xi}), and therefore express the coherence length $\xi$ as
$\frac{\pi}{2b}$.

An alternative argument can be constructed using the Heisenberg uncertainty relationship connecting the mean-free-path to some measure of the smearing in ${\mathbf k}$ i.e. $\xi = \frac{\hbar}{2\Delta k}$. Recalling that $\hbar$ is unity
in atomic units, and identifying the full-width at half-maximum (FWHM) from the 
hyperbolic secant function (which is the derivative of the fitted $\tanh$ function)
to be approximately $1.76b$, we can write,
\begin{eqnarray}
\xi &=& \frac{1}{2\Delta k} \\
        &=& \frac{1}{2 \times 1.76b} \\
        &=& \frac{1}{3.52b}.
\end{eqnarray}
These two estimates of $\xi$ differ by a factor of $\sim 2$.

The results of the fits presented in Fig.~5 of the manuscript are given in Table~\ref{fittable}.

\begin{table}[ht]
  \begin{ruledtabular}
     \begin{tabular}{c c c c}
     Direction & FS sheet &  position '$k_F$' & smearing $b$  \\
     &  & $[\frac{2\pi}{a}]$ & $[\frac{2\pi}{a}]$ \\
     \\
      \hline
    $[100]$ & 1 & 0.42 & 0.14 \\
    $[100]$ & 2 & 0.64 & 0.14 \\
    $[100]$ & 3 & 0.79 & 0.11 \\
    $[110]$ & 1 & 0.45 & 0.14 \\
    $[110]$ & 2 & 0.81 & 0.14 \\
    $[110]$ & 3 & 1.22 & 0.13 \\
     \end{tabular}
  \end{ruledtabular}
  \caption{Fitted parameters to the $\tanh$ functions in Fig.5 of the paper. Note that for the $[110]$ direction, the location
of the third sheet is outside the first Brillouin zone.}
  \label{fittable}
\end{table}

\section*{Impact of resolution}

\begin{figure}
        \begin{center}
        \includegraphics[trim={0cm 1cm 0cm 1cm},angle=0,width=1.00\linewidth]{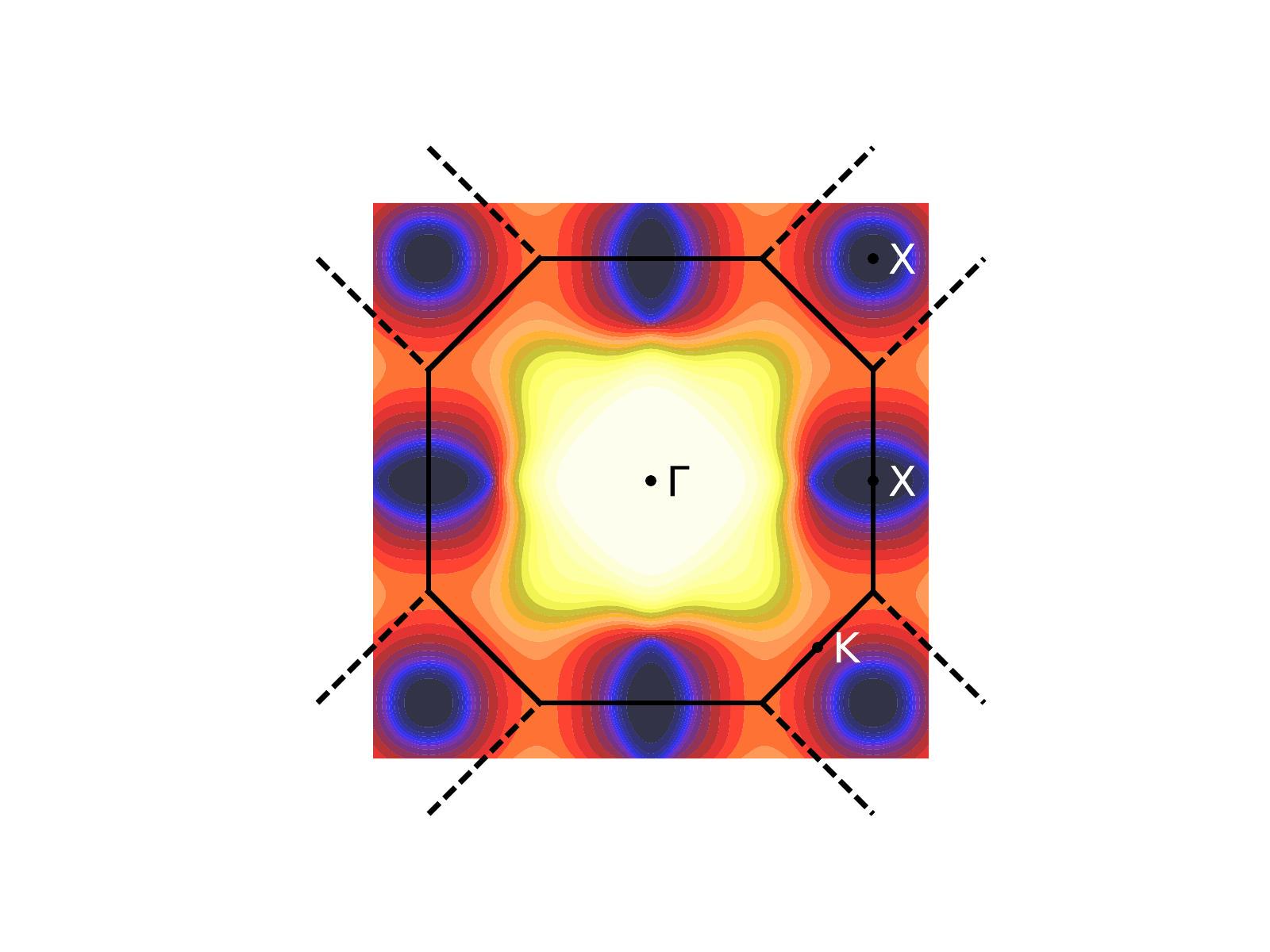}
        \includegraphics[trim={0cm 1cm 0cm 1cm},angle=0,width=1.00\linewidth]{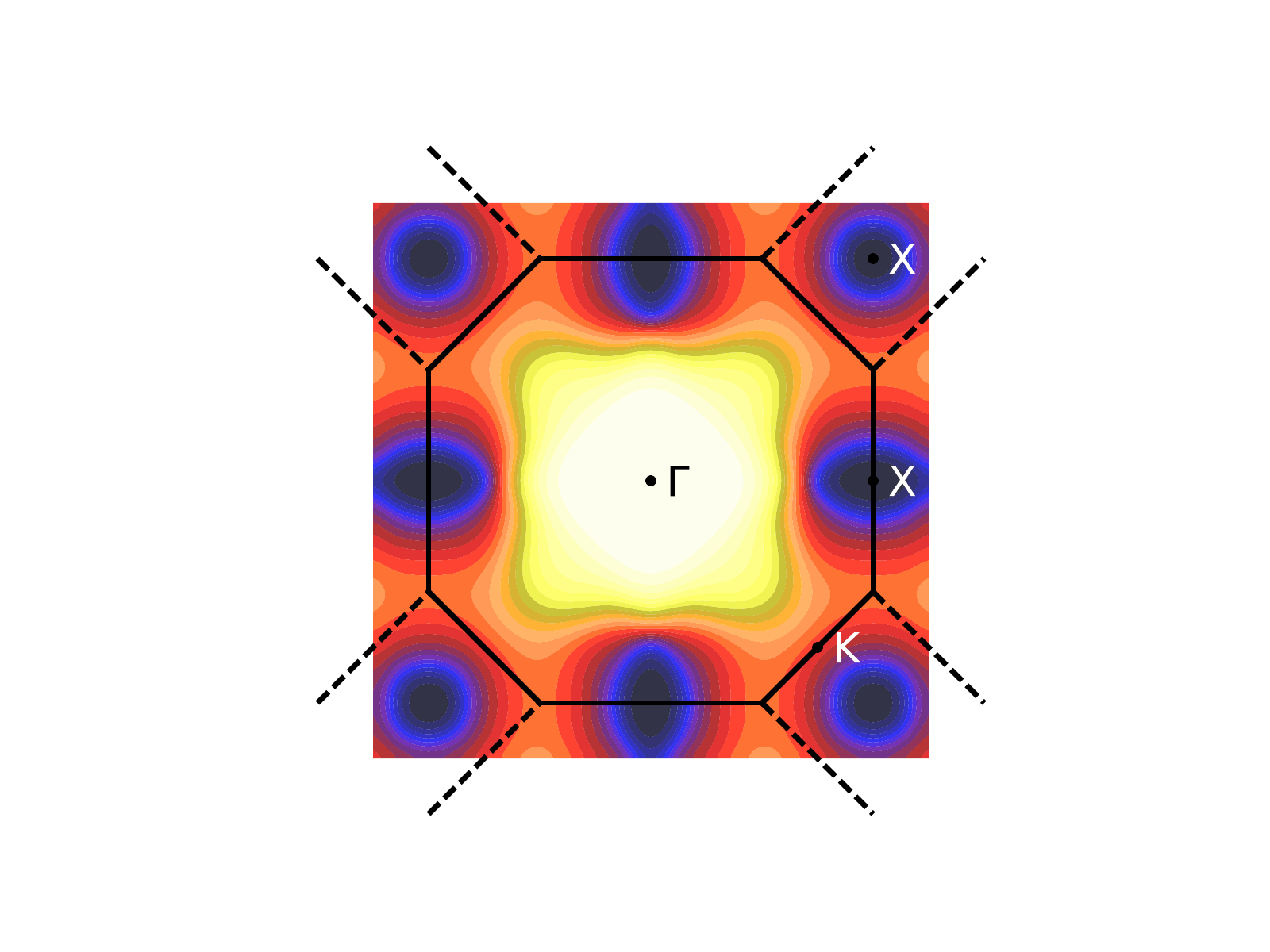}
        \end{center}
\caption{The impact of experimental resolution is relatively small, as can be seen from the $n({\mathbf k})$ in the
(001) plane through $\Gamma$ from the KKR-CPA-DLM calculation with (bottom) and without (top) convolution with the
experimental resolution.}
        \label{resolution}
\end{figure}

\begin{figure}
        \begin{center}
        \includegraphics[angle=0,width=1.00\linewidth]{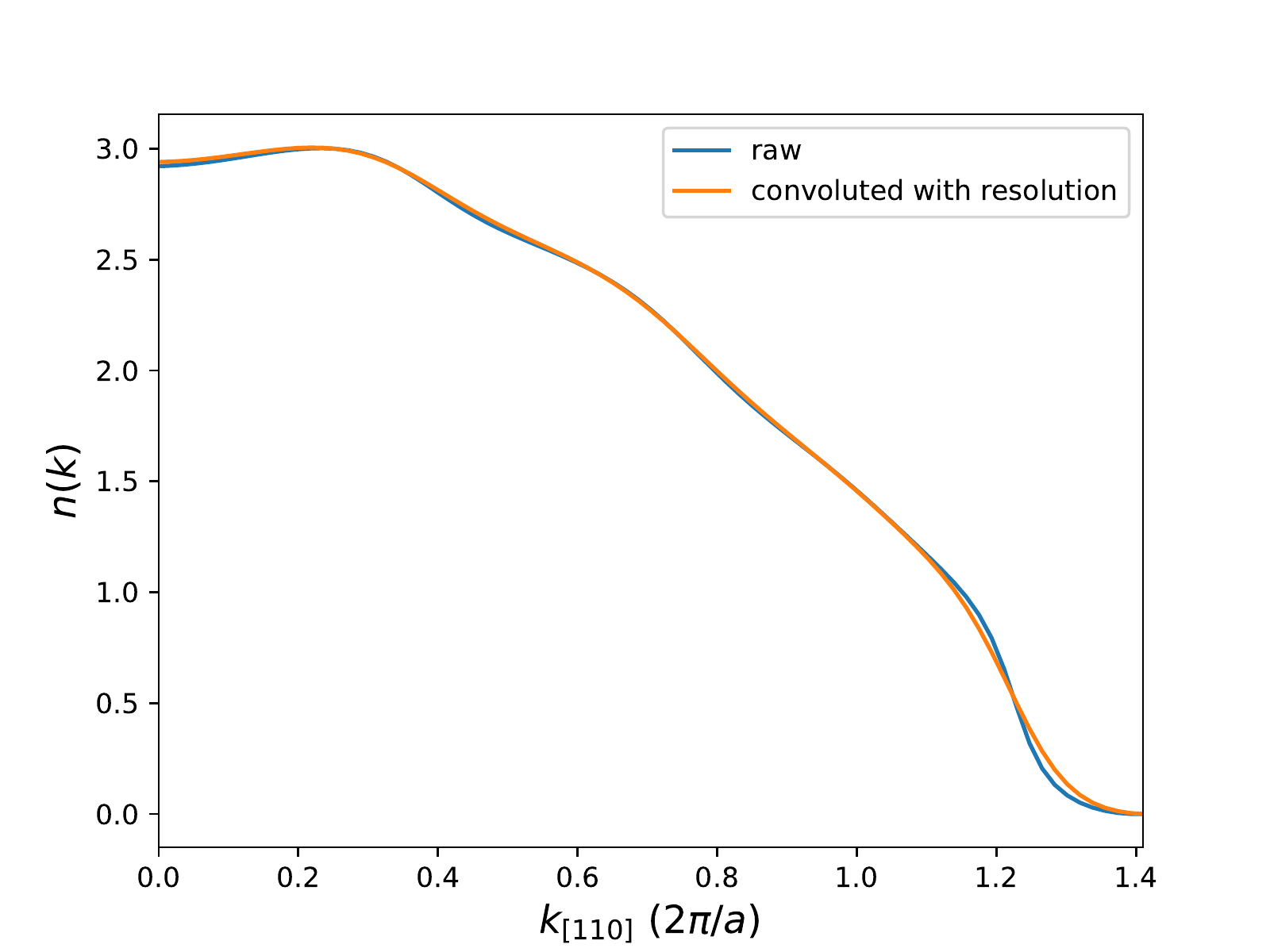}
        \end{center}
\caption{Showing the impact of experimental resolution along the direction $\Gamma$---$X$ in the [110] direction.}
        \label{impactofresolution}
\end{figure}

The $n({\mathbf k})$ is already very smeared due to the extreme compositional chemical disorder, and therefore the impact of
additional smearing due to the finite experimental resolution (FWHM $\sim$ 0.1 a.u.) is in fact rather small. This
is demonstrated in Fig.~\ref{resolution} where $n({\mathbf k})$ obtained from the KKR-CPA-DLM calculation is shown
with and without convolution with the experimental resolution. In Fig.~\ref{impactofresolution}, a cut in the
[110] direction is shown, both with and without convolution to illustrate how small the additional resolution smearing
actually is.
However, to take account fully of the (small) effect of convolution with the experimental
resolution, its impact on Eq.~\ref{tanh} for different smearings $b$ was investigated, 
so that the values extracted from the fits to the experimental data could be
corrected for the finite experimental resolution.

\section*{DLM calculations}

\begin{figure}
        \begin{center}
        \includegraphics[angle=0,width=1.00\linewidth]{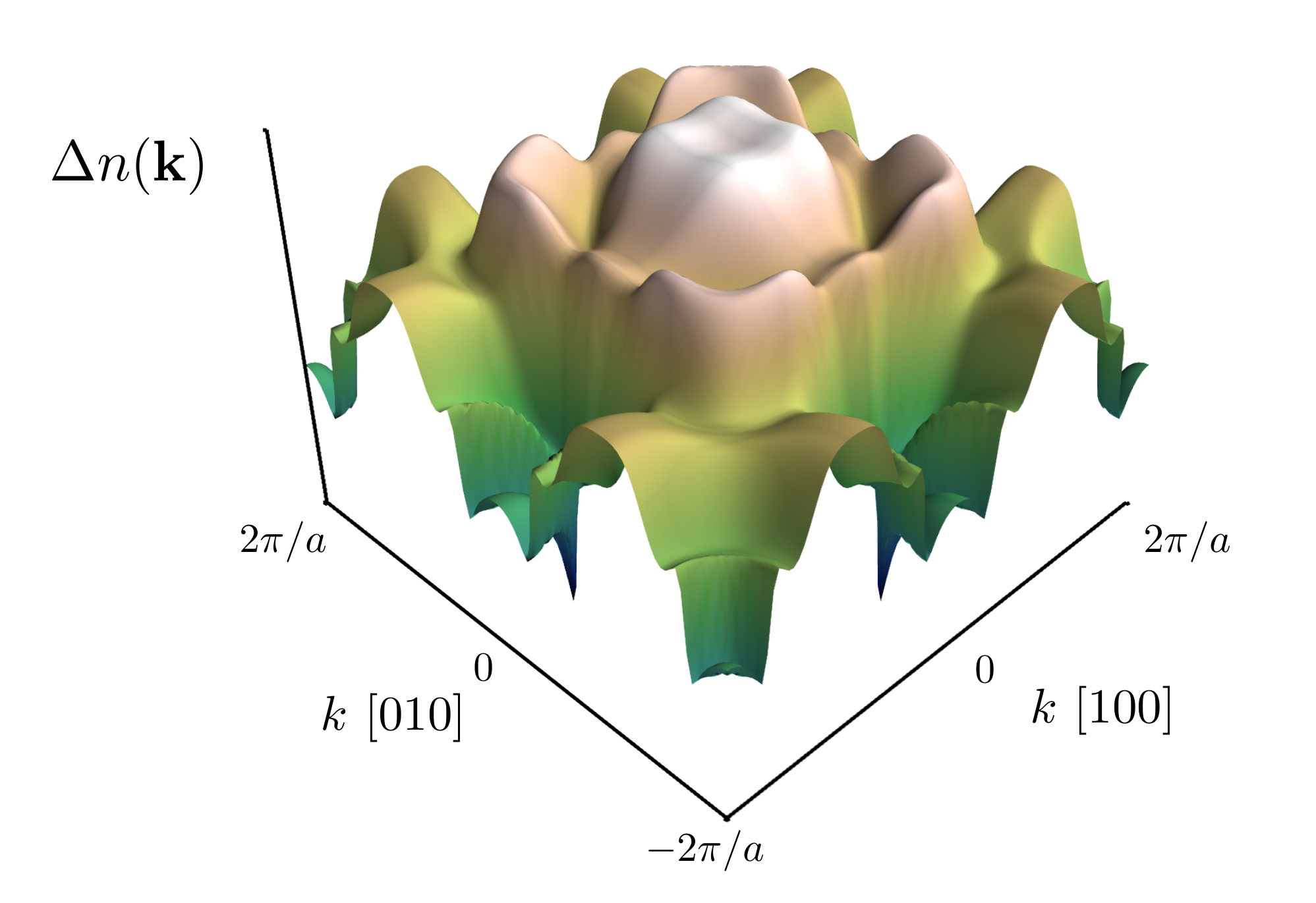}
        \end{center}
\caption{Difference between the $n({\mathbf k})$ obtained for a nonmagnetic and paramagnetic DLM state
(on a (001) plane through $\Gamma$). The structure in the difference appearing around the location of the
Fermi surface discontinuities shows that $n({\mathbf k})$ is even more smeared
when spin disorder is included. The maximum difference is almost 2\%.  }
        \label{dlm}
\end{figure}

With a Curie temperature of approximately 120K \cite{kao:11,jin:16}, it is to be expected that the
``disordered local moment'' (DLM) model would provide a better description of the electronic 
structure
of NiFeCoCr at room temperature. KKR-CPA calculations were performed for both a 
non-magnetic and a DLM state which converged to a state with zero moments on Ni,Co and Cr.
The difference between the occupancies in the (001) plane through the
$\Gamma$ point is shown in Fig.~\ref{dlm}. Quantitatively, the difference is not insignificant
(almost 2\%) and the DLM distribution is more smeared (as evidenced by the structure in the
difference distribution around the Fermi surface discontinuities), consistent with the additional
spin disorder. 

\section*{Comparison of experiment with KKR-CPA-DLM and Elk SQS supercell}

\begin{figure}
        \begin{center}
        \includegraphics[trim={0cm 1cm 0cm 1cm},angle=0,width=1.00\linewidth]{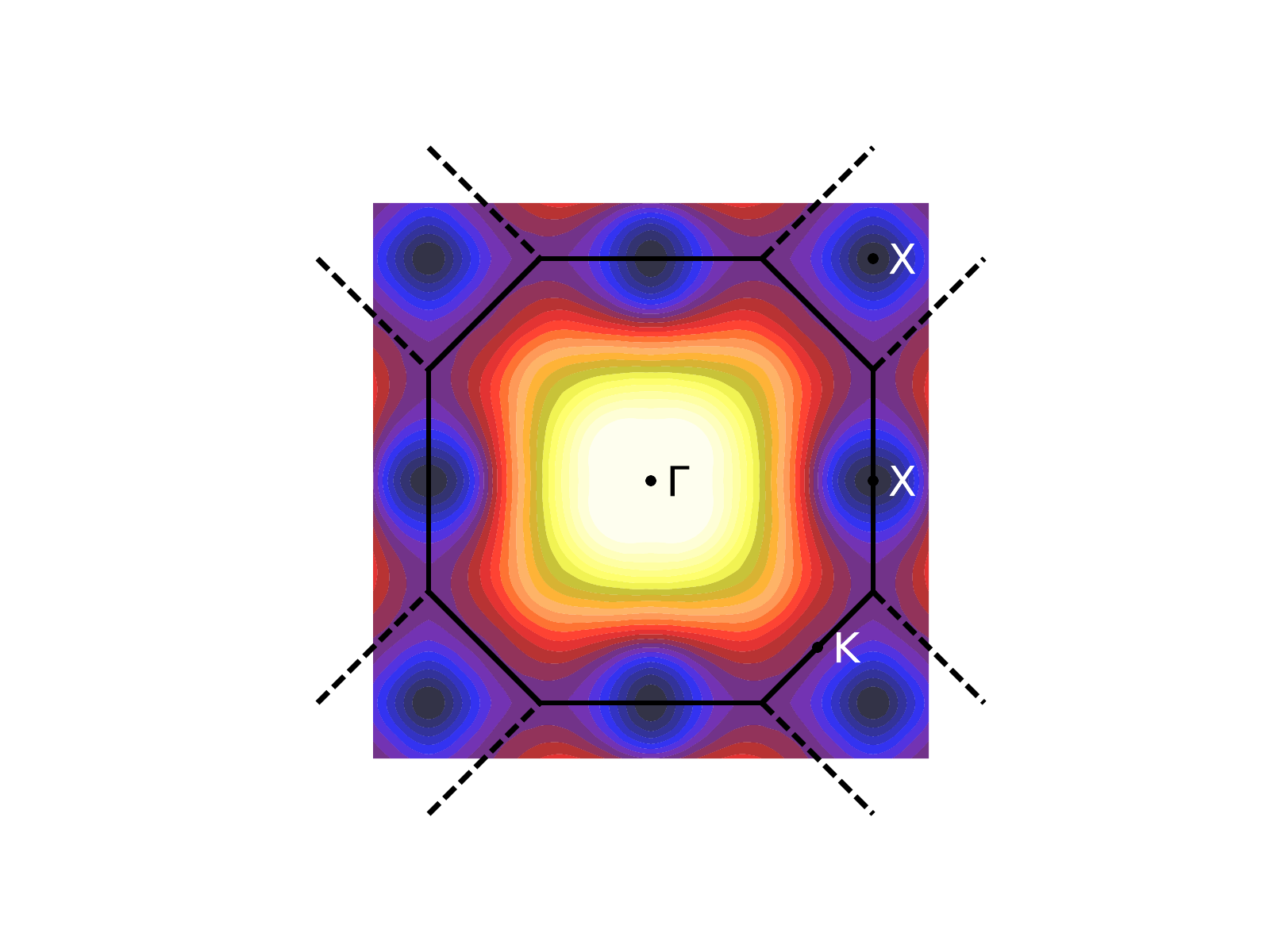}
        \includegraphics[trim={0cm 1cm 0cm 1cm},angle=0,width=1.00\linewidth]{Supp_Figs/001_kkrcpadlm.pdf}
        \includegraphics[trim={0cm 1cm 0cm 1cm},angle=0,width=1.00\linewidth]{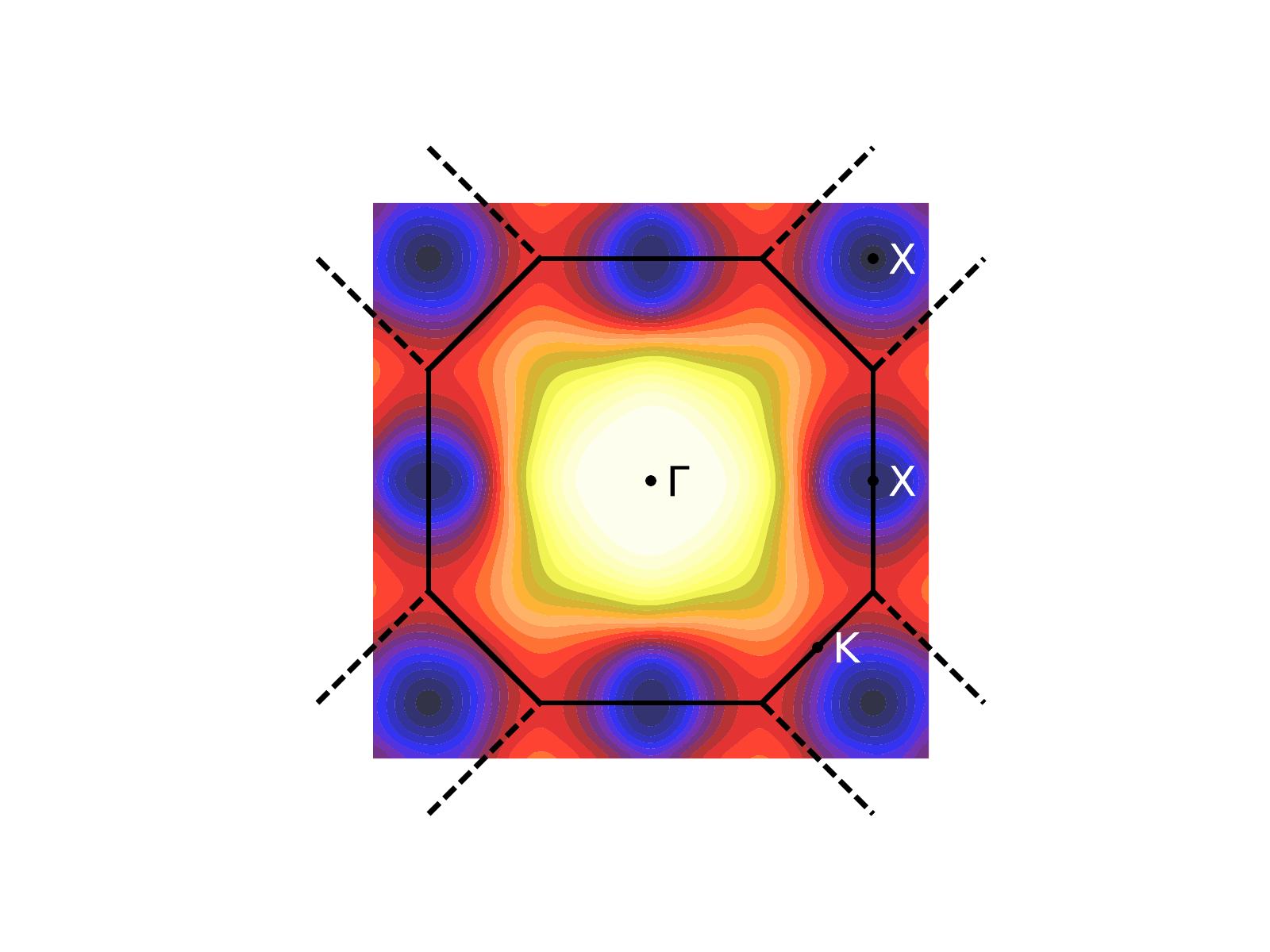}
        \end{center}
\caption{$n({\mathbf k})$ in a (001) plane through $\Gamma$ for the (top) experimental data (middle) KKR-CPA-DLM and
(bottom) Elk SQS supercell. The theoretical data have been convoluted with the experimental resolution.}
        \label{001planes}
\end{figure}

The $n({\mathbf k})$ in a (001) plane through $\Gamma$ are shown in Fig.~\ref{001planes} for the
experimental data (top), the KKR-CPA-DLM calculation (middle) and and Elk special quasirandom structure
(SQS) 32-atom supercell (bottom). Interestingly, the degree of smearing in the experimental data appears to be 
in between that of the KKR-CPA-DLM and the SQS supercell.

\begin{figure}
        \begin{center}
        \includegraphics[angle=0,width=1.00\linewidth]{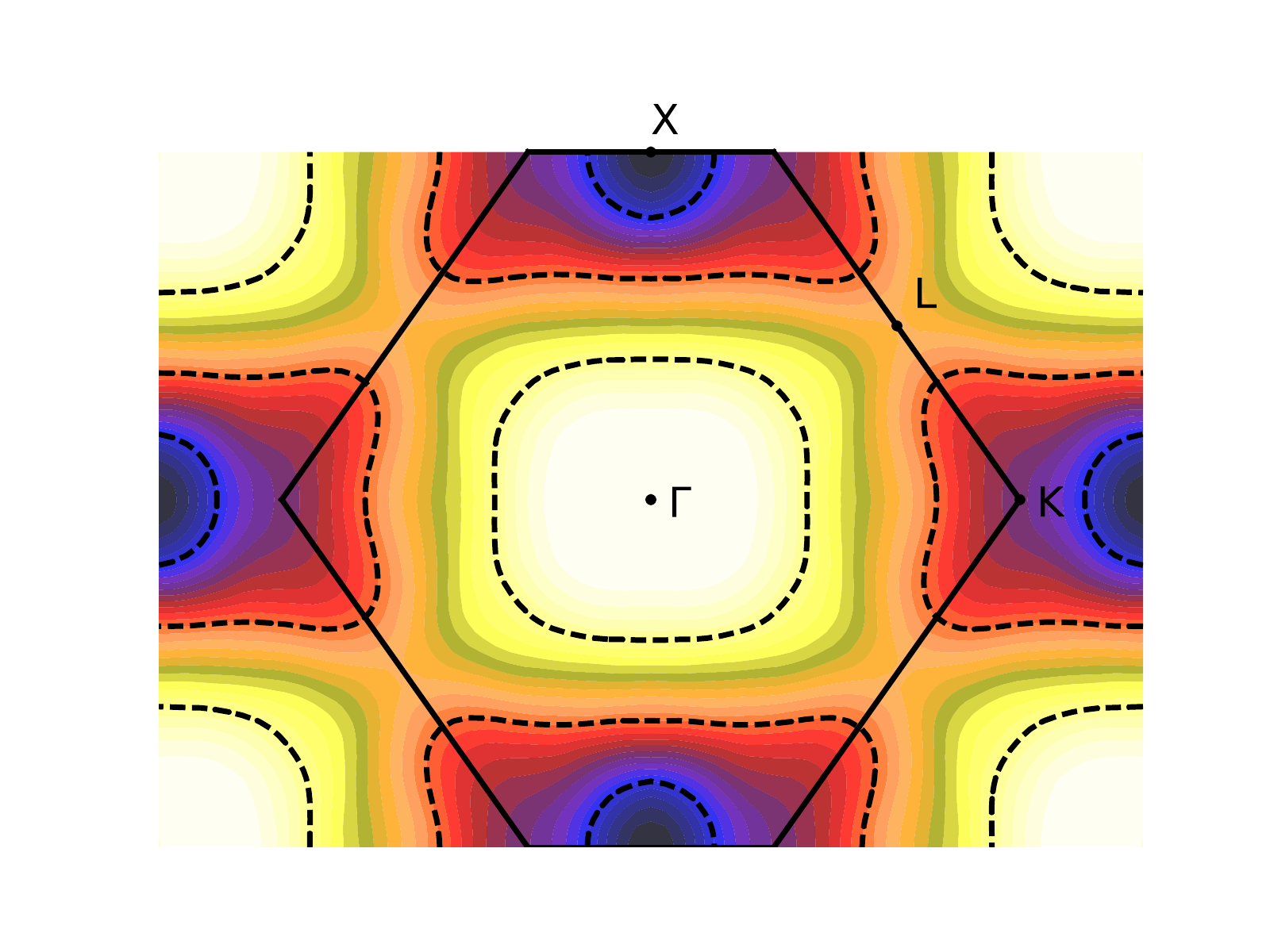}
        \includegraphics[angle=0,width=1.00\linewidth]{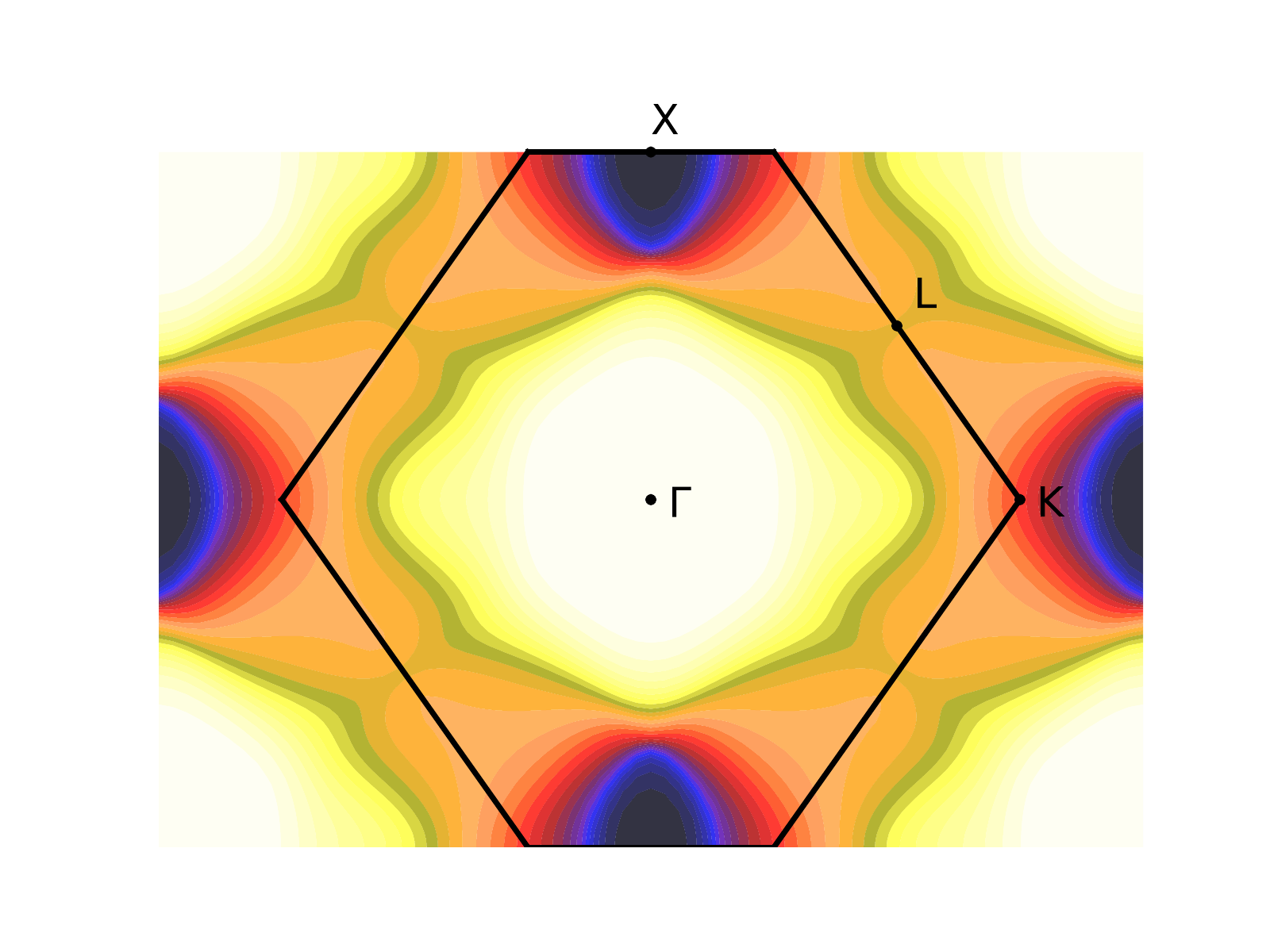}
        \includegraphics[angle=0,width=1.00\linewidth]{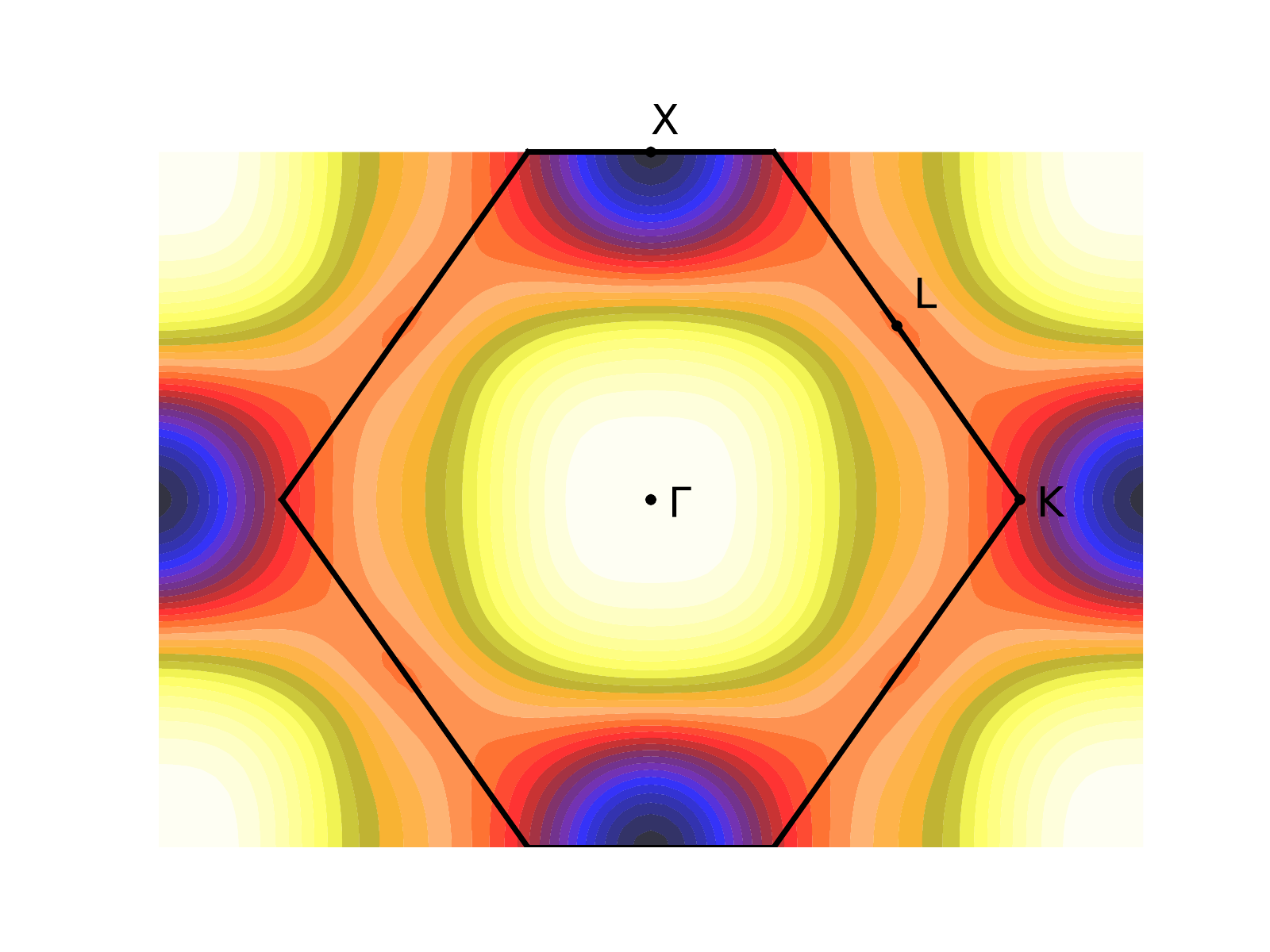}
        \end{center}
\caption{$n({\mathbf k})$ in a (110) plane through $\Gamma$ for the (top) experimental data (middle) KKR-CPA-DLM and
(bottom) Elk SQS supercell. The theoretical data have been convoluted with the experimental resolution. The dashed lines on
the experimental data indicate the isodensity contours i.e. shows the three Fermi surface sheets.}
        \label{110planes}
\end{figure}

\begin{figure}
        \begin{center}
        \includegraphics[angle=0,width=1.00\linewidth]{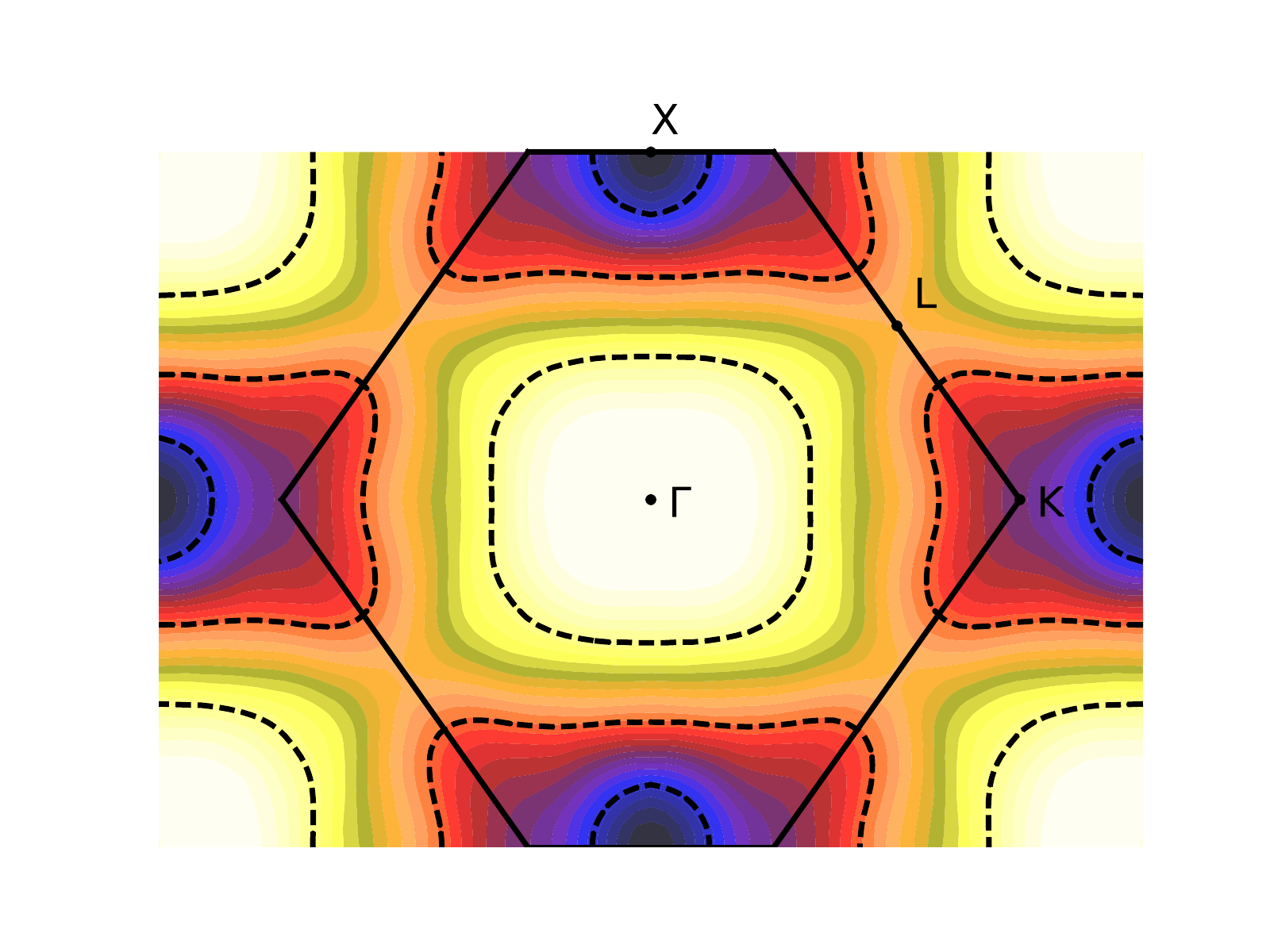}
        \end{center}
\caption{The same experimental data as shown in Fig.~\ref{110planes} but with contours at the slightly different isodensities coming from the
step fitting procedure (0.45,1.48 and 2.57, compared to 0.5,1.5 and 2.5).}
        \label{110shift}
\end{figure}

In Fig.~\ref{110planes}, these are shown for a (110) plane through the $\Gamma$ point.
In Fig.~\ref{110shift}, the same experimental data are plotted as in Fig.~\ref{110planes}, but the contours are at the slightly different isodensities coming from the
step fitting procedure (0.45,1.48 and 2.57, compared to 0.5,1.5 and 2.5), showing that the predicted Fermi surface topology is very similar.

\bibliography{supp}